\documentclass[12pt]{article}

\usepackage[margin = 1in]{geometry}
\usepackage{amsgen,amsmath,amstext,amsbsy,amsopn,amssymb,amsthm}
\usepackage[dvips]{graphicx}
\usepackage[colorlinks=true,linkcolor=blue,citecolor=blue,urlcolor=blue]{hyperref}
\usepackage{color}
\usepackage{xcolor}
\usepackage[numbers]{natbib}
\usepackage{ulem}
\usepackage{dsfont}
\usepackage{float}
\usepackage{subcaption}
\usepackage{setspace}
\usepackage[titlenumbered,ruled,linesnumbered]{algorithm2e}
\usepackage{algpseudocode}
\usepackage{booktabs}
\usepackage{multirow}
\usepackage{makecell} 

\usepackage{mdframed} 
\usepackage{enumitem} 
\usepackage{filecontents} 

\floatstyle{plain}
\newfloat{guidelinebox}{htbp}{loc}
\floatname{guidelinebox}{Box}

\newcommand{\one}{\mathds{1}}
\newcommand{\cee}{\text{CEE}}
\newcommand{\mee}{\text{MEE}}

\newcommand{\var}{\text{Var}}

\newcommand{\eo}{\text{EO}}
\newcommand{\aeo}{\text{AEO}}
\newcommand{\rank}{\text{rank}}
\newcommand{\aaa}{\text{AA}}
\newcommand{\bA}{\bar{A}}
\newcommand{\ba}{\bar{a}}
\newcommand{\RR}{\mathbb{R}}
\newcommand{\PP}{\mathbb{P}}
\newcommand{\II}{\mathbb{I}}
\newcommand{\tp}{\tilde{p}}
\newcommand{\pto}{\stackrel{p}{\to}}
\newcommand{\dto}{\stackrel{d}{\to}}
\newcommand{\w}{\text{w}}

\newcommand{\cH}{\mathcal{H}}
\newcommand{\tcH}{\tilde{\mathcal{H}}}
\newcommand{\tL}{\tilde{L}}

\newcommand{\tc}{\tilde{c}}
\newcommand{\cT}{\mathcal{T}}
\newcommand{\cP}{\mathcal{P}}
\newcommand{\smee}{\text{sMEE}}
\newcommand{\sate}{\text{sATE}}
\newcommand{\loc}{\texttt{loc}}


\usepackage{xr}
\usepackage[capitalize, nameinlink, sort]{cleveref}
\Crefname{thm}{Theorem}{Theorems}
\Crefname{cor}{Corollary}{Corollary}
\Crefname{lem}{Lemma}{Lemmas}
\Crefname{asu}{Assumption}{Assumptions}
\Crefname{rmk}{Remark}{Remarks}
\Crefname{defn}{Definition}{Definitions}
\Crefname{thmlisti}{Theorem}{Theorems}
\Crefname{asulisti}{Assumption}{Assumptions}
\crefname{guidelinebox}{box}{boxes} 
\Crefname{guidelinebox}{Box}{Boxes} 

\usepackage{thmtools}
\declaretheorem[
    name=Theorem,
    Refname={Theorem,Theorems}]{thm}

\declaretheorem[
    name=Assumption,
    Refname={Assumption,Assumptions}]{asu}
\declaretheorem[
    name=Remark,
    Refname={Remark,Remarks}]{rmk}

\usepackage{xpatch}
\xapptocmd\normalsize{%
 \abovedisplayskip=4pt plus 2pt minus 2pt
 \belowdisplayskip=4pt plus 2pt minus 2pt
}{}{}

\title{Micro-randomized Trials with Categorical Treatments: Causal Effect Estimation and Sample Size Calculation}
\author{Jeremy Lin, Tianchen Qian}
\date{Department of Statistics, University of California, Irvine}

\def\spacingset#1{\renewcommand{\baselinestretch}%
{#1}\small\normalsize} \spacingset{1}

\begin{document}

\maketitle

\spacingset{1.2}

\begin{abstract}

Micro-randomized trials (MRTs) are widely used to assess the marginal and moderated effect of mobile health (mHealth) treatments delivered via mobile devices. In many applications, the mHealth treatments are categorical with multiple levels such as different types of message contents, but existing analysis and sample size calculation methods for MRTs only focus on binary treatment options (i.e., prompt vs. no prompt). We extended the causal excursion effect definition and the weighted and centered least squares estimator to MRTs with categorical treatments. Furthermore, we developed a sample size formula for comparing categorical treatment levels, and proved the type I error and power guarantee under working assumptions. We conducted extensive simulations to assess type I error and power under assumption violations, and we provided practical guidelines for using the sample size formula to ensure adequate power in most real-world scenarios. We illustrated the proposed estimator and sample size formula using the HeartSteps MRT.

\vspace{2em}

\end{abstract}

\newpage

\spacingset{1.9}


\section{Introduction}
\label{sec:introduction}

Mobile health (mHealth) interventions aim to promote healthy behavior change by delivering interventions through mobile devices such as smartphones and wearable devices. Treatment in the form of reminders of push notifications, text messages, or audible pings can be delivered to participants to enable healthy behavior change. Micro-randomized trial (MRT) is an experimental design commonly used to inform the development and optimization of mobile health interventions \citep{dempsey2015randomised,liao2016sample}. In an MRT, each participant is repeatedly randomized among intervention options over numerous decision points throughout the trial. This repeated randomization enables causal inference for the marginal and moderated effects of the intervention to learn whether, when, and in what context an intervention option works.

Intervention options refer to all the possible actions that may be delivered at a decision point in an mHealth intervention. Typically, there are more than two intervention options, and this variety arises from the application of different behavioral or interventional theories and variations in framing intervention messages. For example, in HeartSteps, an mHealth intervention app for promoting physical activity, the intervention options include push notifications suggesting either walking or stretching activities, as well as the option of no push notification \citep{klasnja2019efficacy}. In Drink Less, an mHealth intervention app for reducing harmful drinking, the intervention options include messages phrased in various ways, along with the option of no push notification \citep{bell2023notifications}. These types of intervention options, which can take more than two forms, will be referred to as ``categorical treatments'' in this paper. Assessing the differental effects among various intervention options helps optimize the intervention on a more granular level, providing insights not just on whether delivering the intervention is better than not delivering it, but also on which specific intervention option is most effective in a given context. However, existing literature on assessing causal effects for MRTs primarily focuses on binary intervention options (treat vs. not treat) \citep{boruvka2018assessing,qian2021estimating}.

We address this gap by proposing an estimator for the causal effects of categorical treatments using data from MRTs. We focus on the additive causal excursion effect, the primary quantity of interest for MRTs with continuous proximal outcomes. This effect captures the contrast in the potential outcome under two deviations from the MRT treatment policy in a given context \citep{boruvka2018assessing}. We generalize the weighted and centered least squares method proposed by \citep{boruvka2018assessing} to accommodate categorical treatments. We establish the asymptotic normality of the estimator even if a nuisance outcome regression model is misspecified. This robustness is crucial given the high-dimensional nature of the history information, which often leads to model misspecification. We note that an earlier version of this estimator was used in \citet{klasnja2019efficacy}, an MRT with a fixed randomization probability. However, we provide a more general form of the estimator that accommodates randomization probabilities dependent on past history.

Furthermore, we develop a formula for determining the sample size needed to detect a pre-specified differential effect among treatment options, expressed as constraints on linear combinations of the marginal causal excursion effects for the active treatment options. We prove that under specific working assumptions, the sample size formula guarantees the desired power to detect a pre-specified effect while controlling for type I error. Extensive simulation studies demonstrate that certain assumption violations do not impact power, and for assumptions where violations do impact power, we discuss the consequences. We also provide practical guidelines for using the sample size formula. This work extends the sample size calculation literature for MRTs with binary treatment options to enable more granular comparisons among treatment options, which is particularly useful in the planning stage of MRTs \citep{liao2016sample, cohn2023sample}.

\Cref{sec:definition} defines the notation and the causal excursion effect for categorical treatments. \Cref{sec:estimator} introduces the estimator for the causal excursion effect. \Cref{sec:sample-size-formula} provides the sample size formula and practical guidelines for its use. \Cref{sec:application} includes a case study illustrating the estimator and the sample size formula using data from the HeartSteps and Drink Less MRTs.
\Cref{sec:discussion} concludes with a discussion.
A comphresive simulation study for the estimator and the sample size formula is presented in the Supplementary Material \ref{A-sec:simulation-estimator} and \ref{A-sec:simulation-sample-size-formula} due to space limit.

\section{Definition and Assumptions}
\label{sec:definition}

\subsection{Notation}

Consider an MRT that includes data from $n$ individuals, each in the trial for $T$ decision points where treatments are randomized. Variables without the subscript $i$ represent observations from a typical individual. Let $A_t$ denote the treatment assignment at decision point $t$. $A_t$ can take categorical value $A_t \in \{0, 1, \cdots, K\}$, with 0 being the reference level (e.g., no treatment). Let $X_t$ be the contextual information for the individual collected after decision point $t-1$ and up to decision point $t$; $X_1$ includes baseline covariates. Let $I_t \in X_t$ be the availability indicator, where $A_t = 0$ deterministically if $I_t = 0$ (for example, if the individual is driving). The overbar is used to denote a sequence of variables up to a decision point; for instance, $\bA_t = (A_1, \ldots, A_t)$. Information observed up to decision point $t$ is denoted by $H_t = (X_1, A_1, X_2, A_2, \cdots, X_{t-1}, A_{t-1}, X_t) = (\bar{X}_t, \bA_{t-1})$. The randomization probability for $A_t$ can depend on $H_t$ and is denoted by $p_t(k | H_t) = P(A_t = k \mid H_t)$ for $0 \leq k \leq K$. The observed data for a typical individual are represented by $O = (X_1, A_1, \cdots, X_T, A_T, X_{T + 1})$. We assume that the data from different individuals are independent and identically distributed draws from an unknown distribution $\mathcal{P}$. We consider settings where the proximal outcome $Y_{t,\Delta}$ for the treatment at decision point $t$ may be measured over a time window of subsequent $\Delta$ decision points; here $\Delta$ is a fixed positive integer. Specifically, let $Y_{t,\Delta} = y(X_{t+1}, A_{t+1}, \cdots, X_{t + \Delta -1}, A_{t + \Delta -1}, X_{t + \Delta})$ for a given function $y(\cdot)$. We will use $Y_t$ to denote $Y_{t,\Delta = 1}$, the most commonly used proximal outcome in practice. Through this paper, all expected values are taken with respect to $\mathcal{P}$, unless stated otherwise.

For an arbitrary function $f(\cdot)$ of the generic observed data O, We define $\mathbb{P}_n f(O)$ as the sample average $\frac{1}{n} \sum_{i = 1}^n f(O_i)$, where $O_i$ denotes the $i$-th individual's data. We use $\bar{0}_m$ to denote the $1 \times m$ row vector with all entries equal to 0. We use $\RR^{m \times n}$ to denote the space of $m\times n$ real matrices and $\RR^m$ to denote the space of $m \times 1$ real column vectors. We use $\one(\cdot)$ to denote the indicator function. We use $\II_p$ to denote the $p\times p$ identity matrix. For a positive integer $n$, We use $[n]$ to denote the set $\{1,2,\ldots,n\}$. For a vector $\alpha$ and a vector-valued function $f(\alpha)$, let $\dot{f}(\alpha) := \partial f(\alpha) / \partial \alpha^T$ denote the matrix where the $(i,j)$-th entry is the partial derivative of the $i$-th entry of $f$ with respect to the $j$-th entry of $\alpha$. We use $\otimes$ to denote the Kronecker's product between two matrices (or vectors): for $B \in \RR^{d_1 \times d_2}$ and $C \in \RR^{d_3 \times d_4}$, $B \otimes C \in \RR^{d_1d_3 \times d_2d_4}$ is defined as
\begin{align*}
        B \otimes C = \begin{bmatrix}
        b_{1, 1}C &b_{1, 2}C  &\cdots &b_{1, d_2}C\\
        b_{2, 1}C &b_{2, 2}C  &\cdots &b_{2, d_2}C\\
        \vdots  &\vdots   &\vdots &\vdots\\
        b_{d_1, 1}C &b_{d_1, 2}C  &\cdots &b_{d_1, d_2}C\\
    \end{bmatrix}, 
\end{align*}
where $b_{i,j}$ is the $(i,j)$-th entry of $B$. We define $0/0 = 0$.

\subsection{Potential outcomes and causal excursion effect}

To define treatment effects, we adopt the potential outcomes framework (Rubin, 1974: Robins, 1986). For an individual, let $X_t(\ba_{t-1})$ be the information that would have been observed had that individual been assigned to treatment sequence $\ba_{t-1}$. The potential outcomes for a typical individual are $\{X_1, A_1, X_2(a_1), A_2, \cdots, A_T, X_{T+1}(\ba_T): \text{for all } 0 \leq a_s \leq K, s \in [T] \}$.
Denote by $Y_{t, \Delta}(\ba_{t + \Delta -1})$ the potential outcome of $Y_{t, \Delta}$. The potential outcome of the history information $H_t$ is denoted by $H_t(\ba_{t-1}) = \{X_1, A_1, X_2(A_1), A_2, X_3(\ba_2), \cdots X_t(\ba_{t-1})\}$. For any $k \in [K]$ and $t \in [T]$, we define the Causal Excursion Effect (CEE) of treatment $k$ at time $t$ as:
\begin{align}
    \cee_{tk}(S_t) & = E\{Y_{t,\Delta}(\bA_{t-1}, k, \bar{0}_{\Delta - 1})\mid S_t(\bA_{t-1}), I_t(\bA_{t-1}) = 1\} \nonumber \\
    & ~~~~~~~~~~~~~~~~~~~~~ - E\{(Y_{t,\Delta}(\bA_{t-1}, 0, \bar{0}_{\Delta - 1})\mid S_t(\bA_{t-1}), I_t(\bA_{t-1}) = 1)\}, \label{eq:cee-def}
\end{align}
where $\bar{0}_{\Delta - 1}$ represent a length $(\Delta - 1)$ vector of zeros. Expression \eqref{eq:cee-def} represents the difference between the expected outcome under two excursions: getting treatment $k$ at time $t$ and no treatment for the next $\Delta -1$ time points, versus no treatment at time $t$ and no treatment for the next $\Delta - 1$ time points. In both excursions, the treatment assignment up to time $t$ ($\bA_{t-1}$) is random and follows the treatment protocol of the micro-randomized trial. In \eqref{eq:cee-def}, $S_t \subset H_t$ is a set of effect modifiers of interest, and researchers' different choice of $S_t$ leads to different interpretation of CEE. For example, setting $S_t = \emptyset$ assesses a marginal effect that averages over all possible moderators. Setting $S_t = X_t$ assesses effect moderation by current covariates $X_t$. The expectations in \eqref{eq:cee-def} are conditional on a decision point being available ($I_t(\bA_{t-1}) = 1$), because only the effect at the available decision points can be nonparametrically identified and is of scientific interest (it is inappropriate to deliver an intervention when the individual is unavailable).

\subsection{Identification of parameters}

To express the causal excursion effect using the observed data, we make the following assumptions.
\begin{asu}
    \label{asu:causal-assumptions} 
    \normalfont
    \begin{itemize}
        \item[(a)] (SUTVA.) The observed data are equal to the potential outcome under the observed treatment assignment, and one's potential outcomes are not affected by others' treatment assignments. Specifically, $X_t = X_t(\bA_{t-1})$ and $Y_{t, \Delta} = Y_{t,\Delta}(\bA_{t +\Delta-1})$ for all $t \in [T]$.
        \item[(b)] (Positivity.) $P(A_t = k \mid H_t, I_t = 1) > 0$ almost surely for all $0 \leq k \leq K$ and $t \in [T]$.
        \item[(c)] (Sequential ignorability.) For $t \in [T]$, the potential outcomes $\{ X_{t+1}(\ba_t), \cdots, X_{T+1}(\ba_T) : \text{for all } 0 \leq a_s \leq K, s \in [T] \}$ are independent of $A_t$ conditional on $H_t$.
    \end{itemize}
\end{asu}

In a MRT, $A_t$ is sequentially randomized according to known treatment probabilities, $\{p_{t}(k|H_t): 0 \leq k \leq K \}$, which guarantees positivity and sequential ignorability. SUTVA may be violated if the treatment assigned to one individual influences the response of others, such as in a social network. This paper does not consider such possibilities. Under \Cref{asu:causal-assumptions}, CEE in \eqref{eq:cee-def} can be expressed in terms of the observed data distribution as
\begin{align}
    \cee_{tk}(S_t) & = E\bigg[E\bigg\{Y_{t,\Delta} \prod_{j = t + 1}^{t + \Delta - 1} \frac{\one(A_j = 0)}{p_j(0|H_j)} \bigg| A_t = k, H_t\bigg\} \bigg| S_t, I_t = 1 \bigg] \nonumber \\
    & ~~~~~~~~~~~~~~~ - E\bigg[ E \bigg\{Y_{t,\Delta} \prod_{j = t + 1}^{t + \Delta - 1} \frac{\one(A_j = 0)}{p_j(0|H_j)} \bigg| A_t = 0, H_t \bigg\} \bigg| S_t, I_t = 1 \bigg]. \label{eq:cee-identification}
\end{align}
Equation \eqref{eq:cee-identification} follows immediately from Lemma A.1 in \citet{qian2021estimating}.


\section{Estimating CEE for Categorical Treatment}
\label{sec:estimator}

We consider estimating a finite-dimensional unknown parameter $\beta$ in a parametric model for CEE. Specifically, suppose
\begin{align}
    \cee_{tk}(S_t) = f_t(S_t)^T \beta_k \text{ for all } t \in [T] \text{ and } k \in [K], \label{eq:CEE-parametric-model}
\end{align}
where $f_t(\cdot)$ is a prespecified vector-valued function and $\beta_k \in \RR^p$. The $p\times T$ matrix $(f_1,f_2,\ldots,f_T)$ is required to be of rank $p$ in order for $\beta_k$ to be identified. Let $\beta = (\beta_1^T, \ldots, \beta_K^T)^T \in \RR^{Kp}$. This model allows for nonlinear effects: e.g., $f_t(S_t)$ could include basis functions of $t$.

We propose an estimator for $\beta$ that generalizes the weighted and centered least squares approach in \citet{boruvka2018assessing} to the categorical treatment setting. For each $t$, let $\tp_t(k | S_t)$ be arbitrary functions of  $S_t$ such that $\tp_t(k | S_t) > 0$ for $0 \leq k \leq K$ and $\sum_{k=0}^K \tp_t(k | S_t) = 1$. Define $J_t := \frac{\tp_t(A_t|S_t)}{p_t(A_t|H_t)} \prod_{j = t + 1}^{t + \Delta - 1} \frac{\one(A_j = 0)}{p_j(0|H_j)}$. Let $C_k(A_t) := \one(A_t = k) - \tp_t(k | S_t)$. Let $g_t(H_t)^T\alpha$ be a working model for $E(Y_{t,\Delta} \mid H_t, I_t = 1)$.
The proposed estimator $\hat\beta$ is obtained by solving for $(\hat\alpha,\hat\beta)$ that satisfies $\PP_n m(\hat\alpha,\hat\beta) = 0$, where
\begin{align}
    m(\alpha,\beta) := 
    \sum_t^T I_t J_t \bigg\{Y_{t,\Delta} - \sum_{k = 1}^K C_k(A_t) f_t(S_t)^T \beta_k - g_t(H_t)^T \alpha \bigg\} 
    \begin{bmatrix}
        g_t(H_t) \\
        C_1(A_t) f_t(S_t)\\
        \vdots\\
        C_K(A_t) f_t(S_t)
    \end{bmatrix}. \label{eq:ee}
\end{align}
The first term in $J_t$ is similar to a stabilized inverse probability weight in marginal structural models \citep{robins2000marginal}, which is included because of the marginal aspect of \eqref{eq:cee-def}. The numerator probabilities, $\tp_t(k|S_t)$, do not affect the consistency of $\hat\beta$ but should be chosen as close to $p_t(k|H_t)$ as possible for efficiency. For example, one may fit a multinominal logistic regression with response $A_t$ and predictor $S_t$ to predict $\tp_t(k|S_t)$. The second term in $J_t$ is an inverse probability weight due to the $\bar{0}_{\Delta-1}$ in the potential outcomes in \eqref{eq:cee-def}. The centering of $A_t$ in $C_k(A_t)$ produces orthogonality between the estimation of $\beta$ and the estimation of the nuisance parameter $\alpha$, and this leads to the robustness of $\hat\beta$ even when $g_t(H_t)^T\alpha$ is misspecified (see \Cref{thm:CAN}). This robustness property is desirable because $H_t$ can be high-dimensional in an MRT with a large number of time points, making it very difficult to model $E \{Y_{t,\Delta} \mid H_t, I_t = 1 \}$ correctly. We establish the asymptotic property of $\hat\beta$ in \Cref{thm:CAN}, which is proven in the Supplementary Material \ref{A-sec:proof-thm-CAN}.

\begin{thm}[Asymptotic normality.]
    \label{thm:CAN}
    Suppose the parametric CEE model \eqref{eq:CEE-parametric-model} and \Cref{asu:causal-assumptions} hold.
    Suppose that randomization probabilities $\{p_t(k | H_t)\}_{0 \leq k \leq K, t \in [T]}$ are known. Suppose $\beta^0$ is the true value of $\beta$ under the data-generating distribution $\mathcal{P}$. Let $\dot{m}$ be the derivative matrix of $m(\alpha, \beta)$ with respect to $(\alpha, \beta)$. Let $(\hat\alpha, \hat\beta)$ be the solution to $\mathbb{P}_n m(\alpha, \beta) = 0$. Under regularity conditions, we have the following.
    \begin{itemize}
        \item[(i)] There exists $\alpha'\in\RR^q$ such that $\hat\alpha \pto \alpha'$ and $\sqrt{n}(\hat\beta - \beta^0) \dto N(0, M^{-1}\Sigma M^{-1,T})$ where $M := \sum_{t=1}^T E ( I_t J_t D_t D_t^T )$, $\Sigma := \sum_{t=1}^T \sum_{s=1}^T E \{ I_t I_s J_t J_s r_t(\alpha',\beta^0) r_s(\alpha',\beta^0) D_t D_s^T \}$, $r_t(\alpha,\beta) := Y_{t,\Delta} - \sum_{k=1}^K C_k(A_t) f_t(S_t)^T \beta_k - g_t(H_t)^T \alpha$, and $D_t := (C_1(A_t)f_t^T, C_2(A_t)f_t^T, \ldots, C_K(A_t)f_t^T )^T$. The explicit form of $\alpha'$ is given in \eqref{A-eq:thm1-proofuse4} in the Supplementary Material \ref{A-sec:proof-thm-CAN}.
        \item[(ii)] $\hat{M}^{-1} \hat\Sigma \hat{M}^{-1, T}$ is a consistent estimator for the asymptotic variance $M^{-1}\Sigma M^{-1,T}$, with $\hat{M} := \sum_{t=1}^T \PP_n ( I_t J_t D_t D_t^T )$ and $\hat\Sigma := \sum_{t=1}^T \sum_{s=1}^T \PP_n \{ I_t I_s J_t J_s r_t(\hat\alpha,\hat\beta) r_s(\hat\alpha,\hat\beta) D_t D_s^T \}$.
        \item[(iii)] When $\{\tp_t(k | S_t)\}_{0 \leq k \leq K, t \in [T]}$ is estimated either parametrically or nonparametrically, the theorem conclusion still hold with the following modification: $\tp_t(k | S_t)$ replaced by its estimator $\hat\tp_t(k | S_t)$ in $\hat{M}^{-1} \hat\Sigma \hat{M}^{-1, T}$, and $\tp_t(k | S_t)$ replaced by the $L_2$-limit of $\hat\tp_t(k | S_t)$ in $M^{-1}\Sigma M^{-1,T}$.
    \end{itemize}
\end{thm}

\begin{rmk}
    \normalfont
    When the parametric model for CEE \eqref{eq:CEE-parametric-model} is correct, the choice of $\{\tp_t(k | S_t)\}_{0 \leq k \leq K, t \in [T]}$ only affects the asymptotic variance of $\hat\beta$ and does not affect the consistency of $\hat\beta$. However, when model \eqref{eq:CEE-parametric-model} is misspecified, differenct choices of $\{\tp_t(k | S_t)\}_{0 \leq k \leq K, t \in [T]}$ lead to different probability limits of $\hat\beta$. For example, for the immediate effect ($\Delta = 1$), suppose that for each $k \in [K]$ the analysis model asserts a constant marginal effect $\cee_{tk} (\emptyset) = \beta_k$ for all $t \in [T]$, while the true $\cee_{tk}(\emptyset)$ is not constant in $t$. In this case, if for each $0 \leq k \leq K$ we set $\tp_t(k|S_t)$ to be a constant over $t$, then each $\hat\beta_k$ converges in probability to (recall $Y_t := Y_{t,\Delta = 1}$)
    \begin{align*}
        \beta_k'  = \sum_{t = 1}^T \Big[E\{E(Y_t \mid H_t, A_t = k)\mid I_t = 1 \} -  E\{E(Y_t \mid H_t, A_t = 0)\mid I_t = 1 \}\Big] E(I_t) \Big/ \sum_{t=1}^T E(I_t),
    \end{align*}
    which further simplifies to 
    \begin{align*}
        \beta_k'  = \sum_{t = 1}^T \Big[E(Y_t \mid I_t = 1, A_t = k) - E(Y_t \mid I_t = 1, A_t = 0)\Big] E(I_t) \Big/ \sum_{t=1}^T E(I_t)
    \end{align*}
    if the randomization probability $p_t(k|H_t)$ is a constant in $t$.
\end{rmk}



\section{Sample Size Formula with Categorical Treatment}
\label{sec:sample-size-formula}

\subsection{Hypothesis, Test Statistic, and Rejection Region that Controls Type I Error}
\label{subsec:hypothesis-test-statistic-rejection-region}

For sample size calculation, we focus on the most common setting in planning an MRT, where the interest is in the immediate, marginal CEE (i.e., $\Delta = 1$ and $S_t = \emptyset$ in \eqref{eq:cee-def}), and the randomization probability at each decision point is either constant or dependent only on $t$. In this setting, we define $p_t(k) := P(A_t = k \mid I_t = 1) \equiv P(A_t = k \mid H_t, I_t = 1)$, and we define the marginal excursion effect (MEE) as the CEE with $\Delta = 1$ and $S_t = \emptyset$ (recall $Y_t := Y_{t,\Delta=1}$):
\begin{align*}
    \mee_k(t) := E\{Y_t(\bA_{t-1}, k)\mid I_t(\bA_{t-1}) = 1\} - E\{Y_t(\bA_{t-1}, 0)\mid I_t(\bA_{t-1}) = 1\} \text{ for } t \in [T], k \in [K].
\end{align*}
A positive $\mee_k(t)$ indicates that the treatment level $k$ at time $t$ is effective compared to no treatment (assuming that a larger $Y_t$ is better).

Let $\mee_{1:K}(t) = (\mee_1(t), \mee_2(t), \ldots, \mee_K(t))^T$. We consider testing for 
\begin{align}
    \cH_0: L \times \mee_{1:K}(t) = 0 \text{ for all } t \in [T] \text{ v.s. } \cH_1: L \times \mee_{1:K}(t) \neq 0 \text{ for some } t \in [T], \label{eq:null-and-alternative}
\end{align}
where $L \in \RR^{v \times K}$ for some integer $\nu$. The researcher can specify different $L$ to represent different scientific hypotheses. For example, if we set $v = 2$ and $L$ to be a $2 \times K$ matrix with $(1,1)$ and $(2,2)$ entries being $1$ and all other entries being 0, then this gives $\cH_0: \mee_1(t) = \mee_2(t) = 0$ for all $t \in [T]$, a simultaneous test for null effect (compared to the reference level) of both treatment options 1 and 2. If we set $v = 1$ and $L$ be a $1 \times K$ matrix with $(1,1)$-th entry being $1$, $(1,2)$-th entry being $-1$, and all other entries being 0, then this gives $\cH_0: \mee_1(t) = \mee_2(t)$ for all $t \in [T]$, a test for the equivalence of treatment levels 1 and 2.

Nonparametrically testing for all deviations from $L \times \mee_{1:K}(t) = 0$ for each $t$ will result in low power for detecting specific  deviations. Instead we consider a working model for $\mee_k(t)$ that is parametric in $t$. Specifically, we consider the working assumption 
\begin{align}
    \mee_k(t) = f_t^T\beta_k \text{ for } t \in [T] \text{ and } k \in [K], \label{eq:working-assumption-on-MEE}
\end{align}
where $\beta_k \in \RR^p$ and $f_t$ is a predefined $p$-dimensional vector-valued function that only depends on $t$. We focus on calculating sample size for detecting target alternatives that satisfy \eqref{eq:working-assumption-on-MEE}. The particular form of \eqref{eq:working-assumption-on-MEE} that is scientifically plausible is typically decided in consultation with the scientific team. For instance, if the scientific team anticipates that the intevention's effect will not substantially change over time, a constant-in-time $\mee_k(t)$ with $f_t = 1$ would be an appropriate choice. If the scientific team anticipates that the intervention's effect might start near zero, increase gradually early in the study, and potentially decrease later, a quadratic-in-time $\mee_k(t)$ with $f_t = (1, t, t^2)^T$ could be appropriate. We note that although the sample size formula will be derived under the parametric working assumption \eqref{eq:working-assumption-on-MEE}, we show via extensive simulation studies that the sample size formula can yield adequate power even under certain violations of the working assumption (see \Cref{subsec:practical-guideline} and the Supplementary Material \ref{A-sec:simulation-sample-size-formula}).

Under the working assumption \eqref{eq:working-assumption-on-MEE}, we derive in the Supplementary Material \ref{A-subsec:proof-L-tilde-matrix} that the null and the alternative hypotheses in \eqref{eq:null-and-alternative} are equivalent to
\begin{align}
    \tcH_0: \tL \beta = 0 \quad \text{v.s.} \quad \tcH_1: \tL \beta \neq 0, \label{eq:null-and-alternative-tilde}
\end{align}
where $\tL:= L \otimes \II_p$ and $\beta := (\beta_1^T, \ldots, \beta_K^T)^T$. We construct a Wald-type test statistic based on the estimator $\hat\beta$ proposed in \Cref{sec:estimator}. Specifically, let $g_t$ be a predefined $q$-dimensional vector-valued function that only depends on $t$, and let $g_t^T\alpha$ be a working model for $E(Y_t \mid I_t = 1)$. Let $(\hat\alpha, \hat\beta)$ be the estimator that solves $\PP_n m(\alpha,\beta) = 0$, with the definition of $m(\alpha,\beta)$ modified by replacing $f_t(S_t)$, $g_t(H_t)$, $\tp_t(k|S_t)$, $p_t(k|H_t)$, and $J_t$ by $f_t$, $g_t$, $p_t(k)$, $p_t(k)$, and 1, respectively.
The Wald-type test statistic is defined as
\begin{align}
        \cT = n (\tL\hat\beta)^T (\tL\hat{M}^{-1} \hat{\Sigma} \hat{M}^{-1, T} \tL^T)^{-1} (\tL \hat\beta), \label{eq:def-test-stat}
\end{align}
with $\hat{M}$ and $\hat\Sigma$ defined in \Cref{thm:CAN}(ii) but with the same modifications as those for $m(\alpha,\beta)$.

Under \eqref{eq:working-assumption-on-MEE} and $\tcH_0$, it follows from \Cref{thm:CAN} that the large sample distribution of $\cT$ is $\chi^2_l$, a chi-squared distribution with $l = \rank(L)$ degrees of freedom. Thus, a hypothesis test that uses the critical value of the chi-squared distribution will have nominal type I error control asymptotically. To correct the downward bias of the sandwich estimator $\hat{M}^{-1} \hat{\Sigma} \hat{M}^{-1, T}$ when the sample size $n$ is small, we instead use the critical value from a scaled $F$-distributions because $\frac{n - q - l}{l (n - q - l)} \cT$ approximately follows $F_{l, n - q - l}$, an $F$-distribution with degrees of freedom $(l, n - q - l)$ \citep{pan2002small}. Therefore, the rejection region of a test with significance level $\eta$ is 
\begin{align}
    \bigg\{ \cT : \frac{n - q - l}{l (n - q - l)} \cT > F^{-1}_{l, n - q -l} (1 - \eta) \bigg\}, \label{eq:rejection-region}
\end{align}
where $F^{-1}_{l, n - q - l}(1 - \eta)$ is the $(1 - \eta)$-quantile of $F_{l, n - q -l}$. We further incorporated another small sample correction in \cite{mancl2001covariance} by replacing $\hat{\Sigma}$ in (\ref{eq:def-test-stat}) with an adjusted version using the ``hat'' matrix.


\subsection{Sample Size Formula that Guarantees Power Under Working Assumptions}
\label{subsec:sample-size-formula}

Suppose the goal is to find the sample size of the MRT to have at least $(1-b)$ power under some target alternative hypothesis. We define the target alternative using a standardized effect size so that it is more interpretable to domain scientists. Define 
\begin{align}
    \bar\sigma^2 := \frac{1}{T} \sum_{t= 1}^T E\left\{\var(Y_t \mid A_t, I_t = 1)\right\}. \label{eq:sigma_bar}
\end{align}
Define the standardized MEE (sMEE) as
\begin{align}
    \smee_k(t) := \mee_k(t) / \bar\sigma. \label{eq:def-smee}
\end{align}
Consider the target alternative with a parametric model on $\smee_k(t)$ that is compatible with \eqref{eq:working-assumption-on-MEE}: for given $\gamma := (\gamma_1^T,\ldots,\gamma_K^T)$ where each $\gamma_k \in \RR^p$, let 
\begin{align}
    \cH_1^*(\gamma): \smee_k(t) = f_t^T \gamma_k \text{ for } t \in [T] \text{ and } k \in [K]. \label{eq:target-alternative}
\end{align}
The target alternative $\cH_1^*(\gamma)$ is equivalent to $\mee_k(t) = \bar\sigma f_t^T \gamma_k$ for $t\in[T]$ and $k\in[K]$, and it implies $\tL \beta = \bar\sigma \tL\gamma$, a simple hypothesis within the composite $\tcH_1$ in \eqref{eq:null-and-alternative-tilde}. 
Under $\cH_1^*(\gamma)$, it follows from \Cref{thm:CAN} that the test statistic $\cT$ follows approximately $\chi^2_l(\lambda(n))$, a non-central chi-squared distributions with $l = \rank(L)$ degrees of freedom and non-centrality parameter $\lambda(n) := n\bar\sigma^2 (\tL\gamma)^T (\tL M^{-1} \Sigma M^{-1, T} \tL^T)^{-1} (\tL\gamma)$.
Here, $M$ and $\Sigma$ are defined in \Cref{thm:CAN}(i) with the same modifications as those for $m(\alpha,\beta)$ described above display \eqref{eq:def-test-stat}. To improve small sample performance, we again use the $F$-distribution approximation: the scaled test statistic $\frac{n - q - l}{l (n - q - l)} \cT$ approximately follows $F_{l, n - q - l;\lambda(n)}$, a non central $F$-distribution with degrees of freedom $(l, n - q - l)$ and non-centrality parameter $\lambda(n)$ \citep{pan2002small}. In order to have at least $1-b$ power under $\cH_1^*(\gamma)$, the sample size $n$ must satisfy 
\begin{align*}
    P\bigg\{\frac{n-q-l}{l(n-q-l)} \cT > F^{-1}_{l, n-q-l}(1 - \eta)\bigg\} \geq 1-b, \text{ where }
    \frac{n-q-l}{l(n - q - l)} \cT \sim F_{l, n - q - l;\lambda(n)}.
\end{align*}
Therefore, the required sample size is the smallest integer $n$ such that 
\begin{align}
    1 - F_{l, n-q-l;\lambda(n)}\Big\{ F^{-1}_{l, n-q-l}(1 - \eta)\Big\}\geq 1 - b. \label{eq:ss_formula}
\end{align}

Calculating $n$ from \eqref{eq:ss_formula} requires knowing $\lambda(n)$, but $\lambda(n)$ depends on the data generating distribution $\cP$ through $M$ and $\Sigma / \bar{\sigma}^2$, which are typically unknown during trial planning. To make it feasible to compute $n$ from \eqref{eq:ss_formula}, we make the following working assumptions (WA) about $\cP$. For completeness, the working assumption \eqref{eq:working-assumption-on-MEE} is also included below as (WA-a).
\begin{itemize}
    \item[(WA-a)] (Parametric MEE.) For each $k\in[K]$, there exists $\beta_k \in \RR^p$ such that $\mee_k(t) = f_t^T\beta_k$ for $t \in [T]$, where $f_t \in \RR^p$ is known.
    \item[(WA-b)] (Parametric mean outcome.) There exists $\alpha \in \RR^q$ such that $E(Y_t \mid I_t = 1) = g_t^T \alpha$ for $t \in [T]$, where $g_t \in \RR^q$ is known.
    \item[(WA-c)] (Homogeneous variance.) There exists $\sigma^2 > 0$ such that $\sigma^2 = \var(Y_t\mid A_t = k, I_t = 1)$ for $t \in [T]$ and $k \in [K]$.
    \item[(WA-d)] (Known availability probability.) Suppose $E(I_t)= \tau(t)$ for $t \in [T]$, where the value of $\tau(t)$ is known for $t \in [T]$.
    \item[(WA-e)] (No serial correlation in the outcome.) Suppose that for every $(t,s)$ pair with $1 \leq s < t \leq T$, $E(Y_t \mid  I_s = 1, I_t = 1, A_t, A_s, Y_{s} )$ is constant with respect to $Y_s$.  
    \item[(WA-f)] (Exogenous availability process.) Suppose $I_t$ is independent of prior treatment or prior outcomes, i.e $I_t \perp \{A_s, Y_s : 1 \leq s < t\}$ for all $t \in [T]$.
\end{itemize}

Under these working assumptions, we derived a tractable sample size formula presented in \cref{alg:ss-calculator}, where the non-centrality parameter $\lambda$ has an explicit form. \Cref{thm:sample-size-formula} establishes the type I error control and power guarantee of the sample size formula. The derivation of the tractable sample size formula and the proof of \Cref{thm:sample-size-formula} is in the Supplementary Material \ref{A-sec:proof-thm-sample-size-formula}.

\normalem 
\begin{algorithm}[htbp]
    \caption{Sample size calculator for MRT with categorical treatment}
    \label{alg:ss-calculator}
    \SetKwInOut{Input}{Input}
    \SetKwInOut{Output}{Output}
    \Input{$K$: total number of active treatment options (excluding the reference level $0$); \newline
    $T$: total number of decision points per participant; \newline
    $p_t(k) = P(A_t = k \mid I_t = 1)$ for $t \in [T], k \in [K]$: randomization probability; \newline
    $\tau(t) = E(I_t)$ for $t \in [T]$: probability of being available; \newline
    $f_t \in \RR^p$ for $t \in [T]$: the vector in the parametric working model for $\mee_k(t)$; \newline
    $\gamma_k \in \RR^p$ for $k\in [K]$: the coefficients for $f_t$ in the target alternative \eqref{eq:target-alternative}; \newline
    $g_t \in \RR^q$ for $t \in [T]$: the vector in the parametric working model for $E(Y_t \mid I_t = 1)$; \newline
    $L \in \RR^{v \times K}$: the linear contrast matrix in defining $\cH_0$ and $\cH_1$ in \eqref{eq:null-and-alternative}; \newline
    $\eta$: desired type I error; \newline
    $1-b$: desired power.}
    $l \gets \text{rank}(L)$ \\
    $\gamma \gets (\gamma_1^T, \ldots, \gamma_K^T)^T$ \\
    $P_t \gets$ a $K \times K$ matrix whose $(k_1,k_2)$-th entry equals $p_t(k_1)\{1-p_t(k_1)\}$ when $k_1 = k_2$ and $-p_t(k_1)p_t(k_2)$ when $k_1 \neq k_2$\\
    $V \gets \sum_{t=1}^T \tau(t) P_t \otimes (f_t f_t^T)$ \\
    $\text{finished} \gets \text{False}$ \\
    $n \gets 9$ \Comment{Some small $n$ to start with} \\
    \While{not finished}{
        $n \gets n+1$ \\
        $\lambda \gets n \gamma^T V^{-1} \gamma$ \\
        \If{ $F_{l, n-q-l;\lambda}\{F_{l, n-q-l}^{-1}(1-\eta)\} \leq b$ }{$\text{finished} \gets \text{True}$ \Comment{Found smallest $n$ to satisfy \eqref{eq:ss_formula}}}
    }
    \Output{$n$}
\end{algorithm}
\ULforem 

\begin{thm}[Type I error control and power guarantee under working assumptions.]
    \label{thm:sample-size-formula}
    Suppose the randomization probability at each decision point is either constant or dependent only on $t$. Consider the testing procedure for $\cH_0$ vs. $\cH_1$ in \eqref{eq:null-and-alternative} based on test statistic \eqref{eq:def-test-stat} and rejection region \eqref{eq:rejection-region}.
    \begin{itemize}
        \item[(i)] Suppose (WA-a) holds. Then the testing procedure has type I error rate approximately $\eta$.
        \item[(ii)] Suppose (WA-a)--(WA-f) holds. With $n$ calculated from \cref{alg:ss-calculator}, the testing procedure has power at least $1-b$ approximately under the target alternative $\cH_1^*(\gamma)$ in \eqref{eq:target-alternative}, assuming $\tL \gamma \neq 0$.
    \end{itemize}
\end{thm}

We discuss the implications of each of these working assumptions.

(WA-a) assumes that the researcher knows the functional form of $\mee_k(t)$ as a function of $t$. This is equivalent to the working assumption \eqref{eq:working-assumption-on-MEE} or the target alternative \eqref{eq:target-alternative}.

(WA-b) assumes that the researcher knows the functional form of the expected outcome $\eo(t) := E(Y_t \mid I_t = 1)$ as a function of $t$. Note that $E(Y_t \mid I_t = 1)$ averages over the past and current treatments ($A_1, A_2, \ldots, A_t$) and past outcomes ($Y_1, Y_2, \ldots, Y_{t-1}$), and thus $\eo(t)$ depends on the magnitude of current and delayed effects and serial correlation in the outcome.

(WA-c) assumes that the variance of the outcome is constant across all time points and all treatment levels. While this assumption may be implausible, our simulations studies will demonstrate that the sample size formula performs well even when (WA-c) is violated.

(WA-d) assumes that for each decision point, the researcher knows the probability of a participant being available. If (WA-f) is violated, $\tau(t)$ would then reflect the dependence of $I_t$ on previous treatments ($A_s$) and outcomes ($Y_s$)  for all $1 \leq s < t$.

(WA-e) assumes that the outcome is independent of previous outcomes when conditioned on previous treatments. This assumption is introduced to enable an analytic sample size formula. Although this assumption is generally unrealistic in most mobile health applications where outcomes for the same participant measured close in time are often correlated, our simulation studies will show that the sample size formula performs well even when (WA-e) is violated.

(WA-f) assumes that the availability indicator $I_t$ is independent of prior treatments and outcomes. The plausibility of this assumption depends on the study context. For example, (WA-f) is reasonable if a decision point is unavailable due to a technical glitch unrelated to the participant’s behavior or treatment. However, (WA-f) is violated if availability is influenced by participant burden, i.e., if a decision point is more likely to become unavailable under treatment delivery in recent decision points. Our simulation studies will demonstrate that the sample size formula performs well even when (WA-f) is violated. Note that (WA-f) is satisfied in MRTs without availability constraints because $I_t$ will always be 1.

\begin{rmk}[The role of $g_t$ in the sample size formula]
    \label{rmk:gt}
    \normalfont
    \Cref{thm:sample-size-formula} applies to the testing procedure that uses the same $g_t$ for control variables as specified in (WA-b), meaning the $g_t$ that accurately captures the functional form of $\eo(t) = E(Y_t \mid I_t = 1)$. This has two important implications. First, even though $g_t$ does not appear in \cref{alg:ss-calculator} except for its dimension $q$, correctly specifying $g_t$ for conducting the actual test after the MRT is crucial for ensuring power. An incorrect $g_t$ will inflate the variance of $\hat\beta$ and reduce the power of the test, as evidenced by simulation results in the Supplementary Material \ref{A-subsec:simulation-violate-b}. Second, in practice, after conducting the MRT, researchers often use a different and usually more powerful testing procedure by adjusting for additional control variables besides functions of $t$. This can reduce the variance of $\hat\beta$ and lead to more powerful tests. It is not possible to universally determine which of these implications has a more substantial impact on power, as this depends on the unknown data-generating distribution.
\end{rmk}

\subsection{Performance of the Sample Size Formula under Working Assumption Violations and Practical Guidelines}
\label{subsec:practical-guideline}

We performed extensive simulation studies to assess the performance of the sample size formula under ideal conditions when all working assumptions hold, and under scenarios where some working assumptions are violated. In this section, we summarize the performance of the formula in the simulation studies and provide practical guidelines for using the sample size formula. Detailed simulation results can be found in the Supplementary Material \ref{A-sec:simulation-sample-size-formula}.

First, we provide definitions necessary for summarizing the simulation findings. For given $k\in[K]$, define the standardized average treatment effect of treatment level $k$ as 
\begin{align}
    \sate_k := \frac{\sum_{t = 1}^T \smee_k(t) E(I_t)}{ \sum_{t = 1}^T E(I_t)}, \label{eq:ate_k}
\end{align}
the difference between two $\sate$ of treatment level $j$ and $k$ as
\begin{align}
    \Delta \sate_{jk} := \sate_k - \sate_j, \label{eq:delta-ate}
\end{align}
the average expected outcome as 
\begin{align}
    \aeo := \frac{\sum_{t = 1}^T  \eo(t) E(I_t)}{ \sum_{t = 1}^T E(I_t)} \equiv \frac{\sum_{t = 1}^T  E(Y_t \mid I_t = 1) E(I_t)}{ \sum_{t = 1}^T E(I_t)}, \label{eq:aeo}
\end{align}
and the average availability (AA) as 
\begin{align}
    \aaa := \frac{1}{T}\sum_{t = 1}^T \tau(t) \equiv \frac{1}{T}\sum_{t = 1}^T E(I_t). \label{eq:AA}
\end{align}
$\sate_k$ is the standardized MEE of treatment level $k$ (compared to treatment level 0) averaged over time and weighted by availability. $\aeo$ is expected outcome averaged over time and weighted by availability. $\aaa$ is $E(I_t)$ averaged over time. $\sate_k$, $\aeo$, and $\aaa$ summarize the magnitude of $\smee_k(t)$, $\eo(t)$, and $\tau(t)$, respectively. We will use ``pattern'' to refer to how $\smee_k(t)$, $\eo(t)$, and $\tau(t)$ vary over $t$ independently of their magnitude. As we will see, the magnitude and the pattern of $\smee_k(t)$, $\eo(t)$, and $\tau(t)$ impact the performance of the sample size formula in different ways.

We distinguish between two versions of each quantity: one corresponding to the true data-generating distribution (denoted by a superscript $*$) and the other corresponding to the input used in the sample size formula (denoted by a superscript ``w'' for ``working''). For instance, $\sate_k^*$ represents \eqref{eq:ate_k} with expectations calculated according to the true data-generating distribution $\cP$ (which is unknown outside of simulations), and $\sate_k^\w$ represents \eqref{eq:ate_k} with expectations calculated using the input to the sample size formula, assuming all working assumptions hold. Using this notation, (WA-a) corresponds to $\mee_k^\w(t)$ and $\mee_k^*(t)$ being of the same functional form, (WA-b) corresponds to $\eo^\w(t)$ and $\eo^*(t)$ being of the same functional form, and (WA-d) corresponds to $\tau^\w(t) = \tau^*(t)$.

Through the simulation studies presented in the Supplementary Material \ref{A-sec:simulation-sample-size-formula}, we found that the type I error rate is always at the desired level under arbitrary working assumption violations (even when (WA-a) is violated). The sample size formula yields adequately powered MRTs when all the working assumptions hold. The power performance under various working assumption violations is listed in \Cref{table:sample-size-performance}.

In summary, in order to achieve adequate power, one needs to (i) correctly specify the magnitude of MEE (i.e., $\sate_k^\w = \sate_k^*$) and the average availability (i.e., $\aaa^\w = \aaa^*$); (ii) use a constant $\mee_k(t)$ input (i.e., set $f_t = 1$); (iii) use a flexible $\eo(t)$ (e.g., set $g_t = (1,t,t^2)^T$). It helps to be conservative about $\sate$ and $\aaa$, i.e., use the lower end of a range of conjectured values, if such values are either based on prior studies or from domain knokwledge. Once these are satisfied, violating the working assumptions does not hurt the power, except when (WA-c) is violated in that $\var(Y_t \mid A_t)$ depends on $A_t$ and the treatment assignment is imbalanced (i.e., the MRT specifies different probabilities to assign different treatment levels). Practical guidelines for using the sample size formula to ensure adequate power under possible violations of the working assumptions is provided in \Cref{box:practical-guideline}.


\begin{table}[htbp]
    \caption{Sample size formula performance when working assumptions (WA) are violated.}
    \label{table:sample-size-performance}
    \begin{tabular}{|l|ll|l|c|}
    \hline
    WA-violated                 & \multicolumn{2}{l|}{Detail about the violation}                                                                     & Power             \\ \hline
    \multirow{4}{*}{(WA-a)}     & \multicolumn{1}{l|}{\multirow{2}{*}{$\mee_k(t)$ magnitude incorrect}}         & $\sate^\w > \sate^*$                & $\downarrow$      \\ \cline{3-4} 
                                & \multicolumn{1}{l|}{}                                                         & $\sate^\w  < \sate^*$               & $\uparrow$        \\ \cline{2-4} 
                                & \multicolumn{1}{l|}{\multirow{2}{*}{$\mee_k(t)$ pattern incorrect}}           & $\mee_k^\w(t) $ constant            & $\rightarrow$     \\ \cline{3-4} 
                                & \multicolumn{1}{l|}{}                                                         & $\mee_k^\w(t) $ Linear              & $\downarrow^*$    \\ \hline
    \multirow{4}{*}{(WA-b)}     & \multicolumn{1}{l|}{\multirow{2}{*}{$\eo(t)$ magnitude incorrect}}            & $\aeo^\w > \aeo^*$                  & $\rightarrow$     \\ \cline{3-4} 
                                & \multicolumn{1}{l|}{}                                                         & $\aeo^\w <\aeo^*$                   & $\rightarrow$     \\ \cline{2-4} 
                                & \multicolumn{1}{l|}{\multirow{2}{*}{$\eo(t)$ pattern incorrect}}              & $q^\w > q^*$                        & $\rightarrow$     \\ \cline{3-4} 
                                & \multicolumn{1}{l|}{}                                                         & $q^\w <q^*$                         & $\downarrow^*$    \\ \hline
    \multirow{4}{*}{(WA-c)}     & \multicolumn{1}{l|}{\multirow{2}{*}{$\var(Y_t \mid A_t)$  depends on $t$}}    & variance increases with time        & $\uparrow$        \\ \cline{3-4} 
                                & \multicolumn{1}{l|}{}                                                         & variance decreases with time        & $\uparrow$        \\ \cline{2-4} 
                                & \multicolumn{1}{l|}{\multirow{2}{*}{$\var(Y_t \mid A_t)$ depends on $A_t$}}   & same probability assignment         & $\rightarrow$     \\ \cline{3-4} 
                                & \multicolumn{1}{l|}{}                                                         & different probability assignment    & $\downarrow^*$    \\ \hline
    \multirow{4}{*}{(WA-d)}     & \multicolumn{1}{l|}{\multirow{2}{*}{$\tau(t)$ magnitude incorrect}}           & $\aaa^\w > \aaa^*$                  & $\downarrow$      \\ \cline{3-4} 
                                & \multicolumn{1}{l|}{}                                                         & $\aaa^\w < \aaa^*$                  & $\uparrow$        \\ \cline{2-4} 
                                & \multicolumn{1}{l|}{\multirow{2}{*}{$\tau(t)$ pattern incorrect}}             & $\aaa^\w$ constant                  & $\rightarrow$     \\ \cline{3-4} 
                                & \multicolumn{1}{l|}{}                                                         & $\aaa$ non constant                 & $\rightarrow^*$   \\ \hline
    \multirow{2}{*}{(WA-e)}     & \multicolumn{1}{l|}{\multirow{2}{*}{Serial Correlation}}                      & positive serial corellation         & $\rightarrow$     \\ \cline{3-4} 
                                & \multicolumn{1}{l|}{}                                                         & negative serial correlation         & $\rightarrow$     \\ \hline
    (WA-f)                      & \multicolumn{2}{l|}{endogeneous availability process}                                                               & $\rightarrow$     \\ \hline
    \end{tabular}
\end{table}

\begin{guidelinebox}[htbp]
    \caption{Practical guidelines for using the sample size formula.}
    \label{box:practical-guideline}
    \begin{mdframed}[linewidth=1pt, roundcorner=10pt, backgroundcolor=gray!10]
    \spacingset{1}
    \noindent \textbf{For $1-b$, $\eta$, $T$, $p_t(k)$, $L$:}
    \begin{itemize}[leftmargin=2em]
      \item Specify according to the MRT design and the scientific question of interest.
    \end{itemize}

    \noindent \textbf{For $f_t$ and $\gamma_k$:} (recall $\smee_k(t) = f_t^T \gamma_k$)
    \begin{itemize}[leftmargin=2em]
      \item Use a constant pattern ($f_t = 1$) unless strong prior knowledge about a specific form;
      \item Set $\gamma_k = \sate_k$ for a conjectured $\sate_k$ value;
      \item If having a range of conjectured $\sate_k$ values, use the lower bound to be conservative.
    \end{itemize}

    \noindent \textbf{For $g_t$:} (recall $\eo(t) = g_t^T\alpha$)
    \begin{itemize}[leftmargin=2em]
      \item $g_t$ must be at least as complex as $f_t$;
      \item Use a more flexible $g_t$ to be conservative (e.g., $g_t = (1,t,t^2)^T$)
    \end{itemize}

    \noindent \textbf{In addition, for MRT with availability considerations:}
    \begin{itemize}[leftmargin=2em]
      \item Use a constant $\tau(t)$ pattern unless strong prior knowledge about a specific form;
      \item Set $\tau(t) = \aaa$ for a conjectured $\aaa$ value;
      \item If having a range of conjectured $\aaa$ values, use the lower bound to be conservative.
    \end{itemize}

    \noindent \textbf{In addition, for MRT with imbalanced treatment assignment:} (i.e., if $p_t(0),p_t(1),\ldots,p_t(K)$ are not all equal)
    \begin{itemize}[leftmargin=2em]
      \item Increase the sample size by 10--20\% if the variability in the outcome may be affected by prior treatment.
    \end{itemize}
    \end{mdframed}
\end{guidelinebox}

\spacingset{1.9}


\section{Data Example}
\label{sec:application}

We illustrate the use of the CEE estimator and the calculation of sample size using the HeartSteps MRT \citep{klasnja2019efficacy}. We consider the activity suggestion intervention component, which consists of push notifications of two types: walking suggestions (i.e., a suggestion for brief walk activity, which we will refer to as treatment level 1), and anti-sedentary suggestions (i.e., a suggestion for stretching exercises to disrupt sedentary behavior, which we will refer to as treatment level 2). Treatment level 0 means no notification. Each participant was enrolled for 42 days, with 5 decision points per day at pre-specified times evenly spaced across the day, which roughly correspond to morning commute, lunch, mid-afternoon, evening commute, and after dinner. At a decision point, a participant was unavailable ($I_t = 0$) if they were currently walking, driving, or did not have internet connection on their phone. At an available decision point ($I_t = 1$), a participant was randomized with probabilities 0.4, 0.3, and 0.3 to receive nothing (treatment level 0), a walking suggestion (treatment level 1), or an anti-sedentary suggestion (treatment level 2). The proximal longitudinal outcome is the total step count within 30 minutes following each decision point. We applied a log transformation to the 30-minute step count. Because no additional randomizations occured in the 30-minute window by design, we have $\Delta = 1$ and the proximal outcome is denoted by $Y_t$. (Recall that $\Delta$ is the time window length over which the proximal outcome is defined.)

\subsection{Illustration of the CEE estimator}
\label{subsec:application-estimator}

We applied the proposed estimator to separately estimate (a) the marginal CEE (by setting $S_t = \emptyset$ and $f_t(S_t) = 1$ in \eqref{eq:CEE-parametric-model}), (b) the CEE moderated by the decision point index (by setting $S_t = \emptyset$ and $f_t(S_t) = (1,t)^T$), and (c) the CEE moderated by location (by setting $S_t = \loc_t$ and $f_t(S_t) = (1,\loc_t)^T$, where $\loc_t = 1$ if the participant was at home or work and $\loc_t = 0$ otherwise). In particular, we estimated parameters in the following three CEE models:
\begin{align}
    & \text{(a) } E\left\{E(Y_t \mid H_t, A_t = a) - E(Y_t\mid H_t, A_t = 0)\right\} = \beta_a^{\text{marginal}} \quad \text{ for } a = 1,2; \nonumber\\
    & \text{(b) } E\left\{E(Y_t \mid H_t, A_t = a) - E(Y_t\mid H_t, A_t = 0)\right\} = \beta_a^{\text{t-int}} + \beta_a^{\text{t-slp}}\cdot t \quad \text{ for } a = 1,2; \label{eq:application-analysis}\\
    & \text{(c) } E\left\{E(Y_t \mid H_t, A_t = a) - E(Y_t\mid H_t, A_t = 0) \mid \loc_t \right\} = \beta_a^{\text{loc-int}} + \beta_a^{\text{loc-slp}} \cdot \loc_t \quad \text{ for } a = 1,2. \nonumber
\end{align}
Here, ``int'' stands for intercept and ``slp'' stands for slope. $t$ starts at 0 instead of 1 for the interpretability of $\beta_a^{\text{t-int}}$. For all analyses, we included the following control variables in $g_t(H_t)$: the decision point index $t$, $\loc_t$, an indicator for weekday, the lag-1 outcome $Y_{t-1}$, and the log-transformed step count in the 30-minute window preceding the current decision point.

\Cref{tab:application-HeartSteps-analysis} presents the estimated parameters, standard errors, 95\% confidence intervals, and p-values. In this paragraph we exponentiate the coefficients for ease of interpretation. For analysis (a), on average, delivering a walking suggestion, compared to no suggestion, increased the 30-minute step count by 21\% ($ = e^{0.192} - 1$, 95\% CI $= [2\%, 44\%]$, $p = 0.032$). Delivering an anti-sedentary suggestion, compared to no suggestion, does not significantly increase the 30-minute step count. There is no significant marginal difference between delivering a walking suggestion and delivering an anti-sedentary suggestion. For analysis (b), delivering a walking suggestion, compared to no suggestion, increased the 30-minute step count by 107\% ($ = e^{0.729} - 1$, 95\% CI $=[53\%, 181\%]$, $p < 0.001$) at the beginning of the study. This effect decreased linearly over time at a rate of 0.5\% per decision point ($ = 1 - e^{-0.005}$, 95\% CI $=[0.3\%, 0.8\%]$, $p < 0.001$). Neither the initial effect nor the time trend for delivering an anti-sedentary suggestion was significant. Furthermore, at the beginning of the study, the effect of a walking suggestion is significantly different from that of an anti-sedentary suggestion ($p = 0.011$). For analysis (c), when a participant is at home or work, delivering a walking suggestion, compared to no suggestion, increased the 30-minute step count by 51\% ($ = e^{0.413} - 1$, 95\% CI $= [27\%, 103\%]$, $p = 0.007$). When a participant is at home or work, delivering a walking suggestion is significantly more effective than delivering an anti-sedentary suggestion ($p = 0.001$).

\begin{table}[htbp]
    \caption{The HeartSteps analysis result for \Cref{subsec:application-estimator}. The estimates correspond to parameters defined in the three models in \eqref{eq:application-analysis}. \label{tab:application-HeartSteps-analysis}}
    \begin{tabular}[t]{llrrcr}
        \toprule
        & Parameter & Estimate & SE & 95\% CI  & $p$-value\\
        \midrule
        \multirow{3}{*}{\makecell{\textbf{Model (a):} \\ marginal CEE}} & $\beta_1^{\text{marginal}}$    & 0.192 & 0.083 & (\phantom{$-$}0.017, \phantom{$-$}0.367) & 0.032\\
        & $\beta_2^{\text{marginal}}$     & 0.007 & 0.085 & ($-$0.174, \phantom{$-$}0.188) & 0.938\\
        & $\beta_1^{\text{marginal}} - \beta_2^{\text{marginal}}$     & 0.185 & 0.110 & ($-$0.039, \phantom{$-$}0.409) & 0.101 \\
        \midrule
        \multirow{6}{*}{\makecell{\textbf{Model (b):} \\ CEE moderated \\ by the decision \\ point index}} & $\beta_1^{\text{t-int}}$   & 0.729 & 0.144 &(\phantom{$-$}0.425, \phantom{$-$}1.033) & $<0.001$\\
        & $\beta_1^{\text{t-slp}}$   & $-$0.005 & 0.001 &($-$0.008, $-$0.003) & $<0.001$\\
        & $\beta_2^{\text{t-int}}$   & 0.237 & 0.192 &($-$0.171, \phantom{$-$}0.645) & 0.244\\
        & $\beta_2^{\text{t-slp}}$   & $-$0.002 & 0.002 &($-$0.006, \phantom{$-$}0.001) & 0.200\\
        & $\beta_1^{\text{t-int}} - \beta_2^{\text{t-int}}$ &0.492 &0.183 &(\phantom{$-$}0.118, \phantom{$-$}0.866) & 0.011\\
        & $\beta_1^{\text{t-slp}} - \beta_2^{\text{t-slp}}$ & $-$0.003 & 0.002 &($-$0.006, \phantom{$-$}$0.000$) & 0.064 \\
        \midrule
        \multirow{6}{*}{\makecell{\textbf{Model (c):} \\ CEE moderated \\ by location}} & $\beta_1^{\text{loc-int}}$                       & 0.027 & 0.098 &($-$0.183, \phantom{$-$}0.236) & 0.796\\
        & $\beta_2^{\text{loc-int}}$                       & 0.059 & 0.122 &($-$0.202, \phantom{$-$}0.319) & 0.648\\
        & $\beta_1^{\text{loc-int}} + \beta_1^{\text{loc-slp}}$             & 0.413 & 0.143 &(\phantom{$-$}0.120, \phantom{$-$}0.706) & 0.007\\
        & $\beta_2^{\text{loc-int}} + \beta_2^{\text{loc-slp}}$             & $-$0.050 & 0.136 &($-$0.327, \phantom{$-$}0.227) & 0.714\\
        & $\beta_1^{\text{loc-int}} - \beta_2^{\text{loc-int}}$ & $-$0.032 & 0.149 &(\phantom{$-$}0.336, \phantom{$-$}0.272) & 0.831\\
        & \makecell{$\{(\beta_1^{\text{loc-int}} + \beta_1^{\text{loc-slp}})$ \\ $-(\beta_2^{\text{loc-int}} + \beta_2^{\text{loc-slp}})\}$} & 0.463 & 0.124 &(\phantom{$-$}0.211, \phantom{$-$}0.716) & 0.001 \\
        \bottomrule
    \end{tabular}
\end{table}

\subsection{Illustration of the Sample Size Formula}
\label{subsec:application-sample-size}

We illustrate using the sample size formula to determine the sample size of a hypothetical MRT. Suppose that the hypothetical MRT is similar to HeartSteps in the following intervention design and trial design aspects: The hypothetical MRT has two active treatment levels and one control treatment level, with constant randomization probabilities $p_{0t} = 0.4$ and $p_{1t} = p_{2t} = 0.3$ for all $t \in [T]$. Each participant is enrolled for $T = 210$ decision points. Furthermore, we assume for now that the participants are always available for randomization ($\tau(t) = 1$), an assumption that we will relax later. For notation simplicity we will write $\Delta\sate_{12}$ as $\Delta\sate$.

Suppose the data generating distribution of the hypothetical MRT is also similar to HeartSteps, in that $\Delta \sate = 0.053$ and $\aeo = 3$ (recall the definitions in \eqref{eq:delta-ate} and \eqref{eq:aeo}). In practice, these values should be formed based on conversation with domain scientists.

\begin{figure}
    \centering
    \begin{subfigure}[b]{0.45\textwidth}
        \centering
        \caption{}
        \label{fig:HS1}
        \includegraphics[width=\textwidth]{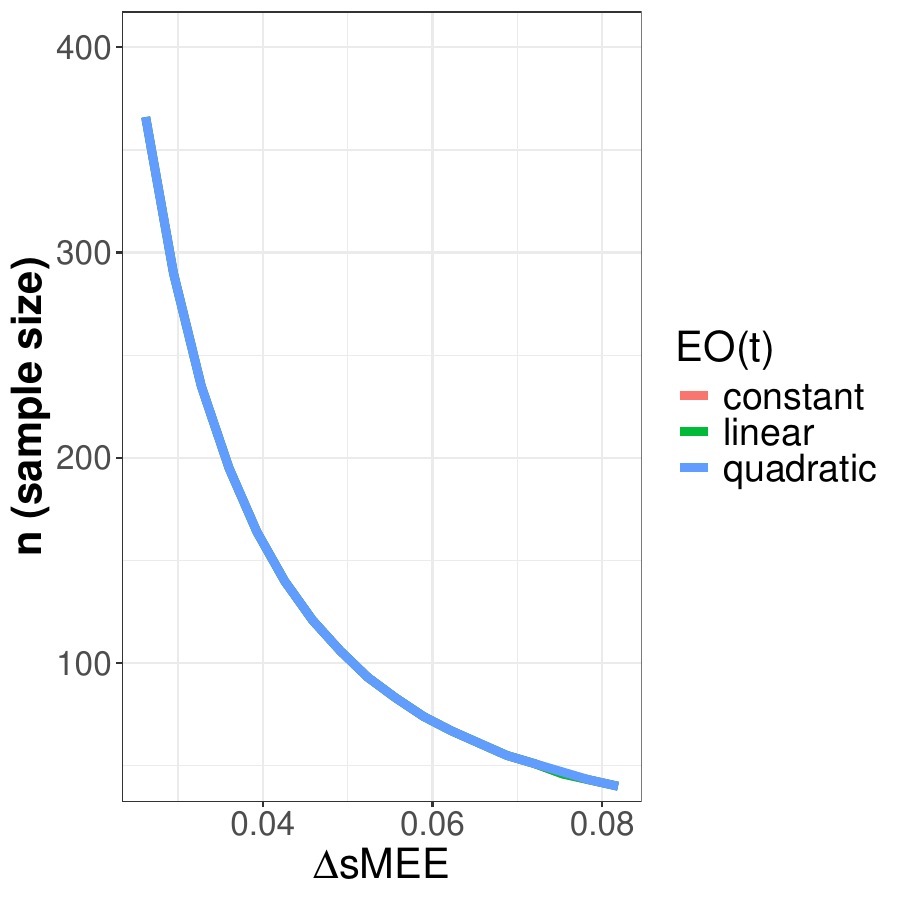}
    \end{subfigure}
    \hfill
    \begin{subfigure}[b]{0.45\textwidth}
        \centering
        \caption{}
        \label{fig:HS2}
        \includegraphics[width=\textwidth]{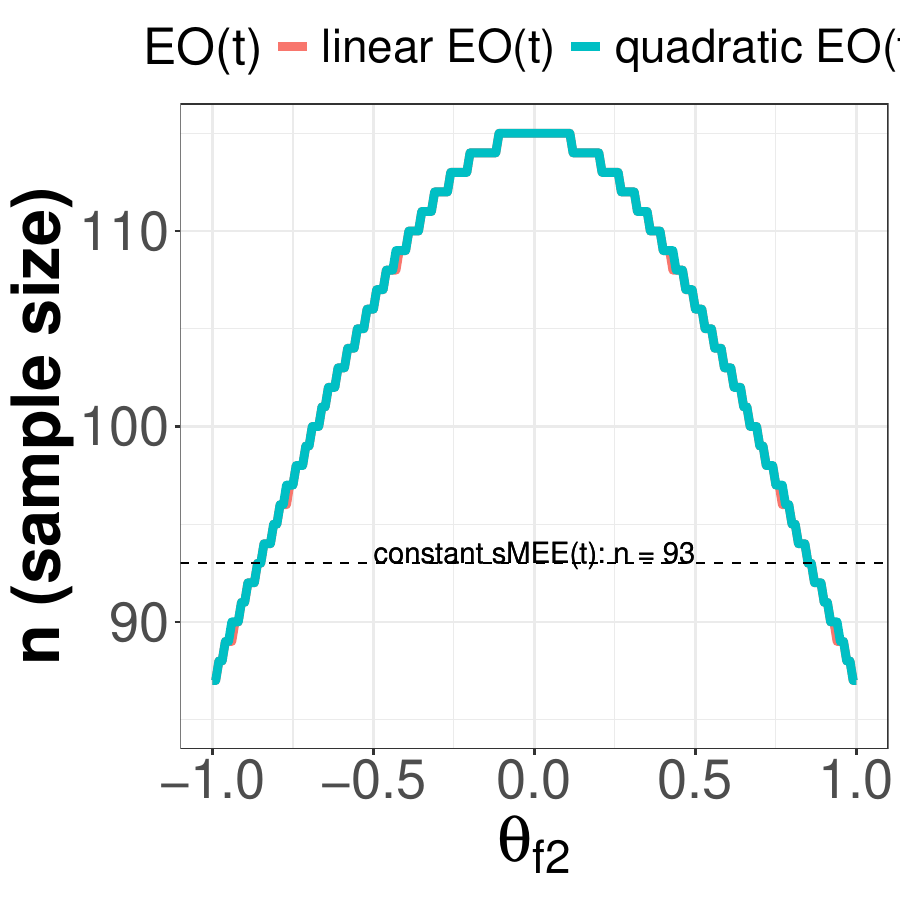}
    \end{subfigure}
    \begin{subfigure}[b]{0.4\textwidth}
        \centering
        \caption{}
        \label{fig:HS3}
        \includegraphics[width=\textwidth]{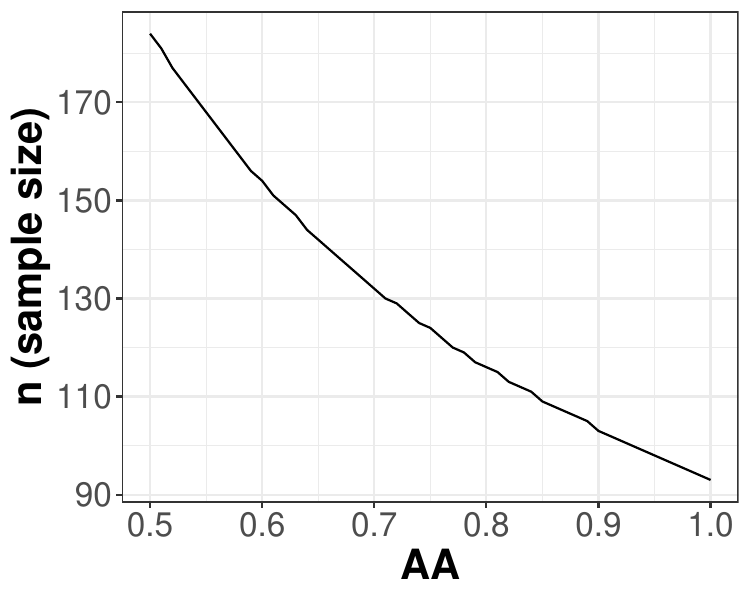}
    \end{subfigure}
    \hfill
    \begin{subfigure}[b]{0.55\textwidth}
        \centering
        \caption{}
        \label{fig:HS4}
        \includegraphics[width=\textwidth]{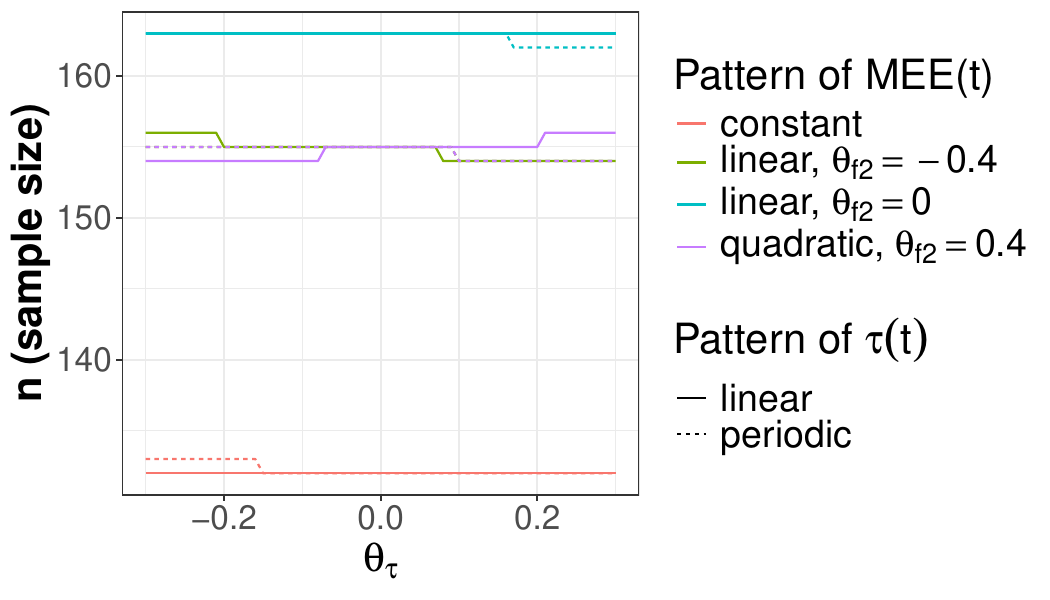}
    \end{subfigure}
    \caption{The dependence of the output sample size $n$ on various inputs for the data example illustration in \cref{subsec:application-sample-size}. \textbf{Panel (a):} $n$ varies monotonically when $\Delta\smee$. \textbf{Panel (b):} $n$ depends on the difference in the slopes of $\mee_1(t)$ and $\mee_2(t)$, parameterized by $\theta_{f2}$. \textbf{Panel (c):} $n$ varies monotonically with the average expected availability, $\aaa$. \textbf{Panel (d):} $n$ essentially does not depend on the pattern of the availability over time, $\tau(t)$.}
\end{figure}

We aim to detect a significant difference between the two active treatment levels. Specifically, we desire to have 80\% power to detect $\Delta \sate = 0.053$ with type I error 0.05. Following the practical guidelines in \Cref{subsec:practical-guideline}, we use constant-over-time $\smee_1(t)$, $\smee_2(t)$, and $\eo(t)$ in the sample size calculation, with the following inputs to \cref{alg:ss-calculator}: $K = 2$, $T = 210$, $p_{0t} = 0.4$, $p_{1t} = p_{2t} = 0.3$, $\tau(t) = 1$, $f_t = 1$, $\gamma_1 = 0.053$, $\gamma_2 = 0$, $g_t = 1$, $L =(1, -1)$, $\eta = 0.05$, $1-b = 0.8$. Note that because $L =(1, -1)$ concerns the contrast between the two active treatments, only the difference between $\gamma_1$ and $\gamma_2$ matters in the sample size calculation, and this particular choice of $\gamma_1 = 0.053$ and $\gamma_2 = 0$ ensures $\smee_1(t) - \smee_2(t) = 0.053$. \cref{alg:ss-calculator} outputs the required sample size $n = 93$. Furthermore, \Cref{fig:HS1} shows how the output sample size $n$ varies with $\Delta \sate$ and $g_t = 1$ when all other inputs are fixed. As expected, $n$ decreases as $\Delta \sate$ increases, and $n$ virtually stays the same with different forms of $g_t$ (i.e., whether $\eo(t)$ is constant, linear, or quadratic in $t$).

Next we fix $\Delta \sate = 0.053$ and $\aeo = 3$, and explore how $n$ changes when $f_t = (1,t)^T$ and $\mee_1(t)$ and $\mee_2(t)$ are not parallel to each other. To operationalize this scenario, we use $\theta_{f1}$ to parameterize the slope of $\mee_1(t)$ and $\theta_{f2}$ to parameterize the difference between the slopes of $\mee_1(t)$ and $\mee_2(t)$. Detailed parameterization is in Supplementary Material \Cref{A-box:detail-gm-0-part2}. \Cref{fig:HS2} shows how much $n$ varies when $\mee_1(t)$ and $\mee_2(t)$ are non-constant and non-parallel. The observation is two-fold. First, inputting a linear MEE (i.e., setting $f_t = (1,t)^T$ instead of $f_t = 1$) generally yields a required sample size larger than when inputting a constant MEE, indicted by the dashed horizontal line. Second, whether the causal effect curves of the two active treatment levels, $\mee_1(t)$ and $\mee_2(t)$, are assumed to be parallel over time or not affects the sample size, and the required sample size is the largest when they are assumed to be parallel, corresponding to $\theta_{f2} = 0$ in the figure. The pattern of $\eo(t)$ is also varied between linear and quadratic and this does not influence the sample size.

Lastly, We we investigate how the assumed expected availabiltiy over time, $\tau(t)$, affects the required sample size $n$. When assuming a constant-over-time $\tau(t)$, the required sample size $n$ decreases as the average availability (AA) increases, as expected (\Cref{fig:HS3}). When assuming a linear or periodic $\tau(t)$, different patterns of $\tau(t)$ lead to virtually the same $n$ (\Cref{fig:HS4}).

\section{Discussion}
\label{sec:discussion}

We considered micro-randomized trials in which the treatment options at each decision point are categorical, comprising more than two levels. We extended the causal excursion effect definition and the corresponding weighted and centered least squares estimator \citep{boruvka2018assessing} to this setting. Furthermore, we provided a sample size formula that accommodates a variety of comparisons among the categorical treatment levels. The sample size formula is proved to guarantee type I error and power under a set of working assumptions. Through extensive simulation studies under scenarios where working assumptions are violated, we showed that the type I error is always guaranteed, and the power is often guaranteed but not always. We provided in \Cref{box:practical-guideline} practical guidelines for specifying the inputs to the sample size formula to ensure adequate power in most real-world scenarios, even when the working assumptions are violated. The proposed estimator and sample size formula were illustrated using the HeartSteps micro-randomized trial.

In deriving the sample size formula, we considered null hypotheses of the form $\tL \beta = 0$. The sample size formula can be straightforwardly generalized to null hypotheses of the form $\tL \beta = \tc$ for a given vector $\tc$. However, the interpretation of such null hypotheses can sometimes be challenging depending on the specific form of $\tL$ and $\tc$. Specifically, some choices of $\tc$ may not correspond to a parsimonious null hypothesis in the original scale on $\mee_{1:K}(t)$; i.e., there may not exist a vector $c$ such that $\tL \beta = \tc$ is equivalent to $L \times \mee_{1:K}(t) = c$. Therefore, we focused on null hypotheses $\tL \beta = 0$ in the paper.

There are several directions for future research. First, recent advances in flexible nonparametric models and machine learning algorithms can be leveraged to model the nuisance functions, potentially improving the efficiency of causal effect estimation. Second, the current estimator and sample size formula can be extended to accommodate binary longitudinal proximal outcomes. Third, developing a sample size formula for detecting significant moderation of causal effects is a valuable direction for future research.

\section*{Supplementary Material}
Supplementary material is available online

\section*{Conflict of Interest}
None declared

\section*{Acknowledgment}
The authors would like to sincerely thank Dr. Tianchen Qian, whose mentorship and guidance have been instrumental in the completion of this project.

\section*{Data Availability}    
The R code for reproducing the power simulation can be downloaded at 
\url{https://github.com/JeremyJosephLin/causal_excursion_mult_trt}
The HeartSteps dataset is available via \url{https://github.com/klasnja/HeartStepsV1}.

\bibliography{references}

\begin{thebibliography}{10}
\providecommand{\natexlab}[1]{#1}
\providecommand{\url}[1]{\texttt{#1}}
\expandafter\ifx\csname urlstyle\endcsname\relax
  \providecommand{\doi}[1]{doi: #1}\else
  \providecommand{\doi}{doi: \begingroup \urlstyle{rm}\Url}\fi

\bibitem[Bell et~al.(2023)Bell, Garnett, Bao, Cheng, Qian, Perski, Potts, Williamson, et~al.]{bell2023notifications}
Lauren Bell, Claire Garnett, Yihan Bao, Zhaoxi Cheng, Tianchen Qian, Olga Perski, Henry~WW Potts, Elizabeth Williamson, et~al.
\newblock How notifications affect engagement with a behavior change app: Results from a micro-randomized trial.
\newblock \emph{JMIR mHealth and uHealth}, 11\penalty0 (1):\penalty0 e38342, 2023.

\bibitem[Boruvka et~al.(2018)Boruvka, Almirall, Witkiewitz, and Murphy]{boruvka2018assessing}
Audrey Boruvka, Daniel Almirall, Katie Witkiewitz, and Susan~A Murphy.
\newblock Assessing time-varying causal effect moderation in mobile health.
\newblock \emph{Journal of the American Statistical Association}, 113\penalty0 (523):\penalty0 1112--1121, 2018.

\bibitem[Cohn et~al.(2023)Cohn, Qian, and Murphy]{cohn2023sample}
Eric~R Cohn, Tianchen Qian, and Susan~A Murphy.
\newblock Sample size considerations for micro-randomized trials with binary proximal outcomes.
\newblock \emph{Statistics in medicine}, 42\penalty0 (16):\penalty0 2777--2796, 2023.

\bibitem[Dempsey et~al.(2015)Dempsey, Liao, Klasnja, Nahum-Shani, and Murphy]{dempsey2015randomised}
Walter Dempsey, Peng Liao, Pedja Klasnja, Inbal Nahum-Shani, and Susan~A Murphy.
\newblock Randomised trials for the fitbit generation.
\newblock \emph{Significance}, 12\penalty0 (6):\penalty0 20--23, 2015.

\bibitem[Klasnja et~al.(2019)Klasnja, Smith, Seewald, Lee, Hall, Luers, Hekler, and Murphy]{klasnja2019efficacy}
Predrag Klasnja, Shawna Smith, Nicholas~J Seewald, Andy Lee, Kelly Hall, Brook Luers, Eric~B Hekler, and Susan~A Murphy.
\newblock Efficacy of contextually tailored suggestions for physical activity: a micro-randomized optimization trial of {H}eart{S}teps.
\newblock \emph{Annals of Behavioral Medicine}, 53\penalty0 (6):\penalty0 573--582, 2019.

\bibitem[Liao et~al.(2016)Liao, Klasnja, Tewari, and Murphy]{liao2016sample}
Peng Liao, Predrag Klasnja, Ambuj Tewari, and Susan~A Murphy.
\newblock Sample size calculations for micro-randomized trials in mhealth.
\newblock \emph{Statistics in medicine}, 35\penalty0 (12):\penalty0 1944--1971, 2016.

\bibitem[Mancl and DeRouen(2001)]{mancl2001covariance}
Lloyd~A Mancl and Timothy~A DeRouen.
\newblock A covariance estimator for gee with improved small-sample properties.
\newblock \emph{Biometrics}, 57\penalty0 (1):\penalty0 126--134, 2001.

\bibitem[Pan and Wall(2002)]{pan2002small}
Wei Pan and Melanie~M Wall.
\newblock Small-sample adjustments in using the sandwich variance estimator in generalized estimating equations.
\newblock \emph{Statistics in medicine}, 21\penalty0 (10):\penalty0 1429--1441, 2002.

\bibitem[Qian et~al.(2021)Qian, Yoo, Klasnja, Almirall, and Murphy]{qian2021estimating}
Tianchen Qian, Hyesun Yoo, Predrag Klasnja, Daniel Almirall, and Susan~A Murphy.
\newblock Estimating time-varying causal excursion effects in mobile health with binary outcomes.
\newblock \emph{Biometrika}, 108\penalty0 (3):\penalty0 507--527, 2021.

\bibitem[Robins et~al.(2000)Robins, Hernan, and Brumback]{robins2000marginal}
James~M Robins, Miguel~Angel Hernan, and Babette Brumback.
\newblock Marginal structural models and causal inference in epidemiology, 2000.

\end{thebibliography}
\bibliographystyle{plainnat}
\makeatletter\@input{xy.tex}\makeatother
\end{document}


\maketitle

\tableofcontents

\newpage

\spacingset{1.9}





\section{Technical Details for \texorpdfstring{\Cref{M-sec:estimator}}{Section 3} on Asymptotic Normality}
\label{sec:proof-thm-CAN}

\subsection{Lemmas}

\begin{lem}
    \label{lem:thm1-proofuse}
    Fix arbitrary $t \in [T]$. Let
    \begin{align*}
        W_t := \prod_{j = t + 1}^{t + \Delta - 1} \frac{\one(A_j = 0)}{p_j(0|H_j)}.
    \end{align*}
    For any $(\alpha,\beta)$, we have
    \begin{align}
        & \sum_{l=0}^K E\bigg[ \bigg\{ Y_{t,\Delta} - \sum_{k = 1}^K C_k(A_t) f_t(S_t)^T \beta_k - g_t(H_t)^T \alpha \bigg\} J_t C_K(A_t) f_t(S_t) ~\Big|~ H_t, I_t = 1, A_t = l \bigg] ~ p_t(l | H_t) \nonumber \\
        = ~& -\tp_t(K|S_t) f_t(S_t) \sum_{k=1}^{K-1} \tp_t(l|S_t) \Big\{E(Y_{t,\Delta} W_t \mid H_t, I_t = 1, A_t = k) \nonumber \\
        & ~~~~~~~~~~~~~~~~~~~~~~~~~~~~~ - E(Y_{t,\Delta} W_t \mid H_t, I_t = 1, A_t = 0) - f_t(S_t)^T \beta_k \Big\} \nonumber \\
        & + \tp_t(K|S_t) \{1 - \tp_t(K|S_t)\} f_t(S_t) \Big\{E(Y_{t,\Delta} W_t \mid H_t, I_t = 1, A_t = K) \nonumber \\
        & ~~~~~~~~~~~~~~~~~~~~~~~~~~~~~ - E(Y_{t,\Delta} W_t \mid H_t, I_t = 1, A_t = 0) - f_t(S_t)^T \beta_K \Big\}. \label{eq:lem1}
    \end{align}
\end{lem}

\begin{proof}[Proof of \cref{lem:thm1-proofuse}]
    We prove the lemma by doing direct algebra. To reduce notation burden, within this proof we use the following short-hand notation: $Y := Y_{t,\Delta}$, $C_k := C_k(A_t)$, $f := f_t(S_t)$, $f\beta_k := f_t(S_t)^T\beta_k$, $g\alpha := g_t(H_t)^T\alpha$, $\tp(k) := \tp_t(k | S_t)$, $p(k) := p_t(k | H_t)$, $W := W_t$, $E_k(\cdot) := E(\cdot \mid H_t, I_t = 1, A_t = k)$. In the derivation below, we will repeatedly use the fact that $E_k(W) = 1$ for all $0 \leq k \leq K$, which follows from the law of iterated expectations. With the short-hand notation, the left hand side of \eqref{eq:lem1} becomes
    \begin{align}
        \sum_{l=0}^K E_l \bigg\{ \bigg( Y - \sum_{k=1}^K C_k f\beta_k - g\alpha\bigg) \frac{\tp(l)}{p(l)} W C_K f \bigg\} p(l) = V_1 + V_2 + V_3, \label{eq:lem1-proofuse1}
    \end{align}
    where
    \begin{align}
        V_1 & := E_0 \bigg\{ \bigg( Y - \sum_{k=1}^K C_k f\beta_k - g\alpha\bigg) \frac{\tp(0)}{p(0)} W C_K f \bigg\} p(0) \nonumber \\
        V_2 & := \sum_{l=1}^{K-1} E_l \bigg\{ \bigg( Y - \sum_{k=1}^K C_k f\beta_k - g\alpha\bigg) \frac{\tp(l)}{p(l)} W C_K f \bigg\} p(l) \nonumber \\
        V_3 & := E_K \bigg\{ \bigg( Y - \sum_{k=1}^K C_k f\beta_k - g\alpha\bigg) \frac{\tp(K)}{p(K)} W C_K f \bigg\} p(K). \nonumber
    \end{align}
    By direct algebra we have
    \begin{align}
        V_1 & = E_0 \bigg[ \bigg\{ Y + \sum_{k=1}^K \tp(k) f\beta_k - g\alpha\bigg\} \tp(0)\{-\tp(K)\} W f \bigg] \nonumber \\
        & = - \bigg\{ E_0(YW) + \sum_{k=1}^K  \tp(k) f\beta_k - g\alpha \bigg\} \tp(0) \tp(K) f. \label{eq:lem1-proofuse-v1}
    \end{align}
    We also have
    \begin{align}
        V_2 & = \sum_{l=1}^{K-1} E_l \bigg\{ \bigg( Y - \sum_{k=1}^K C_k f\beta_k - g\alpha\bigg) \frac{\tp(l)}{p(l)} W C_K f \bigg\} p(l) \nonumber \\
        & = \sum_{l=1}^{K-1} E_l (Y W C_K ) \tp(l) f  - \sum_{l=1}^{K-1} \sum_{k=1}^K E_l ( C_k W C_K ) f\beta_k \tp(l) f - g\alpha \sum_{l=1}^{K-1} E_l(W C_K) \tp(l) f \nonumber \\
        & = V_{21} + V_{22} + V_{23}, \label{eq:lem1-proofuse-v2}
    \end{align}
    where
    \begin{align}
        V_{21} := \sum_{l=1}^{K-1} E_l (Y W C_K ) \tp(l) f = - \tp(K) f \sum_{l=1}^{K-1} E_l (Y W) \tp(l) , \label{eq:lem1-proofuse-v21}
    \end{align}
    \begin{align}
        V_{22} & := - \sum_{l=1}^{K-1} \sum_{k=1}^K E_l ( C_k W C_K ) f\beta_k \tp(l) f \nonumber \\
        & = - \bigg(\sum_{\substack{1 \leq k \leq K-1 \\ 1 \leq l \leq K-1 \\ l \neq k}} + \sum_{\substack{1 \leq k \leq K-1 \\ l = k}} + \sum_{\substack{k = K \\ 1 \leq l \leq K-1}}\bigg) E_l ( C_k W C_K ) f\beta_k \tp(l) f \nonumber \\
        & = \bigg[ - \tp(K) f \sum_{k=1}^{K-1} \sum_{\substack{1 \leq l \leq K-1 \\ l \neq k}} \tp(l) \tp(k)f\beta_k \bigg] + \bigg[ \tp(K) f \sum_{k=1}^{K-1} \{1 - \tp(k)\} \tp(k) f \beta_k \bigg] - \bigg[ \tp(K) f \sum_{k=1}^{K-1} \tp(k) \tp(K) f\beta_K \bigg] \nonumber \\
        & = \bigg[ \tp(K) f \sum_{k=1}^{K-1} \{\tp(0) + \tp(K)\} \tp(k)f\beta_k \bigg] - \bigg[ \tp(K) f \{1 - \tp(0) - \tp(K)\} \tp(K) f\beta_K \bigg] \label{eq:lem1-proofuse-v22}
    \end{align}
    and
    \begin{align}
        V_{23} &: = - g\alpha \sum_{l=1}^{K-1} E_l(W C_K) \tp(l) f = \tp(K)f \sum_{l=1}^{K-1} \tp(l) g\alpha \nonumber \\
        & = \tp(K)f \{1 - \tp(0) - \tp(K)\} g\alpha. \label{eq:lem1-proofuse-v23}
    \end{align}
    Finally, we have
    \begin{align}
        V_3 = \bigg[ E_K(YW)  + \sum_{k=1}^{K-1}\tp(k) f\beta_k - \{1 - \tp(K)\}f \beta_K - g\alpha  \bigg]    \tp(K)\{1-\tp(K)\}f. \label{eq:lem1-proofuse-v3}
    \end{align}

    Putting together \eqref{eq:lem1-proofuse-v1}, \eqref{eq:lem1-proofuse-v21}, \eqref{eq:lem1-proofuse-v22}, \eqref{eq:lem1-proofuse-v23}, and \eqref{eq:lem1-proofuse-v3}, we have that the left hand side of \eqref{eq:lem1} equals
    \begin{align}
        & ~~~~ V_1 + V_{21} + V_{22} + V_{23} + V_3 \nonumber \\
        & = \tp(K) f \bigg[-E_0(YW) \tp(0) - \sum_{l=1}^{K-1} E_l (Y W) \tp(l) + E_K(YW) \{1-\tp(K)\}\bigg] \nonumber \\ 
        & ~~~~ + \tp(K) f \bigg[ - \sum_{k=1}^K \tp(0) \tp(k) f\beta_k + \sum_{k=1}^{K-1} \{\tp(0) + \tp(K)\} \tp(k)f\beta_k - \{1 - \tp(0) - \tp(K)\} \tp(K) f\beta_K \nonumber \\
        & ~~~~~~~~~~~~~~~~ + \{1 - \tp(K)\}\sum_{k=1}^{K-1}\tp(k) f\beta_k - \{1 - \tp(K)\}^2 f \beta_K  \bigg] \nonumber \\
        & + \tp(K) f \bigg[ \tp(0) g\alpha  + \{1 - \tp(0) - \tp(K)\} g\alpha - \{1 - \tp(K)\} g\alpha \bigg]. \label{eq:lem1-proofuse-4}
    \end{align}
    The first bracket in \eqref{eq:lem1-proofuse-4} is
    \begin{align}
        & ~~~~ -E_0(YW) \tp(0) - \sum_{l=1}^{K-1} E_l (Y W) \tp(l) + E_K(YW) \{1-\tp(K)\} \nonumber \\
        & = -E_0(YW) \bigg\{1 - \sum_{l=1}^{K} \tp(l)\bigg\} - \sum_{l=1}^{K-1} E_l (Y W) \tp(l) + E_K(YW) \{1-\tp(K)\} \nonumber \\
        & = - \sum_{l=1}^{K-1} \tp(l) \{E_1(YW) - E_0(YW)\} + \{1-\tp(K)\} \{E_K(YW) - E_0(YW)\}. \label{eq:lem1-proofuse-5}
    \end{align}
    The second bracket in \eqref{eq:lem1-proofuse-4} is
    \begin{align}
        & ~~~~ - \sum_{k=1}^K \tp(0) \tp(k) f\beta_k + \sum_{k=1}^{K-1} \{\tp(0) + \tp(K)\} \tp(k)f\beta_k - \{1 - \tp(0) - \tp(K)\} \tp(K) f\beta_K \nonumber \\
        & ~~~~ + \{1 - \tp(K)\}\sum_{k=1}^{K-1}\tp(k) f\beta_k - \{1 - \tp(K)\}^2 f \beta_K \nonumber \\
        & = \sum_{k=1}^{K-1}\tp(k) f\beta_k - \{1 - \tp(K)\} f\beta_K. \label{eq:lem1-proofuse-6}
    \end{align}
    The third bracket in \eqref{eq:lem1-proofuse-4} is
    \begin{align}
        \tp(0) g\alpha  + \{1 - \tp(0) - \tp(K)\} g\alpha - \{1 - \tp(K)\} g\alpha = 0. \label{eq:lem1-proofuse-7}
    \end{align}
    Putting \eqref{eq:lem1-proofuse-5}, \eqref{eq:lem1-proofuse-6}, and \eqref{eq:lem1-proofuse-7} into \eqref{eq:lem1-proofuse-4} yields \eqref{eq:lem1}. This completes the proof.
\end{proof}

\subsection{Proof of \texorpdfstring{\cref{M-thm:CAN}}{Theorem 1}}

The proof follows from a standard application of the estimating equation theory \citep[e.g., Section 5 of][]{van2000asymptotic}. In Step 1, we will show the unbiasedness of the estimating function $m(\alpha,\beta)$. In Step 2, we will establish the asymptotic normality of $(\hat\alpha,\hat\beta)$. In Step 3, we will derive the explicit form of the asymptotic variance for $\hat\beta$. In Step 4, we will show the consistency of the asymptotic variance estimator. In Step 5, we will show that the theorem conclusion holds when $\tp_t(k|S_t)$ is estimated. 

In the proof, we use the following definition.
\begin{align}
    r_t(\alpha,\beta) := Y_{t,\Delta} - \sum_{k=1}^K C_k(A_t) f_t(S_t)^T \beta_k - g_t(H_t)^T \alpha, \label{eq:rt-def-more-complex}
\end{align}
and
\begin{align}
    \tD_t :=
    \begin{bmatrix}
        g_t(H_t) \\
        C_1(A_t) f_t(S_t)\\
        \vdots\\
        C_K(A_t) f_t(S_t)
    \end{bmatrix}
    \quad \text{and} \quad
    D_t :=
    \begin{bmatrix}
        C_1(A_t) f_t(S_t)\\
        \vdots\\
        C_K(A_t) f_t(S_t)
    \end{bmatrix}. \nonumber
\end{align}

\textbf{Step 1:} We show that there exists some $\alpha' \in \RR^q$ such that $E \{m(\alpha',\beta^0)\} = 0$ with $m(\alpha,\beta)$ being the estimating function defined in \eqref{M-eq:ee} and $\beta^0$ being the true parameter value of $\beta$.

\textbf{Step 1.1:} We show that for arbitrary $\alpha$ and for all $t \in [T]$ the following holds
\begin{align}
    E \bigg[ I_t \bigg\{ Y_{t,\Delta} - \sum_{k = 1}^K C_k(A_t) f_t(S_t)^T \beta^0_k - g_t(H_t)^T \alpha \bigg\} J_t C_K(A_t) f_t(S_t) \bigg] = 0. \label{eq:thm1-proofuse1}
\end{align}
Using iterated expectations we have
\begin{align}
    & ~~~~ E \bigg[ I_t \bigg\{ Y_{t,\Delta} - \sum_{k = 1}^K C_k(A_t) f_t(S_t)^T \beta^0_k - g_t(H_t)^T \alpha \bigg\} J_t C_K(A_t) f_t(S_t) \bigg] \nonumber \\
    & = E \Bigg( I_t \sum_{l=0}^K E \bigg[ \bigg\{ Y_{t,\Delta} - \sum_{k = 1}^K C_k(A_t) f_t(S_t)^T \beta^0_k - g_t(H_t)^T \alpha \bigg\} J_t C_K(A_t) f_t(S_t) \Big | H_t, I_t = 1, A_t = l \bigg] p_t(l|H_t) \Bigg) \nonumber \\
    & = E \Bigg( - I_t \tp_t(K|S_t) f_t(S_t) \sum_{k=1}^{K-1} \tp_t(l|S_t)  \nonumber \\
    & ~~~~~~~~ \times \Big\{E(Y_{t,\Delta} W_t \mid H_t, I_t = 1, A_t = k) - E(Y_{t,\Delta} W_t \mid H_t, I_t = 1, A_t = 0) - f_t(S_t)^T \beta^0_k \Big\} \nonumber \\
    & ~~~~ + I_t \tp_t(K|S_t) \{1 - \tp_t(K|S_t)\} f_t(S_t) \nonumber \\
    & ~~~~~~~~ \times \Big\{E(Y_{t,\Delta} W_t \mid H_t, I_t = 1, A_t = K) - E(Y_{t,\Delta} W_t \mid H_t, I_t = 1, A_t = 0) - f_t(S_t)^T \beta^0_K \Big\} \Bigg), \label{eq:thm1-proofuse2}
\end{align}
where \eqref{eq:thm1-proofuse2} follows from \cref{lem:thm1-proofuse}. Then because of the parametric CEE model \eqref{M-eq:CEE-parametric-model}, it follows immediately from an iterated expectation given $(S_t, I_t)$ that \eqref{eq:thm1-proofuse2} equals 0.

\textbf{Step 1.2:} We show that for there exists $\alpha' \in \RR^q$ such that the following holds
\begin{align}
    \sum_{t=1}^T E \bigg[ I_t \bigg\{ Y_{t,\Delta} - \sum_{k = 1}^K C_k(A_t) f_t(S_t)^T \beta^0_k - g_t(H_t)^T \alpha' \bigg\} J_t g_t(H_t)\bigg] = 0. \label{eq:thm1-proofuse3}
\end{align}
This follows by setting
\begin{align}
    \alpha' = \bigg[\sum_{t=1}^T E \Big\{ I_t J_t g_t(H_t) g_t(H_t)^T \Big\} \bigg]^{-1} \bigg( \sum_{t=1}^T E \Big[ I_t J_t \{Y_{t,\Delta} - \sum_{k=1}^K C_k(A_t) f_t(S_t)^T \beta_k^0\} g_t(H_t) \Big] \bigg), \label{eq:thm1-proofuse4}
\end{align}
assuming the matrix inverse exists by regularity conditions.

Therefore, $E \{m(\alpha',\beta^0)\} = 0$ with $\alpha'$ defined in \eqref{eq:thm1-proofuse4}. This completes Step 1.

\textbf{Step 2:} The asymptotic normality of $(\hat\alpha, \hat\beta)$ follows immediately from Theorems 5.9 and 5.21 of \citet{van2000asymptotic}:
\begin{align}
    \sqrt{n} \left( \begin{bmatrix} \hat\alpha \\ \hat\beta \end{bmatrix} - \begin{bmatrix} \alpha' \\ \beta^0 \end{bmatrix} \right) \dto N\bigg(0, \Big[ E \big\{\dot{m}(\alpha', \beta^0) \big\} \Big]^{-1} \Big[ E \big\{ m(\alpha', \beta^0) m(\alpha', \beta^0)^T \big\} \Big] \Big[ E \big\{\dot{m}(\alpha', \beta^0) \big\} \Big]^{-1,T} \bigg). \label{eq:thm1-proofuse5}
\end{align}

\textbf{Step 3:} Derive the explicit form of the asymptotic variance for $\hat\beta$, i.e., the lower $Kp \times Kp$ principal submatrix of the asymptotic variance in \eqref{eq:thm1-proofuse5}.
By definition, $m(\alpha,\beta) = \sum_{t=1}^T I_t J_t \tD_t r_t(\alpha,\beta)$, which implies that
\begin{align}
    \dot{m}(\alpha,\beta) = \sum_{t=1}^T I_t J_t \tD_t \dot{r}_t(\alpha,\beta) = - \sum_{t=1}^T I_t J_t \tD_t \tD_t^T. \nonumber
\end{align}
Therefore,
\begin{align}
    E \big\{\dot{m}(\alpha', \beta^0) \big\} & = - \sum_{t=1}^T E(I_t J_t \tD_t \tD_t^T) \nonumber \\
    & = - \sum_{t=1}^T E\left\{I_t J_t \begin{bmatrix} g_t(H_t)g_t(H_t)^T & g_t(H_t) D_t^T \\ D_t g_t(H_t)^T & D_t D_t^T \end{bmatrix} \right\}. \label{eq:thm1-proofuse6}
\end{align}
The off-diagonal term in \eqref{eq:thm1-proofuse6} have expectation 0, because it is easy to verify that $E \{J_t C_k(A_t) \mid H_t, I_t = 1\} = 0$ for any $k \in [K]$. This implies that 
\begin{align}
    E \big\{\dot{m}(\alpha', \beta^0) \big\} = - \begin{bmatrix} \sum_{t=1}^T E \{ I_t J_t g_t(H_t)g_t(H_t)^T \} & 0 \\ 0 & \sum_{t=1}^T E ( I_t J_t D_t D_t^T ) \end{bmatrix}. \label{eq:thm1-proofuse7}
\end{align}
This implies that only the lower $Kp \times Kp$ principal submatrix of $\{ m(\alpha', \beta^0) m(\alpha', \beta^0)^T \}$ matters when calculating the asymptotic variance of $\hat\beta$. By the definition of $m(\alpha,\beta)$, this submatrix has the form
\begin{align}
    \sum_{t=1}^T \sum_{s=1}^T E \{ I_t I_s J_t J_s r_t(\alpha',\beta^0) r_s(\alpha',\beta^0) D_t D_s^T \}. \label{eq:thm1-proofuse8}
\end{align}
Plug \eqref{eq:thm1-proofuse7} and \eqref{eq:thm1-proofuse8} into \eqref{eq:thm1-proofuse5} and we have the asymptotic variance for $\hat\beta$:
\begin{align}
    \Big\{\sum_{t=1}^T E ( I_t J_t D_t D_t^T )\Big\}^{-1} \Big[\sum_{t=1}^T \sum_{s=1}^T E \{ I_t I_s J_t J_s r_t(\alpha',\beta^0) r_s(\alpha',\beta^0) D_t D_s^T \} \Big] \Big\{\sum_{t=1}^T E ( I_t J_t D_t D_t^T ) \Big\}^{-1, T}. \label{eq:thm1-proofuse9}
\end{align}

\textbf{Step 4:} Show that the asymptotic variance \eqref{eq:thm1-proofuse9} can be consistently estimated by replacing all the $E$ with $\PP_n$ and replacing the in-probability limits of the estimators by the estimators themselves in \eqref{eq:thm1-proofuse9}. This can be shown using standard empirical process arguments; see, for example, Appendix D of \citet{bao2023estimating}.

\textbf{Step 5:} The theorem conclusion (asymptotic normality and consistent variance estimator) holds when $\tp_t(k|S_t)$ is estimated either parametrically or nonparametrically. This holds because $\tp_t(k|S_t)$ is not used in idenfifying the parameter $\beta$ and is only used for improving efficiency---in other words, the estimating function for $\beta$ is always unbiased regardless of the choice of $\tp_t(k|S_t)$. For parametrically estimated $\tp_t(k|S_t)$, the fact that the theorem still holds is due to \citet[Section 5.2]{lok2021estimating}. For nonparametrically estimated $\tp_t(k|S_t)$, the fact that the theorem still holds is due to \cite{newey1994asymptotic} or \citet[Theorem 5.1]{cheng2023efficient}.
This completes the proof of \cref{M-thm:CAN}.

\section{Technical Details for \texorpdfstring{\Cref{M-sec:sample-size-formula}}{Section 4} on Sample Size Formula}
\label{sec:proof-thm-sample-size-formula}

\subsection{Derivation of \texorpdfstring{$\tL$}{L tilde} in \texorpdfstring{\eqref{M-eq:null-and-alternative-tilde}}{(7)}}
\label{subsec:proof-L-tilde-matrix}

We prove that when \eqref{M-eq:working-assumption-on-MEE} hold, i.e., when $\mee_k(t) = f_t^T\beta_k$ for $t \in [T]$ and $k \in [K]$, the following are equivalent:
\begin{align}
    L \times \mee_{1:K}(t) = 0 \text{ for } t \in [T] \quad \text{and} \quad \tL \beta = 0. \label{eq:L-tilde-proofuse-0}
\end{align}
Recall that $L \in \RR^{\nu \times K}$, $\mee_{1:K}(t) := (\mee_1(t), \mee_2(t), \ldots, \mee_K(t))^T$, and $\tL:= L \otimes \II_p$, where $\II_k$ denotes a $k \times k$ identity matrix.

Let $L_{i,.}$ denote the $i$-th row of matrix $L$ and $l_{ij}$ the $(i,j)$-th the element of $L$. The $i$-th row of $L \times \mee_{1:K}(t) = 0$ is
\begin{align}
    [l_{i1},\ldots,l_{iK}] \left[\begin{matrix} \mee_1(t) \\ \vdots \\ \mee_K(t)  \end{matrix}\right] = 0. \label{eq:L-tilde-proofuse-1}
\end{align}
Because $\mee_k(t) = f_t^T\beta_k$, \eqref{eq:L-tilde-proofuse-1} is equivalent to
\begin{align}
    [l_{i1},\ldots,l_{iK}] \left[\begin{matrix} f_t^T\beta_1 \\ \vdots \\ f_t^T\beta_K  \end{matrix}\right] = 0. \label{eq:L-tilde-proofuse-2}
\end{align}
The left hand side of \eqref{eq:L-tilde-proofuse-2} is equal to
\begin{align*}
    & ~~~~ [l_{i1},\ldots,l_{iK}] \left[\begin{matrix} f_t^T\beta_1 \\ \vdots \\ f_t^T\beta_K  \end{matrix}\right] \\
    & = l_{i1}f_t^T\beta_1 + \cdots +l_{iK}f_t^T\beta_K \\
    & = f_t^T (l_{i1}\beta_1 + \cdots + l_{iK}\beta_K) \\
    & = f_t^T \left[\begin{matrix} l_{i1}\beta_{11} + \cdots + l_{iK}\beta_{K1} \\ l_{i1}\beta_{12} + \ldots + l_{iK}\beta_{K2} \\ \vdots \\ l_{i1}\beta_{1p} + \cdots + l_{iK}\beta_{Kp} \end{matrix}\right] \\
    & = f_t^T ([l_{i1},\ldots,l_{iK}] \otimes \II_p) \beta. 
\end{align*}
Therefore, \eqref{eq:L-tilde-proofuse-2} (and in turn \eqref{eq:L-tilde-proofuse-1}) is equivalent to $f_t^T ([l_{i1},\ldots,l_{iK}] \otimes \II_p) \beta = 0$. This implies that $L \times \mee_{1:K}(t) = 0$ is equivalent to $f_t^T (L\otimes \II_p) \beta = 0$. Therefore, $L \times \mee_{1:K}(t) = 0$ for all $t \in [T]$ is equivalent to
\begin{align}
    \left[\begin{matrix} f_1^T \\ \vdots \\ f_T^T \end{matrix}\right] (L\otimes \II_p) \beta = 0. \label{eq:L-tilde-proofuse-3}
\end{align}
And because the first matrix in \eqref{eq:L-tilde-proofuse-3} is full rank (a requirement we stated at the beginning of \Cref{M-sec:estimator}), this implies the equivalence \eqref{eq:L-tilde-proofuse-0}. This completes the proof.




\subsection{Additional Definitions}

Recall that in \cref{M-thm:sample-size-formula}, we assume that the randomization probability depends at most on the decision point index and is thus written as $p_t(k) := P(A_t = k) \equiv P(A_t = k \mid H_t)$. We also assume that the proximal outcome window $\Delta = 1$ and thus $Y_{t,\Delta}$ is simply written as $Y_t$. We will use $\beta^0$ and $\alpha^0$ to denote the true parameter values corresponding to (WA-a) and (WA-b).

We introduce additional definitions to reduce notation burden. We will sometimes use $\PP$ to denote expectation to reduce notation burden. In particular, $\PP f := E(f)$. We will write $\sum_{t=1}^T$ as $\sum_t$, $\sum_{t=1}^T \sum_{s=1}^T$ as $\sum_{t,s}$, and $\sum_{1 \leq t, s \leq T: t \neq s}$ as $\sum_{t\neq s}$.

Define
\begin{align}
    r_t(\alpha,\beta) := Y_t - \sum_{k=1}^K C_k(A_t) f_t^T \beta_k - g_t^T \alpha, \label{eq:rt-def}
\end{align}
where $Y_t$ and $A_t$ denote random variables of a generic subject (and thus the subscript $i$ is omitted). Define $\sigma^2_t : = \var(Y \mid I_t = 1, A_t)$, which does not depend on $A_t$ due to (WA-c). (Note that (WA-c) further implies that $\sigma^2_t = \sigma^2$ that does not depend on $t$, but we will not collapse $\sigma^2_t$ into $\sigma^2$ until necessary, to clearly show where each assumption is used.)

We will omit $(\alpha,\beta)$ when they are evaluated at $(\alpha^0, \beta^0)$: i.e., $r_t := r_t(\alpha^0, \beta^0), m := m(\alpha^0, \beta^0)$. 

In addition, define
\begin{align}
    C(A_t) := \begin{bmatrix} C_1(A_t) \\ C_2(A_t) \\ \vdots \\ C_K(A_t) \end{bmatrix}, \quad
    \tD_t :=
    \begin{bmatrix}
        g_t \\
        C_1(A_t) f_t\\
        \vdots\\
        C_K(A_t) f_t
    \end{bmatrix}
    \quad \text{and} \quad
    D_t :=
    \begin{bmatrix}
        C_1(A_t) f_t\\
        \vdots\\
        C_K(A_t) f_t
    \end{bmatrix}. \nonumber
\end{align}

\subsection{A Weaker Working Assumption (WA-g)}

We present a working assumption (WA-g) that is an implication of (WA-a)--(WA-f), and we prove \cref{M-thm:sample-size-formula}(ii) under (WA-a), (WA-b), (WA-c), (WA-d), and (WA-g). (WA-g) is more technical and harder to interpret, so we presented its sufficient conditions (WA-e) and (WA-f) in the paper.

Consider the following working assumption:
\begin{itemize}
    \item[(WA-g)] $E\{r_t(\alpha^0,\beta^0) ~ r_s(\alpha^0,\beta^0) \mid I_t = 1, I_s = 1, A_t, A_s\}$ is constant in $A_t, A_s$ for all $1 \leq s < t \leq m$.
\end{itemize}

(WA-g) is weaker than (WA-e) plus (WA-f) in the sense that when (WA-a) and (WA-b) hold, (WA-e) and (WA-f) implies (WA-g). This can be established in the same way as Lemma B.2 in \citet{cohn2023sample} except that their exponential function will be replaced by the identity function.

\subsection{Lemmas}

Note that all the lemmas in this subsection concern the particulra setting considered in the sample size calculation \Cref{M-sec:sample-size-formula}, where the interest is in the immediate, marginal CEE (i.e., $\Delta = 1$ and $S_t = \emptyset$ in \eqref{M-eq:cee-def}), and the randomization probability at each decision point is either constant or dependent only on $t$. We do not include this specification in each lemma statement to avoid repetition.

\begin{lem}
    \label{lem:thm2-proofuse1}
    We have the following results:
    \begin{align}
        \PP I_t C_k(A_t) & = 0, \\
        \PP I_t C_k(A_t) C_l(A_t) & = - E(I_t) p_t(k) p_t(l), \\
        \PP I_t C_k(A_t)^2 & = E(I_t) p_t(k) \{1 - p_t(k)\}.
    \end{align}
\end{lem}

\begin{proof}[Proof of \cref{lem:thm2-proofuse1}]
    The first equation and the third equation follow immediately by iterated expectation conditional on $I_t = 1$.

    The second equation follows from
    \begin{align}
        \PP I_t C_k(A_t) C_l(A_t) & = E(I_t) E[ \{\one(A_t = k) - p_t(k)\} \{\one(A_t = l) - p_t(l)\} \mid I_t = 1] \nonumber \\
        & = - E(I_t) p_t(k) p_t(l). \nonumber
    \end{align}
\end{proof}

\begin{lem}
    \label{lem:thm2-proofuse2}
    Under (WA-a), (WA-b), and (WA-c), we have
    \begin{align}
        \sum_{t=1}^T \PP I_t r_t^2(\alpha^0,\beta^0) D_t D_t^T = \sum_{t=1}^T E(I_t) \sigma^2_t P_t \otimes (f_t f_t^T).
    \end{align}
\end{lem}

\begin{proof}[Proof of \cref{lem:thm2-proofuse2}]
    First, (WA-a), (WA-b), and the fact that the randomization probability depends at most on $t$ implies that $E(r_t \mid I_t = 1, A_t) = 0$. Therefore, for any $0 \leq l \leq K$ we have
    \begin{align}
        & ~~~~ E(r_t^2 \mid I_t = 1, A_t = l) = \var(r_t \mid I_t = 1, A_t = l) \nonumber \\
        & = \var\Big\{Y_t - \sum_{k=1}^K C_k(A_t) f_t^T \beta_k - g_t^T \alpha \mid I_t = 1, A_t = l \Big\} \nonumber \\
        & = \var(Y \mid I_t = 1, A_t = l) = \sigma^2_t, \label{eq:lem2-thm2-proofuse2}
    \end{align}
    and where the last equality (and the fact that $\sigma^2_t$ does not depend on $l$) is due to (WA-c).
    
    Second, for any $1 \leq k \neq l \leq K$, we have
    \begin{align}
        \PP I_t r_t^2 C_k(A_t) C_l(A_t) & = E(I_t) E\{r_t^2 C_k(A_t) C_l(A_t) \mid I_t = 1\} \nonumber \\
        & = E(I_t) E(r_t^2 \mid I_t = 1, A_t = k) \{1 - p_t(k)\}\{-p_t(l)\} p_t(k) \nonumber \\
        & ~~~~ + E(I_t) E(r_t^2 \mid I_t = 1, A_t = l) \{- p_t(k)\}\{1 - p_t(l)\} p_t(l) \nonumber \\
        & ~~~~ + E(I_t) E(r_t^2 \mid I_t = 1, A_t \notin \{k, l\} ) \{- p_t(k)\} \{- p_t(l)\} \{1 - p_t(k) - p_t(l)\} \nonumber \\
        & = - E(I_t) \sigma^2_t p_t(k) p_t(l), \label{eq:lem2-thm2-proofuse3}
    \end{align}
    where the last equality follows from \eqref{eq:lem2-thm2-proofuse2}.

    Third, for any $1 \leq k \leq K$, we have
    \begin{align}
        \PP I_t r_t^2 C_k^2(A_t) & = E(I_t) E\{r_t^2 C_k^2(A_t) \mid I_t = 1\} \nonumber \\
        & = E(I_t) E(r_t^2 \mid I_t = 1, A_t = k) \{1 - p_t(k)\}^2 p_t(k) \nonumber \\
        & ~~~~ + E(I_t) E(r_t^2 \mid I_t = 1, A_t \neq k) \{- p_t(k)\}^2 \{1 - p_t(k)\} \nonumber \\
        & = E(I_t) \sigma^2_t p_t(k)\{1 - p_t(k)\}. \label{eq:lem2-thm2-proofuse4}
    \end{align}
    
    Finally, it is easy to verify that
    \begin{align}
        D_t D_t^T = \begin{bmatrix} C_1^2(A_t) & \ldots & C_1(A_t) C_K(A_t) \\ \vdots & & \vdots \\ C_K(A_t)C_1(A_t) & \ldots & C_K^2(A_t) \end{bmatrix} \otimes (f_t f_t^T). \label{eq:lem2-thm2-proofuse5}
    \end{align}
    So putting together \eqref{eq:lem2-thm2-proofuse3}, \eqref{eq:lem2-thm2-proofuse4}, and \eqref{eq:lem2-thm2-proofuse4} yields
    \begin{align}
        \PP I_t r_t^2(\alpha^0,\beta^0)^2 D_t D_t^T = E(I_t) \sigma^2_t P_t \otimes (f_t f_t^T). \nonumber
    \end{align}
    Thus the lemma statement follows. This completes the proof.
\end{proof}

\begin{lem}
    \label{lem:thm2-proofuse3}
    Under (WA-a), (WA-b), and (WA-g), we have for any $1 \leq t \neq s \leq T$
    \begin{align}
        \PP I_t r_t(\alpha^0,\beta^0) r_s(\alpha^0,\beta^0) D_t D_s^T = 0.
    \end{align}
\end{lem}

\begin{proof}[Proof of \cref{lem:thm2-proofuse3}]
    For any $1 \leq k, l \leq K$ (including the case of $k = l$), we have
    \begin{align}
        \PP I_t I_s r_t r_s C_k(A_t) C_l(A_s) & = E(I_t I_s) E\{r_t r_s C_k(A_t) C_l(A_s) \mid I_t = 1, I_s = 1\} \nonumber \\
        & = E(I_t I_s) E\{E(r_t r_s \mid I_t = 1, I_s = 1, A_t, A_s) C_k(A_t) C_l(A_s) \mid I_t = 1, I_s = 1\} \nonumber \\
        & = E(I_t I_s) E\{ c_{t,s} C_k(A_t) C_l(A_s) \mid I_t = 1, I_s = 1\} \label{eq:lem3-thm2-proofuse1} \\
        & = 0, \label{eq:lem3-thm2-proofuse2}
    \end{align}
    where \eqref{eq:lem3-thm2-proofuse1} follows from (WA-g) and \eqref{eq:lem3-thm2-proofuse2} follows from the fact that the randomization probability depends at most on $t$.

    In addition, it is easy to verify that
    \begin{align}
        D_t D_s^T = \begin{bmatrix} C_1(A_t) C_1(A_s) & \ldots & C_1(A_t) C_K(A_s) \\ \vdots & & \vdots \\ C_K(A_t)C_1(A_s) & \ldots & C_K(A_t) C_K(A_s) \end{bmatrix} \otimes (f_t f_t^T). \label{eq:lem3-thm2-proofuse3}
    \end{align}
    Equations \eqref{eq:lem3-thm2-proofuse2} and \eqref{eq:lem3-thm2-proofuse3} imply the lemma. This completes the proof.
\end{proof}

\subsection{Proof of \texorpdfstring{\cref{M-thm:sample-size-formula}}{Theorem 2} Under Weaker Working Assumptions}

\cref{M-thm:sample-size-formula}(i) is already established in \cref{M-subsec:hypothesis-test-statistic-rejection-region}. We now prove \cref{M-thm:sample-size-formula}(ii) under working assumptions (WA-a), (WA-b), (WA-c), (WA-d), and (WA-g). Given the derivation in \cref{M-subsec:sample-size-formula} up to \cref{M-thm:sample-size-formula}, it suffices to show that under these working assumptions the asymptotic variance of $\hat\beta$, $M^{-1}\Sigma M^{-1,T}$, simplifies to $\bar\sigma^2 V$ with $V$ defined in \cref{M-alg:ss-calculator}.

\cref{M-sec:sample-size-formula} considers the setting where the randomization probability depends at most on $t$ and thus can be written as $p_t(k)$, and thus we can set $\tp_t(k) = p_t(k)$. Therefore, the estimating equation for $(\alpha,\beta)$ defined in \eqref{M-eq:ee} becomes
\begin{align}
    m(\alpha,\beta) & = \sum_{t=1}^T I_t \Big\{Y_t - \sum_{k=1}^K C_k(A_t) f_t^T\beta_k - g_t^T\alpha\Big\}
    \begin{bmatrix}
        g_t \\
        C_1(A_t) f_t\\
        \vdots\\
        C_K(A_t) f_t
    \end{bmatrix}
    = \sum_{t=1}^T I_t r_t(\alpha,\beta) \tD_t. \nonumber
\end{align}

Now we calculate the asymptotic variance for $\hat\beta$ under the working assumptions. By \cref{M-thm:CAN}, the asymptotic variance for $(\hat\alpha,\hat\beta)$ has the form
\begin{align}
    \{\PP \dot{m}(\alpha^0, \beta^0)\}^{-1} ~ \{\PP m(\alpha^0, \beta^0) m(\alpha^0, \beta^0)^T\} ~ \{\PP \dot{m}(\alpha^0, \beta^0)\}^{-1,T}. \label{eq:thm2-proofuse1}
\end{align}
We compute each term in \eqref{eq:thm2-proofuse1}.

\textbf{Step 1: Compute $\{\PP \dot{m}(\alpha^0, \beta^0)\}^{-1}$.} We have $\dot{m}(\alpha,\beta) = \sum_t I_t \tD_t \dot{r}_t(\alpha, \beta) = - \sum_t I_t \tD_t \tD_t^T$. Therefore,
\begin{align}
    \PP \dot{m}(\alpha, \beta)
    & = - \PP \sum_t I_t \tD_t \tD_t^T = - \PP \sum_t I_t \veccol{g_t}{D_t} \vecrow{g_t^T}{D_t^T} \nonumber \\
    & = - \PP \sum_t I_t \begin{bmatrix} g_t g_t^T & g_t D_t^T \\ D_t g_t^T & D_t D_t^T \end{bmatrix} \nonumber \\
    & = - \sum_t  \begin{bmatrix} E(I_t) g_t g_t^T & \PP  g_t E(I_t D_t^T) \\ E(I_t D_t) g_t^T & E(I_t D_t D_t^T) \end{bmatrix}. \label{eq:thm2-proofuse2}
\end{align}
\cref{lem:thm2-proofuse1} implies that $g_t E(I_t D_t^T) = E(I_t D_t) g_t^T = 0$ and $E(I_t D_t D_t^T) = E(I_t) P_t \otimes f_t f_t^T$, where $P_t \in \RR^{K \times K}$ whose $(k_1,k_2)$-th entry equals $p_t(k_1)\{1-p_t(k_1)\}$ when $k_1 = k_2$ and $-p_t(k_1)p_t(k_2)$ when $k_1 \neq k_2$. Therefore, for any $(\alpha,\beta)$ we have
\begin{align}
    \{\PP \dot{m}(\alpha, \beta)\}^{-1} = - \begin{bmatrix} \{\sum_t E(I_t) g_t g_t^T\}^{-1} & 0 \\ 0 & \{\sum_t E(I_t) P_t \otimes (f_t f_t^T)\}^{-1} \end{bmatrix}. \label{eq:thm2-proofuse3}
\end{align}

\textbf{Step 2: Compute $\PP m(\alpha^0, \beta^0) m(\alpha^0, \beta^0)^T$.} Because the goal is to derive the asymptotic variance for $\hat\beta$ (the lower $Kp \times Kp$ principal submatrix of \eqref{eq:thm2-proofuse1}) and because the off-diagonal blocks of $\{\PP \dot{m}(\alpha^0, \beta^0)\}^{-1}$ is 0 (derived in \eqref{eq:thm2-proofuse3}), it suffices to compute the lower $Kp \times Kp$ principal submatrix of $\PP m(\alpha^0, \beta^0) m(\alpha^0, \beta^0)^T$. We have
\begin{align}
    \PP m(\alpha^0, \beta^0) m(\alpha^0, \beta^0)^T & = \PP \Big(\sum_t I_t r_t \tD_t\Big) \Big(\sum_s I_s r_s \tD_s\Big)^T \nonumber \\
    & = \PP \sum_{t,s} I_t I_s r_t r_s \veccol{g_t}{D_t} \vecrow{g_s^T}{D_s^T} \nonumber \\
    & = \PP \sum_{t,s} I_t I_s r_t r_s \begin{bmatrix} g_t g_s^T & g_t D_s^T \\ D_t g_s^T & D_t D_s^T \end{bmatrix}. \nonumber
\end{align}
Therefore, the lower $Kp \times Kp$ principal submatrix of $\PP m(\alpha^0, \beta^0) m(\alpha^0, \beta^0)^T$ is
\begin{align}
    \PP \sum_{t,s} I_t I_s r_t r_s D_t D_s^T & = \sum_{t} \PP I_t r_t^2 D_t D_t^T + \sum_{t\neq s} \PP I_t I_s r_t r_s D_t D_s^T \nonumber \\
    & = \sum_{t=1}^T E(I_t) \sigma^2_t P_t \otimes (f_t f_t^T) + 0 \label{eq:thm2-proofuse4} \\
    & = \sigma^2 \sum_{t=1}^T E(I_t) P_t \otimes (f_t f_t^T), \label{eq:thm2-proofuse5}
\end{align}
where \eqref{eq:thm2-proofuse4} follows from \cref{lem:thm2-proofuse2} and \cref{lem:thm2-proofuse3}, and \eqref{eq:thm2-proofuse5} follows from (WA-c).

\textbf{Step 3: Computing the asymptotic variance of $\hat\beta$.} Plugging \eqref{eq:thm2-proofuse3} and \eqref{eq:thm2-proofuse5} into \eqref{eq:thm2-proofuse1} implies that the asymptotic variance of $\hat\beta$ is
\begin{align}
    \sigma^2 \left\{\sum_{t=1}^T E(I_t) P_t \otimes (f_t f_t^T)\right\}^{-1}.
\end{align}

This completes the proof.

\section{Simulation Results on Consistency and Asymptotic Normality}
\label{sec:simulation-estimator}

We evaluate the consistency and asymptotic normality of the estimator $\hat\beta$ for the CEE model \eqref{M-eq:CEE-parametric-model}, proposed in \Cref{M-sec:estimator}. We focus on the setting where the length of the excursion $\Delta = 1$ and participants are always available ($I_t \equiv 1$). In the generative model, we set the total number of decision points for each individual $T = 15$. A time-varying covariate $Z_t$, which is independent of all variables and takes three values 0, 1, and 2 with equal probability, is generated.

The number of active treatment levels is $K = 2$, and the categorical treatment $A_t \in \{0,1,2\}$ is generated from a categorical distribution with $P(A_t = 0) = 0.2$, $P(A_t = 1) = 0.5$, $P(A_t = 2) = 0.3$. The outcome $Y_t = Y_{t,\Delta = 1}$ is generated by $Y_t  = 0.2 \one(Z_t = 0) + 0.5 \one(Z_t = 1) + 0.4 \one(Z_t = 2) + \one(A_t = 1) (0.1  + 0.3 Z_t) + \one(A_t = 2) (0.45 + 0.1 Z_t) + \epsilon_t$.
    
We consider two sets of estimands in the simulation. When setting $S_t = \emptyset$ in \eqref{M-eq:cee-def}, the fully marginal CEEs are $\cee_{t1}(\emptyset) = 0.4$ and $\cee_{t2}(\emptyset) = 0.55$. When setting $S_t = Z_t$, the moderated CEEs are $\cee_{t1}(Z_t) = 0.1 + 0.3 Z_t$ and $\cee_{t2}(Z_t) = 0.45 + 0.1 Z_t$. We use the estimating function \eqref{M-eq:ee} to estimate both CEEs by setting different $S_t$'s. The working model for $E(Y \mid H_t)$ is misspecified as $g_t(H_t)^T\alpha = \alpha_0 + \alpha_1 Z_t$. The numerator probabilities $\tp_t(k \mid S_t)$ are set to be equal to the constant randomization probabilities and thus $J_t \equiv 1$.

\Cref{tbl:est_no_moderator} shows the simulation results for estimating the fully marginal CEEs ($S_t = 1$) and \Cref{tbl:est_moderator} shows the simulation results for estimating the moderated CEEs ($S_t = Z_t$). The bias, root mean squared error, standard deviation, and 95\% confidence interval coverage probability are based on 1000 replicates. In both settings, the proposed method yields consistent estimators with nominal coverage probability. 

\begin{table}[htbp]
\caption{Simulation results for the proposed estimator for fully marginal and moderated CEEs.}
\label{tab:consistency}
 \begin{subtable}[h]{1\textwidth}
\resizebox{\textwidth}{!}{
\begin{tabular}{c|cccc|cccc}

\multicolumn{1}{c}{ } & \multicolumn{4}{c}{$\beta_1$} & \multicolumn{4}{c}{$\beta_2$} \\
Sample size & Bias & RMSE & SD & CP & Bias & RMSE & SD & CP\\
\hline
15 & -0.002 & 0.219 & 0.241 & 0.955 & 0.015 & 0.238 & 0.252 & 0.958\\
20 & 0.001 & 0.193 & 0.204 & 0.966 & 0.003 & 0.208 & 0.223 & 0.959\\
25 & 0.001 & 0.174 & 0.186 & 0.954 & -0.006 & 0.189 & 0.201 & 0.956\\
30 & 0.003 & 0.158 & 0.170 & 0.945 & 0.001 & 0.171 & 0.181 & 0.948\\
35 & 0.005 & 0.148 & 0.149 & 0.952 & 0.004 & 0.160 & 0.162 & 0.958\\
40 & -0.003 & 0.141 & 0.142 & 0.962 & -0.005 & 0.152 & 0.155 & 0.959\\
45 & 0.000 & 0.132 & 0.137 & 0.949 & 0.000 & 0.142 & 0.149 & 0.959\\
50 & -0.002 & 0.125 & 0.129 & 0.945 & -0.002 & 0.135 & 0.143 & 0.945\\
\hline
\end{tabular}}
 \caption{\label{tbl:est_no_moderator} Estimator for constant marginal excursion effect where $\beta_1$ and $\beta_2$ is the treatment effect for treatment 1 and 2 respectively; SD, standard deviation; RMSE, root mean square; CP, 95\% confidence interval coverage probability.}
\vspace{1em}
 \end{subtable}
  \begin{subtable}[h]{1\textwidth}
\resizebox{\textwidth}{!}{
\begin{tabular}{cc|cccc|cccc}

\multicolumn{2}{c}{ } & \multicolumn{4}{c}{Intercept} & \multicolumn{4}{c}{$Z_t$} \\

Trt & Sample size & Bias & RMSE & SD & CP & Bias & RMSE & SD & CP\\
\cline{1-10}
1 & 15 & 0.012 & 0.361 & 0.360 & 0.959 & -0.011 & 0.287 & 0.287 & 0.965\\
 & 20 & -0.002 & 0.313 & 0.313 & 0.959 & 0.004 & 0.245 & 0.245 & 0.966\\
 & 25 & -0.005 & 0.286 & 0.286 & 0.948 & 0.004 & 0.216 & 0.216 & 0.968\\
 & 30 & -0.004 & 0.264 & 0.264 & 0.948 & 0.006 & 0.201 & 0.201 & 0.943\\
 & 40 & -0.002 & 0.224 & 0.224 & 0.955 & 0.004 & 0.172 & 0.172 & 0.959\\
 & 50 & 0.001 & 0.194 & 0.194 & 0.956 & -0.004 & 0.153 & 0.153 & 0.962\\
  \cline{1-10}
2 & 15 & 0.023 & 0.396 & 0.395 & 0.963 & -0.009 & 0.325 & 0.324 & 0.959\\
 & 20 & 0.014 & 0.335 & 0.335 & 0.962 & -0.011 & 0.263 & 0.263 & 0.959\\
 & 25 & -0.011 & 0.305 & 0.305 & 0.965 & 0.004 & 0.233 & 0.233 & 0.948\\
 & 30 & -0.009 & 0.287 & 0.287 & 0.954 & 0.008 & 0.220 & 0.220 & 0.948\\
 & 40 & 0.002 & 0.245 & 0.245 & 0.951 & 0.005 & 0.190 & 0.189 & 0.955\\
 & 50 & 0.008 & 0.214 & 0.214 & 0.950 & -0.008 & 0.164 & 0.164 & 0.956\\ \hline
\end{tabular}}
\caption{\label{tbl:est_moderator} Estimator for excursion effect with $S_t = Z_t$ for 2 levels of treatment; SD, standard deviation; RMSE, root mean square; CP, 95\% confidence interval coverage probability.}
   \end{subtable}
\end{table}

\section{Simulation Results on Performance of Sample Size Formula}
\label{sec:simulation-sample-size-formula}

\subsection{Generative Models}

We assess the performance of the sample size calculator and the testing procedure proposed in \Cref{M-sec:sample-size-formula} in terms of type I error and power, when all working assumptions hold and when some working assumptions are violated. For all simulations, the desired type I error rate is 0.05 and the desired power is 0.8. In the generative models, the number of active treatment levels is $K = 2$ so $A_t \in \{0,1,2\}$, and we test for $\cH_0: \mee_1(t) = \mee_2(t)$ for all $t \in [T]$ vs. $\cH_1: \mee_1(t) \neq \mee_2(t)$ for some $t \in [T]$. For notation simplicity we will write $\Delta\sate_{12}$ as $\Delta\sate$.


We consider a simple generative model (GM-0) and three generative models with more complicated features: a generative model where the error variance depends on time and current treatment (GM-EV), a generative model where the outcomes are serially correlated (GM-SC), and a generative model where the availability process depends on previous outcomes and previous treatments, i.e., is endogeneous (GM-EA). We denote parameters associated with the true generative model by superscript ``*'', and the input of the sample size formula, i.e., the working model, by superscript ``w''.

GM-0 is characterized by a set of parameters : $\{p^*_{kt}: k=0,1,2\}, \alpha^*, g_t^*, \{\beta_k^*:k=1,2\}, f_t^*$, and $\tau^*(t)$ for $t \in [T]$. $p^*_{kt}$ is the randomization probability for treatment level $k$: $p^*_{kt} = P(A_t = k \mid I_t =1)$, and recall that $P(A_t = 0 \mid I_t =0) = 1$. $\alpha^*$ and $g_t^*$ determine the mean proximal outcome under no treatment. $\beta_k^*$ and $f_t^*$ determine the marginal excursion effect of treatment $k$. $\tau(t)^*$ is the time-varying availability pattern. For each individual, their observations are generated sequentially over $t \in [T]$:  $I_t \sim \text{Bernoulli}(\tau^*(t))$. $A_t = 0$ if $I_t = 0$. $A_t \mid I_t = 1$ is generated from a categorical distribution with support $\{0, 1, 2\}$ and probabilities $\{p^*_{kt}: k=0,1,2\}$. $Y_t$ is generated as $Y_t  = g_t^{*T} \alpha^* + \one(A_t = 1) f_t^{*T} \beta_1^* + \one(A_t = 2) f_t^{*T} \beta_2^* + \epsilon_t$, where $\epsilon_t \sim N(0,1)$. Details about the specifications of the parameters are given in \Cref{box:detail-gm-0-part1} and \Cref{box:detail-gm-0-part2}.

For GM-EV and GM-SC, $I_t$ and $A_t$ are generated the same way as in GM-0, and $Y_t$ is generated differently. In GM-EV, $Y_t$ is generated as $Y_t = g_t^{*T} \alpha^* + \one(A_t = 1) f_t^{*T} \beta_1^* + \one(A_t = 2) f_t^{*T} \beta_2^* + r(t) \times s(A_t) \epsilon_t$, where $\epsilon_t \sim N(0, 1)$. $r(t)$ and $s(A_t)$ make the error variance depend on time and current treatment and are chosen so that $\bar{\sigma}^2 = 1$ with $\bar{\sigma}^2$ defined in \eqref{M-eq:sigma_bar}. The exact form of $r(t)$ and $s(A_t)$ is given in \Cref{box:detail-gm-ev}. For GM-SC, $Y_t$ is generated as $Y_t = g_t^{*T} \alpha^* + \one(A_t = 1) f_t^{*T} \beta_1^* + \one(A_t = 2) f_t^{*T} \beta_2^* + \nu_1 \epsilon_{it-1} + \nu_0 \epsilon_t$, where $\epsilon_t \sim N(0,1)$. The term $\nu_1 \epsilon_{it-1}$ makes the outcomes serially correlated, and $\nu_1$ takes value between 0 and 1. The term $\nu_0$ is calculated such that the error variance $\bar{\sigma}^2 = 1$.

For GM-EA, $I_t$ is generated from a Bernoulli distribution with success probability
\begin{align*}
    \tau(t-1) + \nu_2 \{\one(A_{t-1} = 1) - p_{1(t-1)}  + \one(A_{t-1} = 2) - p_{2(t-1)} \} + \nu_3 \text{Trunc}(\epsilon_{t-1}),
\end{align*}
where $\text{Trunc}(x) = x \one(|x| \leq 1) + \text{sign}(x) \one(|x|>1)$. The $\nu_2$-term and $\nu_3$-term make $I_t$ depend on previous treatment and previous outcome, and $\nu_2$ and $\nu_3$ take values between $-0.2$ and $0.2$. $A_t$ and $Y_t$ are generated the same way as in GM-0.

\begin{guidelinebox}[htbp]
    \caption{Details about GM-0: $\tau(t)$ and $\eo(t)$.}
    \label{box:detail-gm-0-part1}
    \begin{mdframed}[linewidth=1pt, roundcorner=10pt, backgroundcolor=gray!10]
    \spacingset{1}

    \noindent \textbf{Parameterization of $\tau(t)$:}

    Given AA, we consider two parameterization of $\tau(t)$: constant and linear 
    \begin{itemize}
        \item Constant $\tau(t)$: We set $\tau(t) = AA$ for all $t$.
        \item Linear $\tau(t)$ parameterized by $\theta_{\tau}$: We set $\tau^*(t)$ to vary linearly in t and is parameterized by $\theta_\tau$ such that $\tau(1) = AA^* +\theta_\tau$ and $\tau(T) = AA - \theta_\tau$. In other words, $\tau(t) = \frac{-2}{T - 1} \theta_\tau t + AA -  \frac{T+1}{T-1} \theta_\tau$.
    \end{itemize}

    \noindent \textbf{Parameterization of $\eo(t)$:}

    Given $\aeo(t)$ and $\tau(t)$, we consider three parameterizations of $\eo(t)$: constant, linear, and quadratic. 
    \begin{itemize}
        \item Constant $\eo(t)$: We set $\eo(t) = \aeo$ for all t. 
        \item Linear $\eo(t)$ parameterized by $\theta_g$; We set $\eo(t) = \alpha_0 + \alpha_1 t$, where $(\alpha_0, \alpha_1)$ is determined as follows. A positive (negative) $\theta_g$ indicates a linearly increasing (decreasing) $\eo(t)$. In particular, if $\theta_g = 0$ we set $\alpha_1 = 0$ and $\alpha_0 = \aeo$. If $\theta_g \in (-1, 0)\cup (0,1)$, we set 
        \begin{align} 
            \frac{\alpha_0 + \alpha_1}{\alpha_0 + T \alpha_1} = \frac{1 + \theta_g}{1 - \theta_g}
        \end{align} 
        and thus 
        \begin{align*}\label{theta g}
            \alpha_1 = \frac{2 \theta_g}{ 1- T - \theta_g(1 - T)} \alpha_0,
        \end{align*}
        and then we solve for $\alpha_0$ and $\alpha_1$ from equation \eqref{theta g} using the given $\aeo$ and $\tau(t)$.
        \item quadratic $\eo(t)$ parameterized by $\theta_g$; We set $\aeo(t) = \alpha_0 + \alpha_1 t + \alpha_2 t^2$ where $(\alpha_0, \alpha_1, \alpha_2$ is determined as follows. We constraint $\aeo(1) = \aeo(t)$ and a positive (negative) $\theta_g$ indicates the parabola $\aeo(t)$ opens upward (downward). In particular for $\theta_g \in (-1, 1)$, we set : 
        \begin{align}
            \alpha_0 + \alpha_1 + \alpha_2 &= \alpha_0 + \alpha_1 T + \alpha_2 T^2 \nonumber \\
            \frac{\eo\{(T+1)/2\}}{\eo(1)} &= \frac{\alpha_0 + \frac{T+1}{2}\alpha_1 + \frac{(T+1)^2}{4} \alpha_2}{\alpha_0 + \alpha_1 + \alpha_2} = \frac{1 + \theta_g}{ 1 - \theta_g} 
        \end{align}
    \end{itemize}
    \end{mdframed}
\end{guidelinebox}

\begin{guidelinebox}[htbp]
    \caption{Details about GM-0: $\mee_k(t)$.}
    \label{box:detail-gm-0-part2}
    \begin{mdframed}[linewidth=1pt, roundcorner=10pt, backgroundcolor=gray!10]
    \spacingset{1}
    \noindent \textbf{Parameterization of $\mee_k(t)$:}

    Given $\ate_1$, $\ate_2$, and $\tau(t)$, we consider two parameterizations of $\mee_1(t)$ and $\mee_2(t)$: constants and linear, 
    \begin{itemize}
        \item Constant $\mee_k(t)$: We set $\mee_k(t) = \ate_k$ for all $t \in [T]$. 
        \item Linear $\mee_k(t)$ parameterized by $\theta_{f1}$ and $\theta_{f2}$, where $\theta_{f1}$ parameterize the slope of $\mee_1(t)$, while $\theta_{f2}$ parameterize the difference between the slope of two treatment level, i.e $\theta_{f2} = 0$ indicates that the two treatment have a parallel slope. We set $\mee_1(t) = \beta_1 + \beta_2 t$, where $(\beta_1, \beta_2)$ is determined as follows: a positive (negative) $\theta_{f1}$ indicates a linearly increasing (decreasing) $\mee_1(t)$. In particular, if $\theta_{f1} = 0$ we set $\beta_2 = 0$ and $\beta_1 = \ate$. If $\theta_{f1} \in (-1, 0)\cup (0,1)$, we set 
        \begin{align} 
            \frac{\beta_1+ \beta_2}{\beta_1 + T \beta_2} = \frac{1 + \theta_{f1}}{1 - \theta_{f1}}.
        \end{align} 
        We set $\mee_2(t) = \mee_1(t) + \beta_3 + \beta_4 t$, when $\theta_{f2} = 0$, we set $\beta_4 = 0$ and $\beta_3 = \ate_2(t)$. If $\theta_{f2} \in (-1, 0)\cup (0,1)$, we set 
        
        \begin{align} 
            \frac{\beta_3 + \beta_4}{\beta_3 + T \beta_4} = \frac{1 + \theta_{f2}}{1 - \theta_{f2}}.
        \end{align} 
        and then we solve for $\beta_1$, $\beta_2$, $\beta_3$ and $\beta_4$ from equation \eqref{M-eq:ate_k} using the given $\ate_k$ and $\tau(t)$.
    \end{itemize}
    \end{mdframed}
\end{guidelinebox}

\begin{guidelinebox}[htbp]
    \caption{Details about GM-EV: $r(t)$ and $s(A_t)$.}
    \label{box:detail-gm-ev}
    \begin{mdframed}[linewidth=1pt, roundcorner=10pt, backgroundcolor=gray!10]
    \spacingset{1}

    Here we describe the generative model where the error variance is not constant in time ($t$) and treatment assignment ($A_t$). $Y_{it}$ is generated as 
    \begin{align*}
        Y_t = g_t^{*T} \alpha^* + \one(A_t = 1) f_t^{*T} \beta_1^* + \one(A_t = 2) f_t^{*T} \beta_2^* + r(t) \times s(A_t) \epsilon_t
    \end{align*}

    \noindent\textbf{Parameterization of $r(t)$:}

    Denote $r(t)$ as a linear function of t such that $ r(1) = \sqrt{1 + \frac{\theta_r}{\bar{\sigma}^2}}$ and $r(T) = \sqrt{1 - \frac{\theta_r}{\bar{\sigma}^2}}$, where $\theta_r$ measures the effect of time on the variance of $Y_t$, and $\theta_r = 0$ implies that variance doesn't depend on time. 

    \noindent\textbf{Parameterization of $s(A_t)$:}

    Denote $s(A_t) = \sqrt{a + \one_{(A_t = 1)} \theta_s + \one_{(A_t = 2)} b}$, where $a$ and $b$ are selected so that $\E[s(A_t)]= 1$, and thus 
    \begin{align*}
        b = -\left( \frac{p_t(1) - 1}{p_t(2) - 1}\right) \theta_s,
    \end{align*}
    and  $a = 1 - \theta_s - b$. Where $\theta_s$ is the parameter that measures the effect of treatment level on the variance of $Y_t$, and  $\theta_s = 0$ implies that variance doesn't depend on treatment level.
    \end{mdframed}
\end{guidelinebox}

\spacingset{1.9}

\subsection{Type I error when working assumptions do or do not hold}
\label{subsec:simulation-result-type-i-error}

\begin{figure}[htbp]
\centering
\includegraphics[width=1\textwidth]{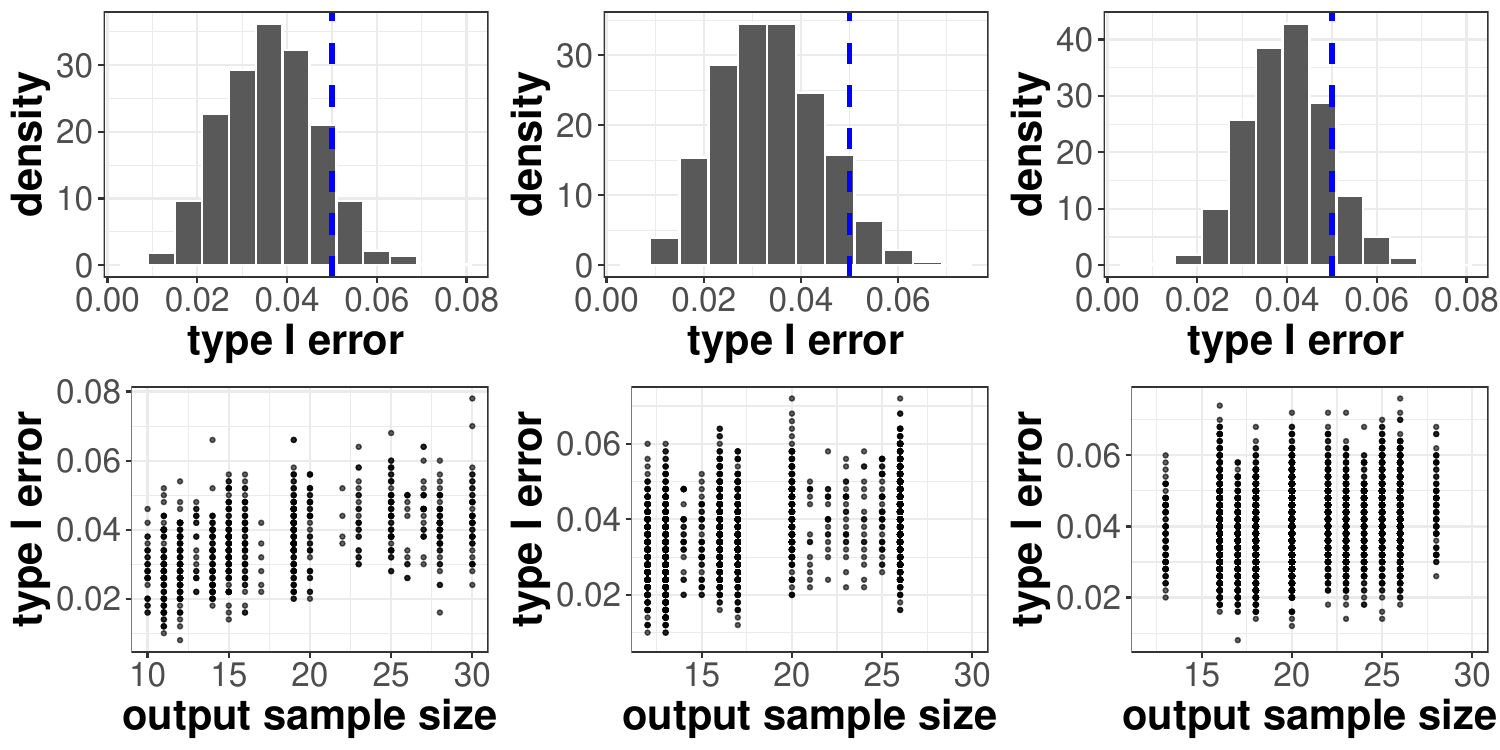}
  \caption{\textbf{Left plots:} type I error when all working assumptions hold. \textbf{Middle plots:} type I error when (WA-a) is violated. \textbf{Right plots:} type I error when other WAs are violated but (WA-a) holds. Each dot in the bottom plots represent the type I error under a specific generative model obtained from 1,000 simulation replicates. Histograms in the top plots capture the distribution of empirical type I error from simulations using the various generative models.}
  \label{fig:typeierror_hist}
\end{figure}

To assess the type I error numerically, we considered three sets of generative models: (a) all working assumptions hold; (b) (WA-a) is violated; and (c) one or more of (WA-b)-(WA-f) is violated but (WA-a) hold. Note that \Cref{M-thm:sample-size-formula}(i) guarentees asymptotic type I error control for (a) and (c) but not (b). Simulation results show that type I error is always controlled at the 0.05 level in all three scenarios (\Cref{fig:typeierror_hist}), including when (WA-a) is violated (middle panels of \Cref{fig:typeierror_hist}). We note that the small sample correction helped to improve the type I error control (empirical type I error without small sample correction is not shown in the figure).

Specifically, for (a) we considered 1,000 settings (each corresponding to a specific parameterization of the data generating distribution) when all working assumptions hold using GM-0; \Cref{box:sim-all-wa-hold-1000-settings} lists the 1,000 settings. For (b) we misspecified separately the magnitude of $\mee(t)$ or the pattern of $\mee(t)$ using GM-0; \Cref{subsec:simulation-violate-a} on power contains the specific form of misspecification. For (c) we used various generative models from GM-EV, GM-SC, and GM-EV.

\begin{guidelinebox}[htbp]
    \caption{The 1,000 simulation settings when all working assumptions hold.}
    \label{box:sim-all-wa-hold-1000-settings}
    \begin{mdframed}[linewidth=1pt, roundcorner=10pt, backgroundcolor=gray!10]
    \spacingset{1}

    The 1,000 simulation settings are from the factorial design of the following factors ($1000 = 25 \times 4 \times 2 \times 5$): settings under of a combination of constant/linear $\mee_k(t)$ with various $\theta_f$ values, constant/linear/quadratic $g_t$ with various $\theta_g$ values (subject to the constraint that $f_t$ is a subset of $g_t$), constant/linear $\tau(t)$ with various $\theta_{\tau}$ values, and a variety of values for $\ate^*_1$, $\ate^*_2$, $\text{AEO}^*$, and $\aaa^*$.

    \begin{itemize}
        \item  Factor 1: patterns of  $f_t$, $g_t$, and $p_t$ (25 levels). The 25 levels ,each ensuring that $f_t$ is a subset of $g_t$, are listed below:
        \begin{enumerate}
            \item (1 level) constant $p_t = (0.33, 0.33, 0.33)$; constant $f_t$; constant $g_t$
            \item (3 level) constant $p_t = (0.33, 0.33, 0.33)$; constant $f_t$; linear $g_t$ with $\theta_g = -0.3, 0, 0.3$ 
            \item (3 level) constant $p_t = (0.33, 0.33, 0.33)$; constant $f_t$; quadratic $g_t$ with $\theta_g = -0.3, 0, 0.3$ 
            \item (9 level) constant $p_t = (0.33, 0.33, 0.33)$; linear $f_t$ with  with $\theta_g = -0.3, 0, 0.3$; linear $g_t$ with $\theta_g = -0.3, 0, 0.3$
            \item (9 level) constant $p_t = (0.33, 0.33, 0.33)$; linear $f_t$ with  with $\theta_g = -0.3, 0, 0.3$; quadratic $g_t$ with $\theta_g = -0.3, 0, 0.3$ 
        \end{enumerate}
        \item Factor 2: sATE of two different treatment effects (4 levels).
        \begin{enumerate}
            \item $\sate_1 = 1.2$ or $1.3$
            \item $\sate_2 = 1.4$ and $1.5$
        \end{enumerate}
        \item Factor 3: $\aeo = 0.2$ or 0.4 (2 levels).
        \item Factor 4: pattern of $\tau(t)$ and value of AA (5 levels). The 5 levels are listed below:
        \begin{enumerate}
            \item AA = 1, constant $\tau(t)$.
            \item AA = 0.5, pattern of $\tau(t)$ is constant, linear(decreasing from 0.7 to 0.3).
            \item AA = 0.8, pattern of $\tau(t)$ is constant, linear(decreasing from 0.7 to 0.3).
        \end{enumerate}
    \end{itemize}
    \end{mdframed}
\end{guidelinebox}

\spacingset{1.9}

\subsection{Power when all working assumptions hold}
\label{subsec:simulation-all-hold}

\begin{figure}[htbp]
\centering
\includegraphics[width=1\textwidth]{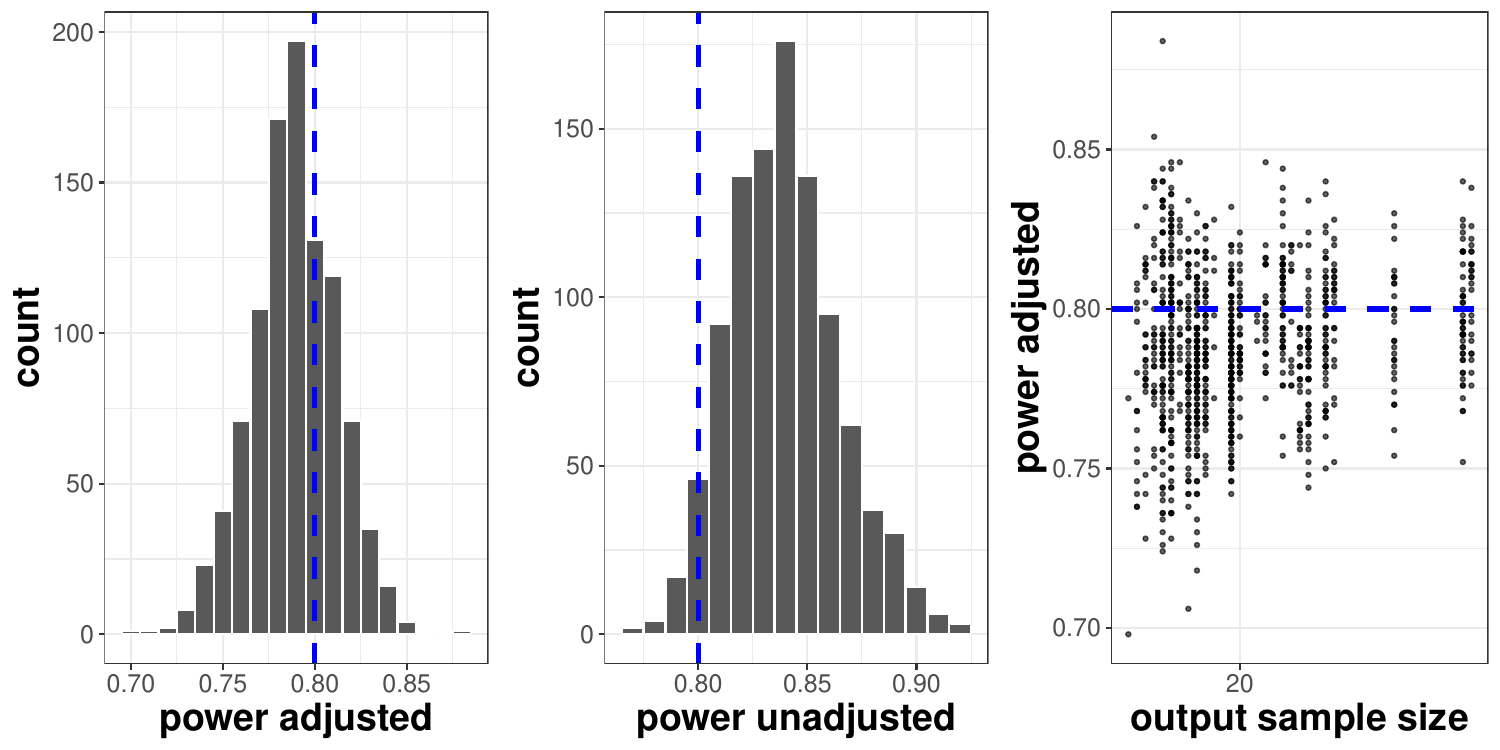}
  \caption{Adjusted and unadjusted power when all working assumptions hold. The first two figures show the histogram of the adjusted and unadjusted power under 1,000 settings. The last figure shows the adjusted power under each setting against the sample size $n$ obtained from the sample size formula.}
  \label{fig:power_working_assumption}
\end{figure}

Using GM-0, we considered 1,000 simulations under a variety of settings with different patterns and magnitudes of $\mee^*_k(t)$, $\sate^*_k$, $\aeo^*$, and $\aaa^*$, assuming all working assumptions hold. That is, when $\smee^*_k(t) = \smee^\w_k(t)$, $\sate^*_k = \sate^\w_k$, $\aeo^* = \aeo^\w$, and $\aaa^* = \aaa^\w$ (see \Cref{box:sim-all-wa-hold-1000-settings} for more detailed simulation settings). In all 1,000 settings we considered, the power is consistently around the desired level (\Cref{fig:power_working_assumption}). The power is slightly greater than 0.8 before applying the small sample correction and slightly below the desired 0.8 after the correction is applied. 

\subsection{Power when (WA-a) is violated}
\label{subsec:simulation-violate-a}

Using GM-0, we consider two scenarios where (WA-a) can be violated: incorrect magnitude of $\mee_k(t)$ and incorrect time-varying pattern of $\mee_k(t)$. 

\begin{figure}[htbp]
\centering
    \includegraphics[width=1\textwidth]{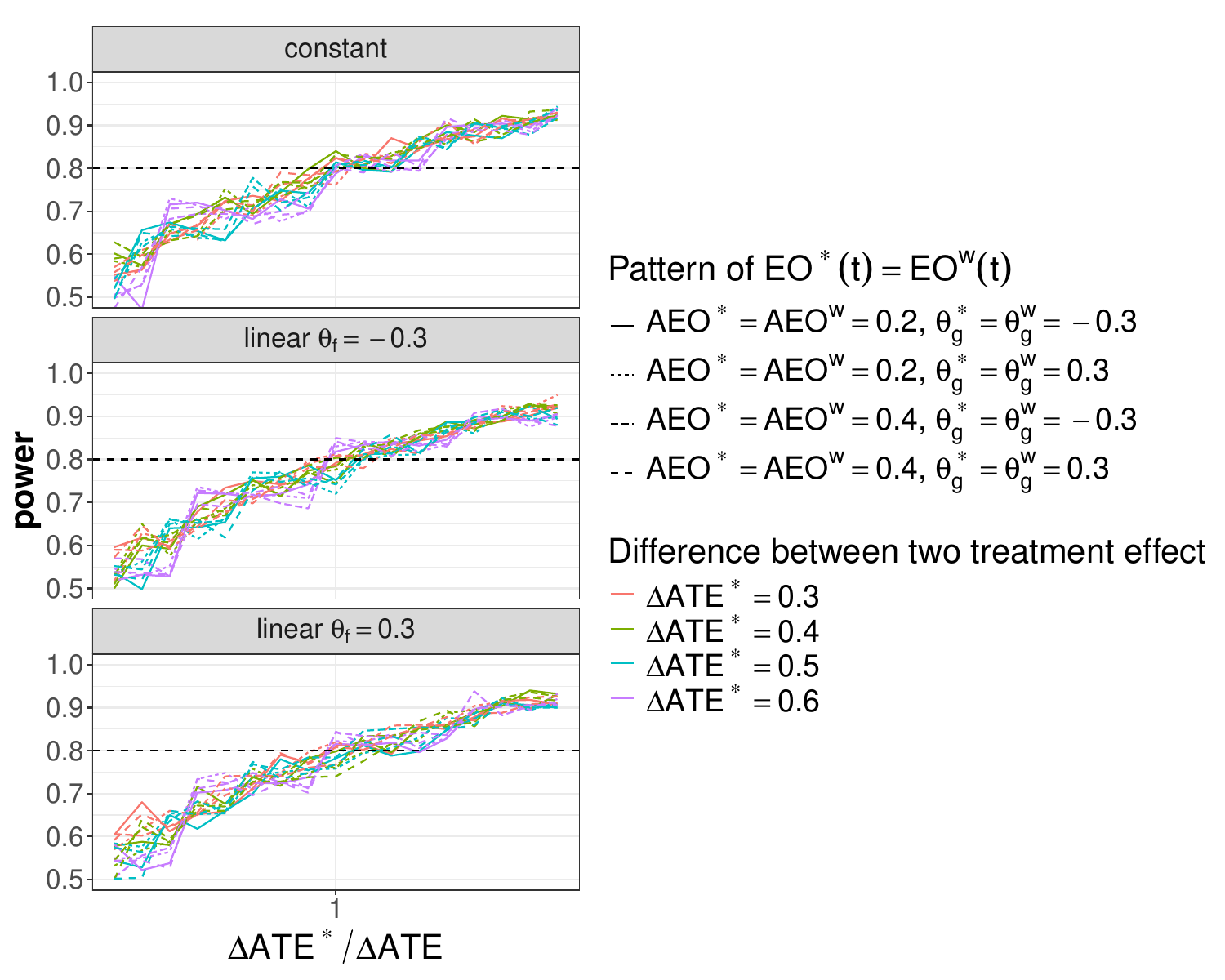}
    \caption{Power when (WA-a) is violated in that $\Delta \ate^\w \neq\Delta \text{ATE}^*$ while the working $\mee_k^\w(t)$ is specified to be of the same time-varying pattern as $\mee_k^*(t)$. $\theta_f$ captures the slope of $\mee_k^*(t)$ and the details can be found in \Cref{box:detail-gm-0-part2}.}
    \label{fig:viol miss ATE}
\end{figure}

In the first scenario, where the magnitude between two treatment effects ($\Delta \sate := \sate_2 - \sate_1$) is misspecified while the working $\smee_k^\w(t)$ is specified to be of the same time-varying pattern as $\smee_k^*(t)$ (constant or linear), if $\Delta \sate^\w > \Delta \sate^*$ then the MRT is under-powered, and if $\Delta \sate^\w < \Delta \sate^*$ then MRT is overpowered (\Cref{fig:viol miss ATE}).

In the second scenario, where the pattern of $\mee_k(t)$ is misspecified (i.e., the working $\mee_k^\w(t)$ is specified to be of a different time-varying pattern from $\mee_k^*(t)$) while the magnitude is correct (i.e., $\Delta \sate^\w = \Delta \sate^*$), we considered linear and constant patterns for both working and true $\mee_k(t)$. The take-away is that specifying constant working $\mee_k^\w(t)$ always yields adequate power, regardless of whether the true $\mee_k^*(t)$ is constant or not; specifying a linear working $\mee_k^\w(t)$ sometimes leads to inadequate power. The detailed result is as follows.

\begin{figure}[htbp]
    \centering
    \begin{subfigure}[b]{0.49\textwidth}
        \centering
        \includegraphics[width=\textwidth]{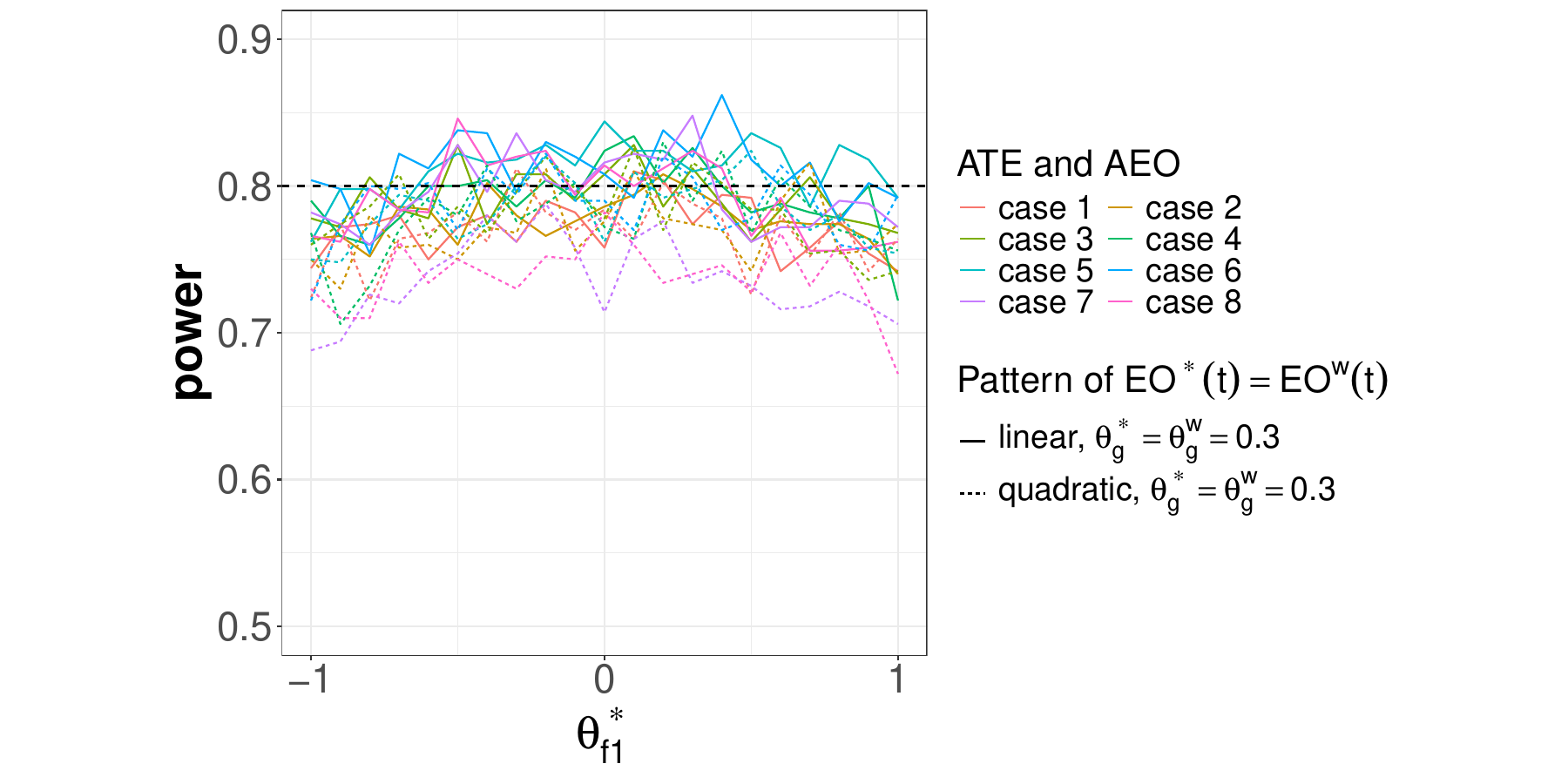}
        \caption{}
        \label{fig:mee_working_constant}
    \end{subfigure}
    \hfill
    \begin{subfigure}[b]{0.49\textwidth}  
        \centering 
        \includegraphics[width=\textwidth]{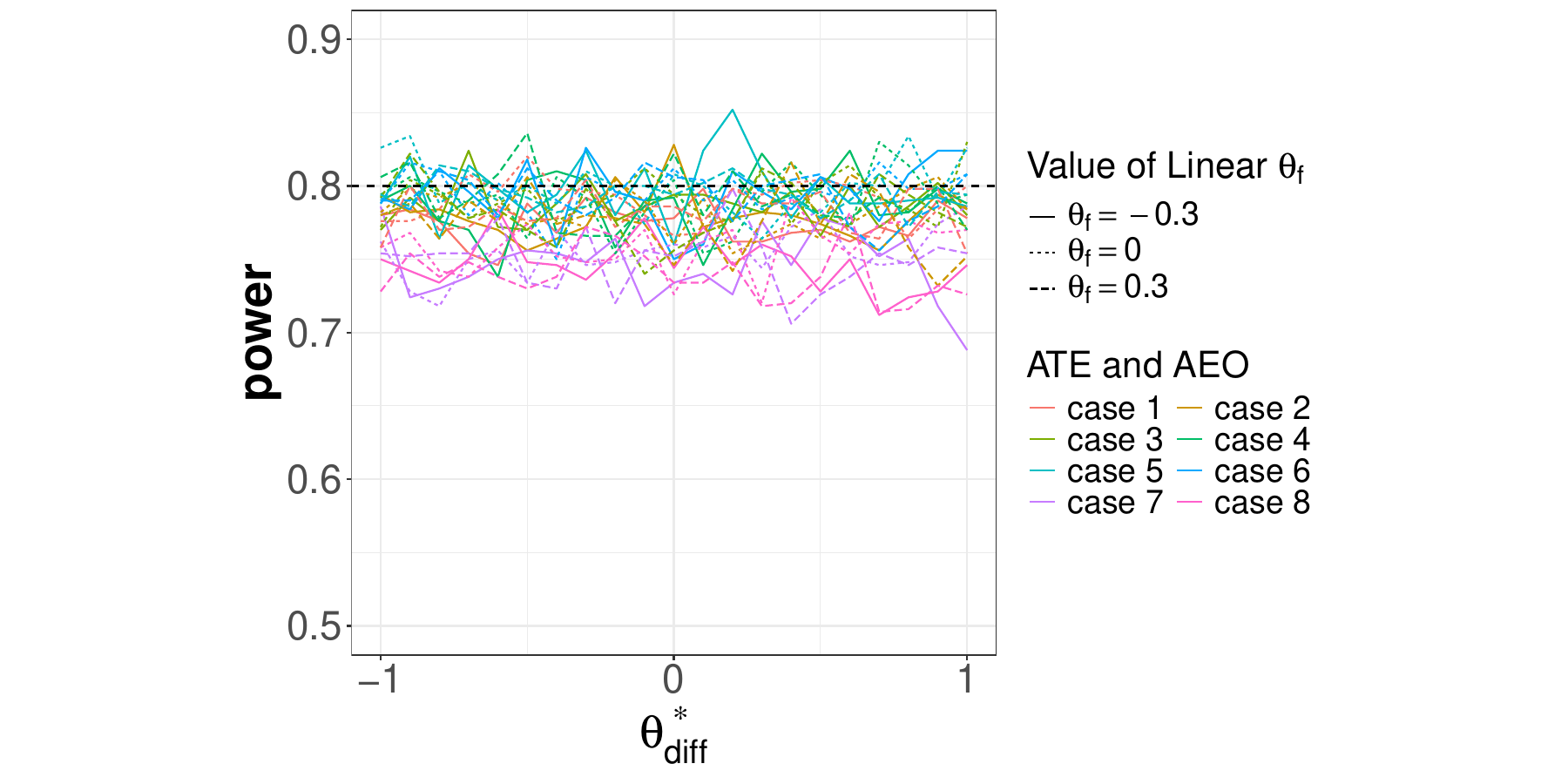}
        \caption{}  
        \label{fig:mee_working_constant_mis_ATE}
    \end{subfigure}
    
    \begin{subfigure}[b]{0.49\textwidth}   
        \centering 
        \includegraphics[width=\textwidth]{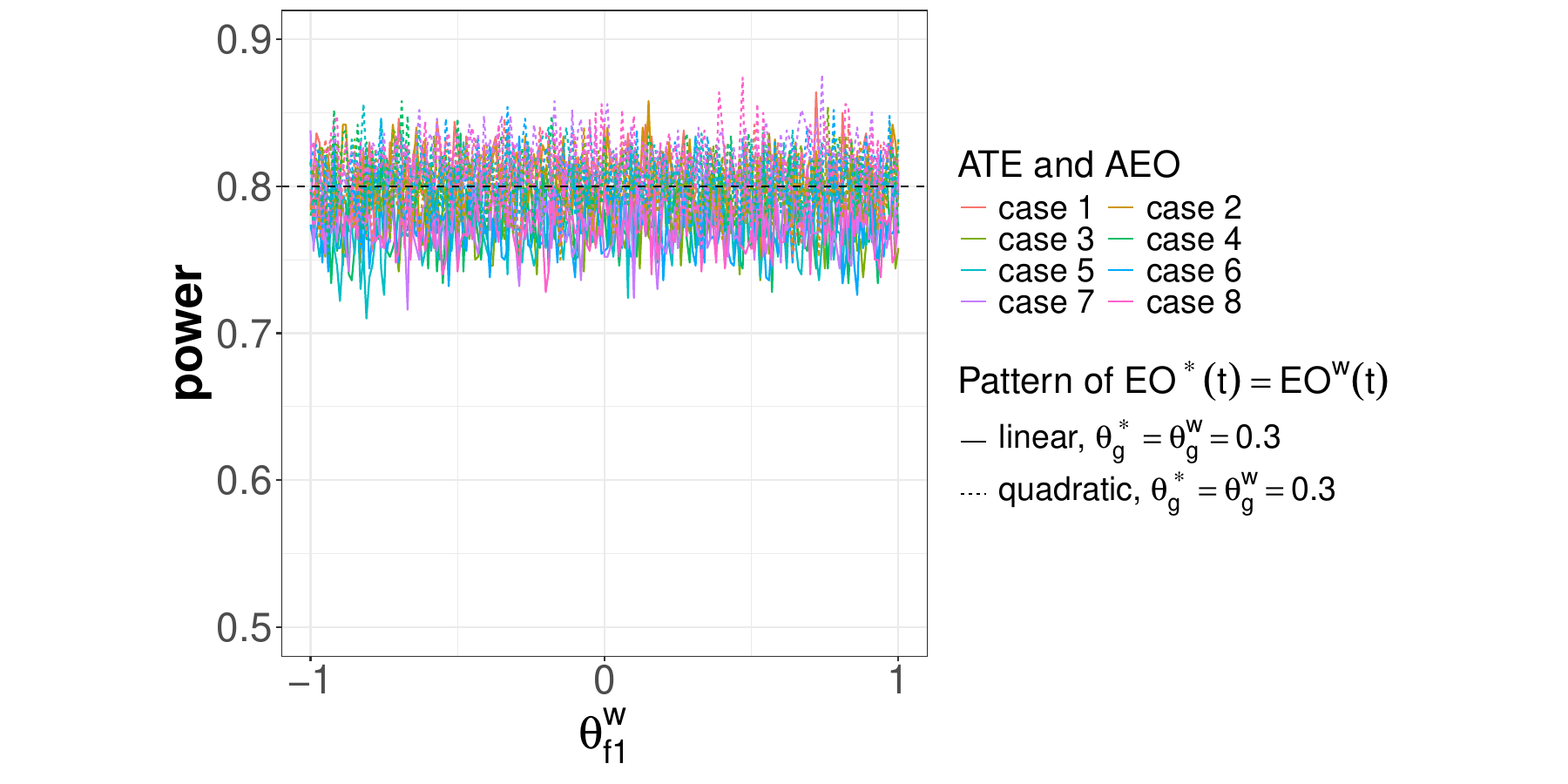}
        \caption{}
        \label{fig:mee_working_linear_parallel slope}
    \end{subfigure}
    \hfill
    \begin{subfigure}[b]{0.49\textwidth}   
        \centering 
        \includegraphics[width=\textwidth]{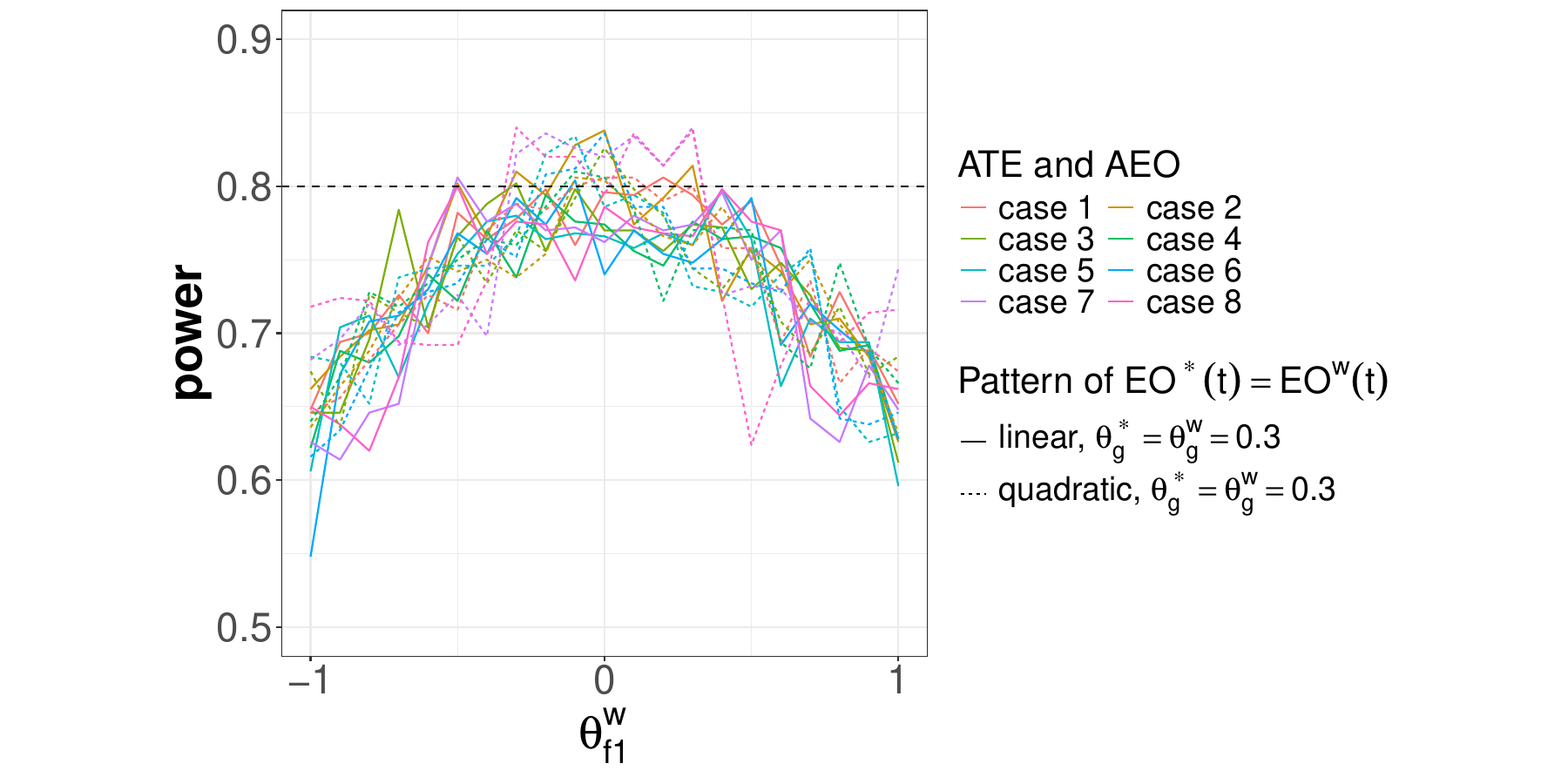}
        \caption{}   
        \label{fig:mee_working_mis_slope}
    \end{subfigure}
    \begin{subfigure}[b]{0.49\textwidth}   
        \centering 
        \includegraphics[width=1\textwidth]{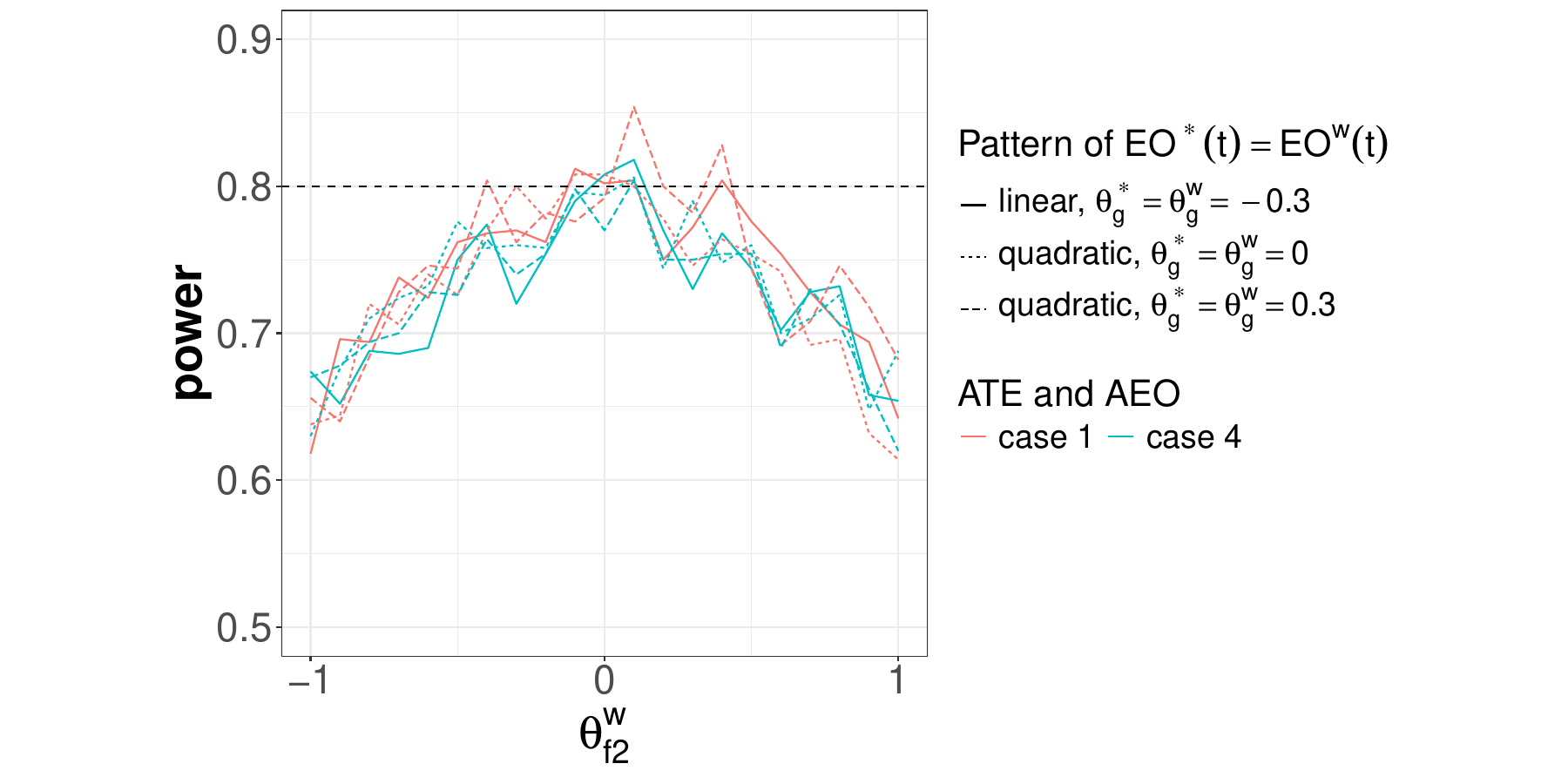}
        \caption{}
        \label{fig:mee_true_parallel_slope}
    \end{subfigure}
    \hfill
    \begin{subfigure}[b]{0.49\textwidth}   
        \centering 
        \includegraphics[width=1\textwidth]{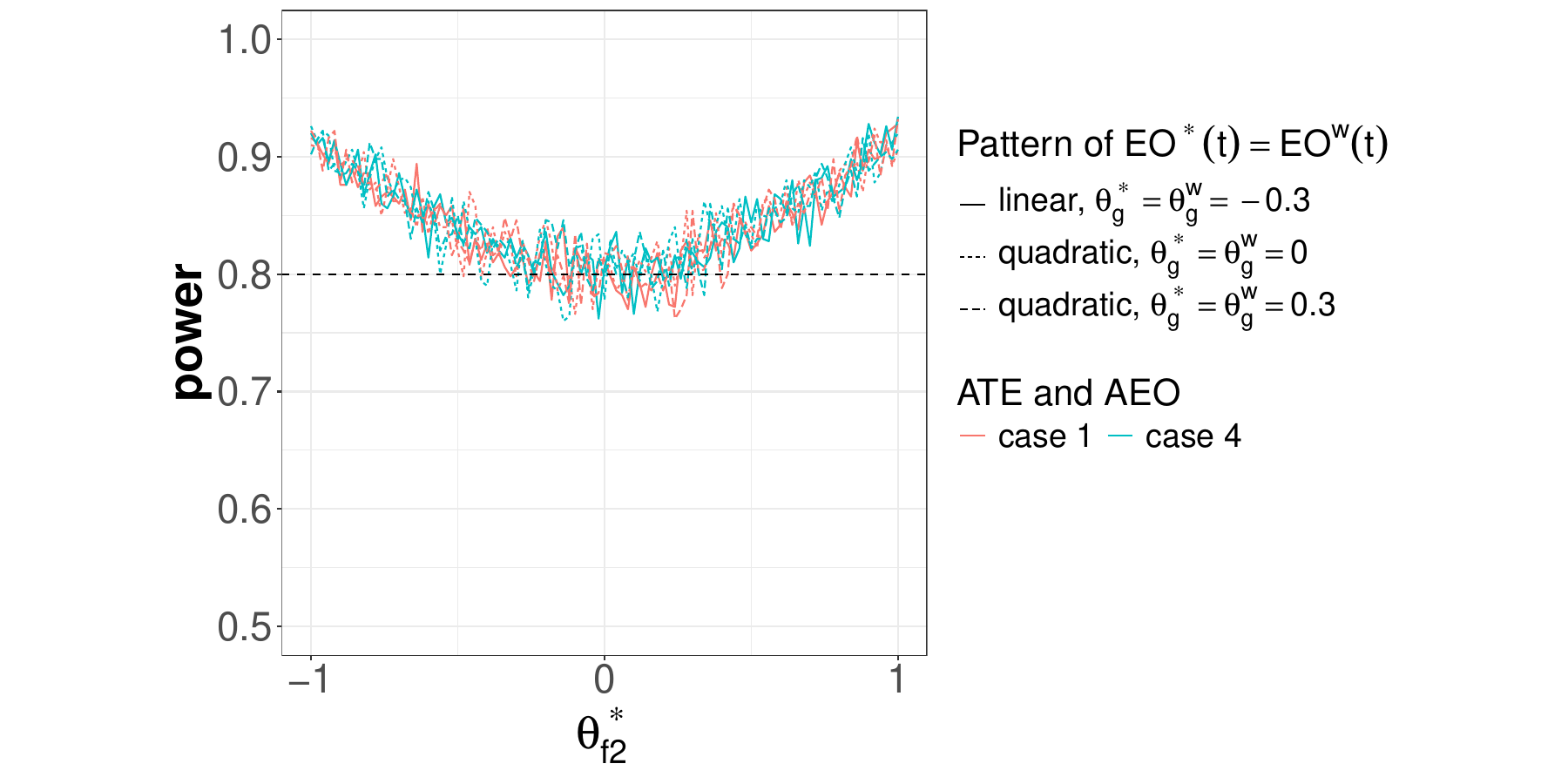}
        \caption{}
        \label{fig:mee_working_parallel slope}
    \end{subfigure}
    
    {\scriptsize case 1 : $\Delta ATE^* = \Delta ATE^w = 0.3$, $AEO^* = AEO^w = 0.2$, case 2 : $\Delta ATE^* = \Delta ATE^w = 0.3$, $AEO^* = AEO^w = 0.4$, \\
    case 3 : $\Delta ATE^* = \Delta ATE^w = 0.4$, $AEO^* = AEO^w = 0.2$, case 4 : $\Delta ATE^* = \Delta ATE^w = 0.4$, $AEO^* = AEO^w = 0.4$,\\
    case 5 : $\Delta ATE^* = \Delta ATE^w = 0.5$, $AEO^* = AEO^w = 0.2$, case 6 : $\Delta ATE^* = \Delta ATE^w = 0.5$, $AEO^* = AEO^w = 0.4$,\\ case 7 : $\Delta ATE^* = \Delta ATE^w = 0.6$, $AEO^* = AEO^w = 0.2$, case 8 : $\Delta ATE^* = \Delta ATE^w = 0.6$, $AEO^* = AEO^w = 0.4$ \par}
    \caption{Power when (WA-a) is violated in that the time-varying pattern of $\mee_k^\w(t)$ is different from that of $\mee_k^*(t)$ while $\Delta ATE^\w = \Delta ATE^*$. \textbf{Panels (a) and (b):} the true $\mee_k^*(t)$ is linear in $t$ but the working $\mee_k^\w(t)$ is constant. \textbf{Panel (a):} $\ate_1^* = \ate_1^\w$ and $\ate_2^* = \ate_2^\w$. \textbf{Panel (b):} $\ate_1^* \neq \ate_1^\w$ and $\ate_2^* \neq \ate_2^\w$, yet $\Delta \ate^* = \Delta \ate^\w$. \textbf{Panels (c) and (d):} the true $\mee_k^*(t)$ is constant but the working $\mee_k^\w(t)$ is linear in $t$. \textbf{Panel (c):} $\Delta\mee^\w(t) = \Delta\mee^*(t)$. \textbf{Panel (d):} $\Delta\mee^\w(t) \neq \Delta\mee^*(t)$ yet $\Delta ATE^\w = \Delta ATE^*$. \textbf{Panels (e) and (f):} both the true $\mee_k^*(t)$ and the working $\mee_k^\w(t)$ are linear in $t$. \textbf{Panel (e):} the true $\mee^*_1(t)$ and $\mee^*_2(t)$ are parallel but the working $\mee^\w_1(t)$ and $\mee^\w_2(t)$ are not parallel. \textbf{Panel (f):} the true $\mee^*_1(t)$ and $\mee^*_2(t)$ are not parallel but the working $\mee^\w_1(t)$ and $\mee^\w_2(t)$ are parallel.}
\end{figure}

When the true $\mee_k^*(t)$ is linear in $t$ but the working $\mee_k^\w(t)$ is constant, we considered two cases: one where the average effects for both active treatment levels ($\sate_k$) are correctly specified (\Cref{fig:mee_working_constant}), and a second case where the $\sate_k$ is misspecified but the effect difference ($\Delta \sate$) is correctly specified (\Cref{fig:mee_working_constant_mis_ATE}). In both cases, the MRT is adequately powered.

When the true $\mee_k^*(t)$ is constant, but the working $\mee_k^\w(t)$ is linear in $t$, we considered two cases: one where $\Delta\mee^\w(t) = \Delta\mee^*(t)$ (specifically, $\mee_1^\w(t)$ and $\mee_2^\w(t)$ are parallel, i.e., $\theta_{f2} = 0$ with $\theta_{f2}$ defined in \Cref{box:detail-gm-0-part2}), and a second case where $\Delta\mee^\w(t) \neq \Delta\mee^*(t)$ yet $\Delta \sate^\w = \Delta \sate^*$ (specifically, $\mee_1^\w(t)$ and $\mee_2^\w(t)$ are not parallel, , i.e., $\theta_{f2} \neq 0$). In the first case, the power is roughly 80 percent (\Cref{fig:mee_working_linear_parallel slope}). In the second case, the MRT is always underpowered (\Cref{fig:mee_working_mis_slope}). 

When both true $\mee^*_k(t)$ and working $\mee^\w_k(t)$ are linear in $t$, we consider the two cases: one where the true $\mee^*_1(t)$ and $\mee^*_2(t)$ are parallel but the working $\mee^\w_1(t)$ and $\mee^\w_2(t)$ are not parallel, and a second case where the true $\mee^*_1(t)$ and $\mee^*_2(t)$ are not parallel but the working $\mee^\w_1(t)$ and $\mee^\w_2(t)$ are parallel. In the first case,, the MRT is always underpowered (\Cref{fig:mee_true_parallel_slope}). In the second case, the MRT is always overpowered (\Cref{fig:mee_working_parallel slope}).

\subsection{Power when (WA-b) is violated}
\label{subsec:simulation-violate-b}

Using GM-0, we consider two scenarios where (WA-b) can be violated: the magnitude of the average-over-time mean outcome is misspecified ($\aeo^* \neq \aeo^\w$) but the time-varying pattern of the mean outcome is correct (e.g., both $\eo^*(t)$ and $\eo^\w(t)$ are linear in $t$), or the magnitude is correct ($\aeo^* = \aeo^\w$) but the time-varying pattern is incorrect ($\eo^*(t) \neq \eo^\w(t)$).

\begin{figure}[htbp]
    \centering
    \begin{subfigure}[b]{0.475\textwidth}
        \centering
        \includegraphics[width=\textwidth]{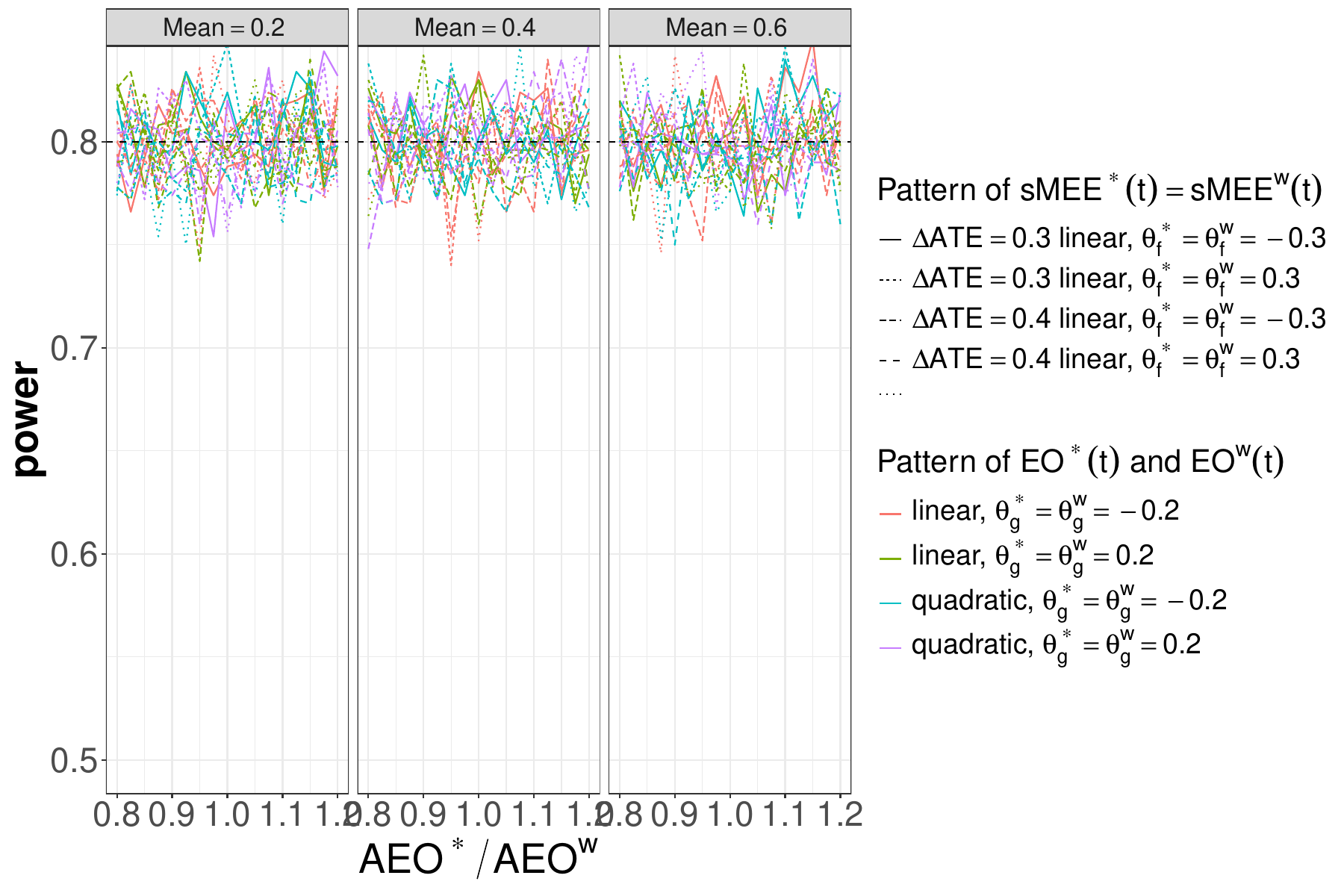}
        \caption{}
        \label{fig:mis_aeo}
    \end{subfigure}
    \hfill
    \begin{subfigure}[b]{0.475\textwidth}  
        \centering 
        \includegraphics[width=\textwidth]{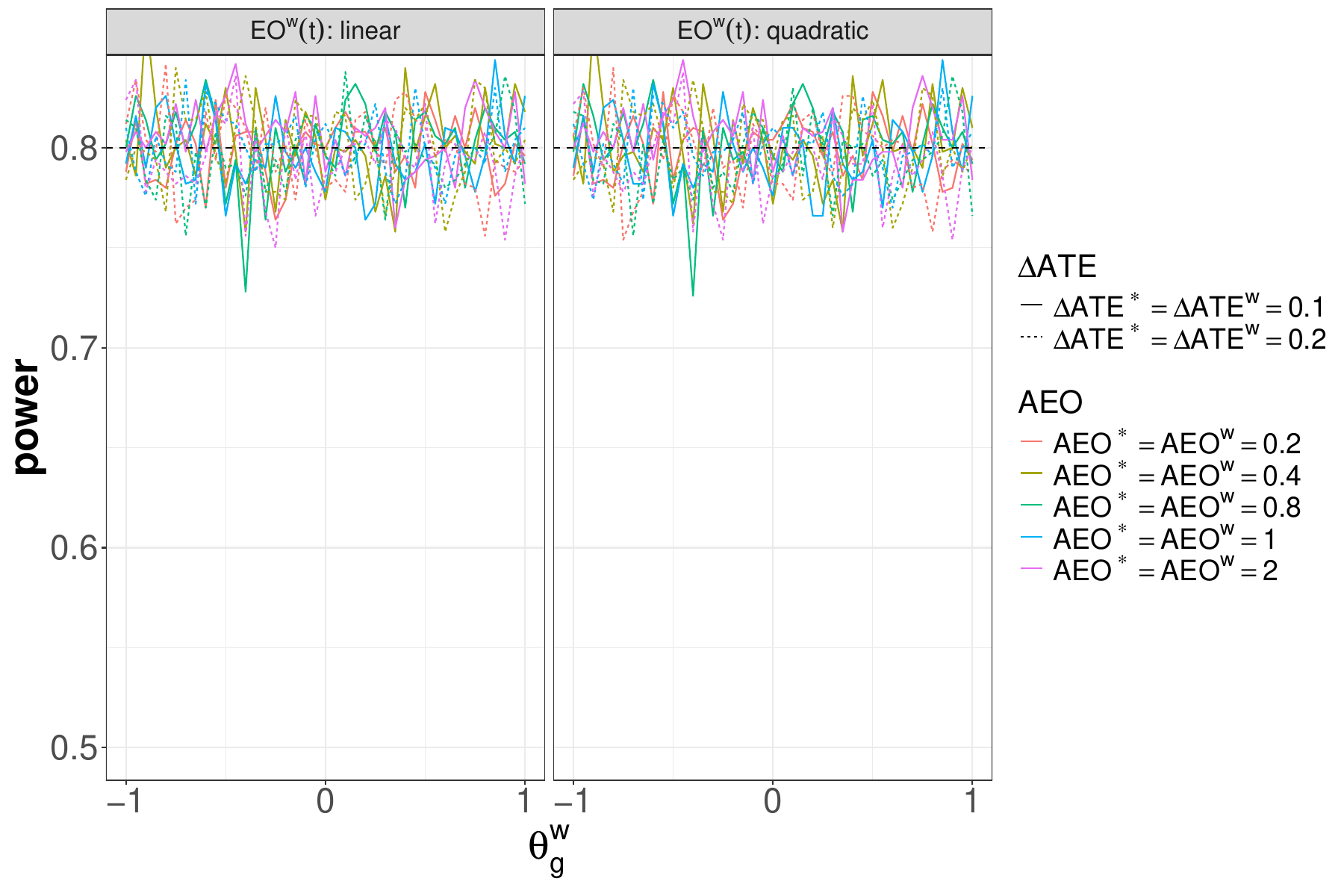}
        \caption{}  
        \label{fig:gt_true_constant}
    \end{subfigure}

    \begin{subfigure}[b]{0.475\textwidth}   
        \centering 
        \includegraphics[width=\textwidth]{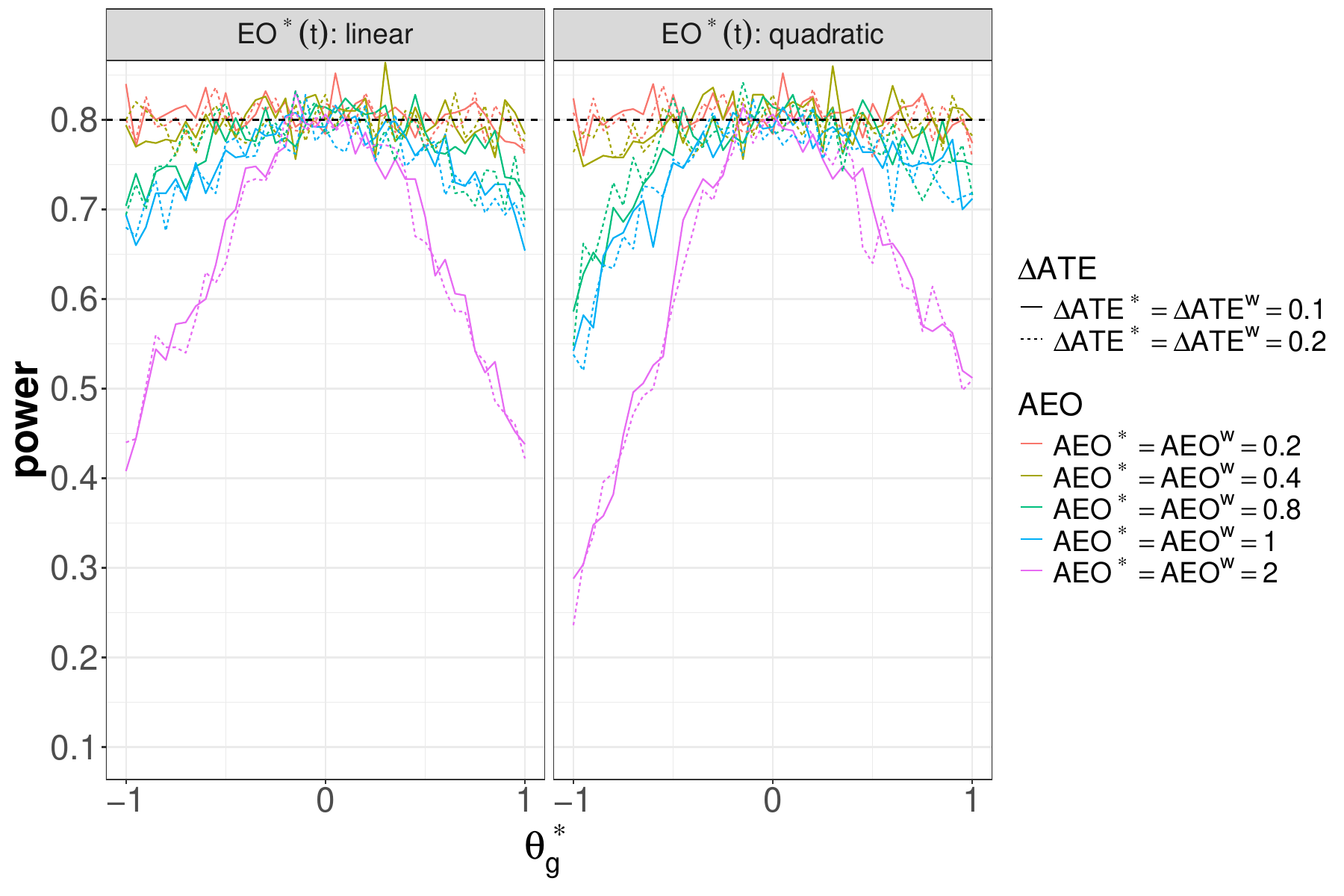}
        \caption{}
        \label{fig:gt_working_constant}
    \end{subfigure}
    \hfill
    \begin{subfigure}[b]{0.475\textwidth}   
        \centering 
        \includegraphics[width=\textwidth]{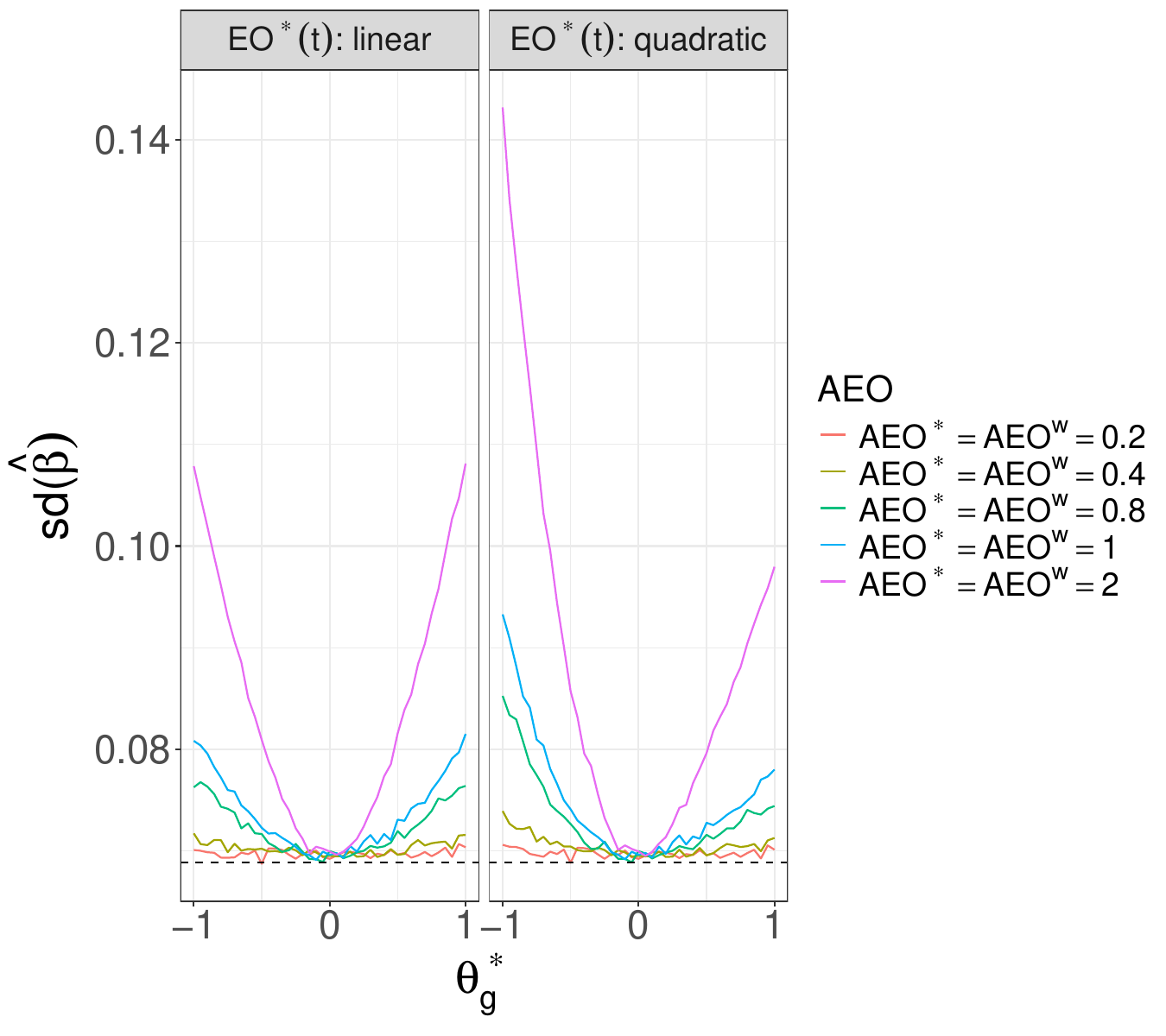}
        \caption{}   
        \label{fig:gt_working_constant_se}
    \end{subfigure}
    \caption{Power when (WA-b) is violated. \textbf{Panels (a):} the magnitude of the average-over-time mean outcome is misspecified ($\aeo^* \neq \aeo^\w$) but the time-varying pattern of the mean outcome is correct (e.g., both $\eo^*(t)$ and $\eo^\w(t)$ are linear in $t$). \textbf{Panel (b):} the working $\eo^\w(t)$ is linear or quadratic in $t$ but the true $\eo^*(t)$ is constant. \textbf{Panel (c):} the working $\eo^\w(t)$ is constant but the true $\eo^*(t)$ is linear or quadratic in $t$. \textbf{Panel (d):} the standard deviation of $\hat\beta$ for the same setting as Panel (c), explaining the low power in Panel (c).}
\end{figure}

In the first scenario, misspecifying the magnitude of the mean outcome so that $\aeo^* \neq \aeo^\w$ does not affect power (\Cref{fig:mis_aeo}), because the only input to \Cref{M-alg:ss-calculator} that is related to (WA-b) is the dimension of $g_t$. 

\begin{figure}[htbp]
    \centering
    \includegraphics[width=1\textwidth]{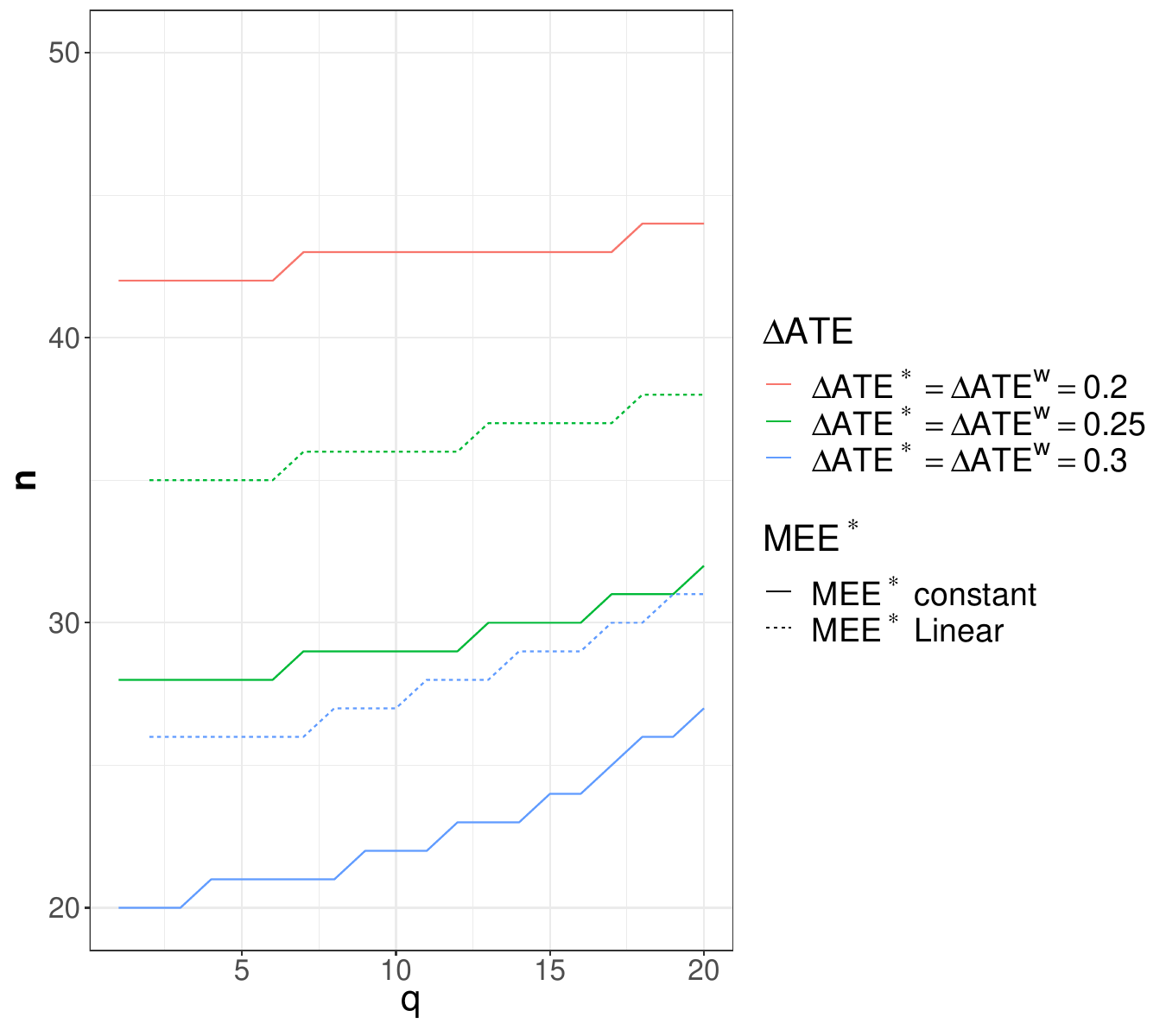}
    \caption{Dependence of the output sample size $n$ on $q$ (the degrees of freedom of the input $g_t$ to \Cref{M-alg:ss-calculator}), stratified by various $\mee_k(t)$ specifications.}
    \label{fig:n versus q}
\end{figure}

In the second scenario, when the working $\eo^\w(t)$ is linear or quadratic in $t$ but the true $\eo^*(t)$ is constant, the power is adequate (\Cref{fig:gt_true_constant}). When the working $\eo^\w(t)$ is constant but the true $\eo^*(t)$ is linear or quadratic in $t$, the power is too low (\Cref{fig:gt_working_constant}). The low power in the last case is not due to the under-specified degrees of freedom $q$ for $g_t$, but rather due to the increased standard deviation in $\hat\beta$ (\Cref{fig:gt_working_constant_se}) as the $g_t$ used in computing the test statistic does not adequately capture the variability in $\eo^*(t)$, which is required by (WA-b). After all, \Cref{M-thm:sample-size-formula} only guarantees adequate power when the $g_t$ used in computing the test statistic is the same as the $g_t$ used as the input to \Cref{M-alg:ss-calculator} \textit{and} the $g_t$ in the true $\eo^*(t)$. \Cref{fig:n versus q} shows how the output sample size $n$ from \Cref{M-alg:ss-calculator} depends on $q$ (the degrees of freedom of the input $g_t$), \textit{assuming that the $g_t$ adequately captures the variability in $\eo^*(t)$, i.e., assuming (WA-b) holds}. Indeed, $q$ alone does not substantially change the output sample size unless $q$ gets extremely large (e.g., $q > 10$).

\begin{figure}[htbp]
    \centering
    \begin{subfigure}[b]{0.44\textwidth}
        \centering
        \includegraphics[width=\textwidth]{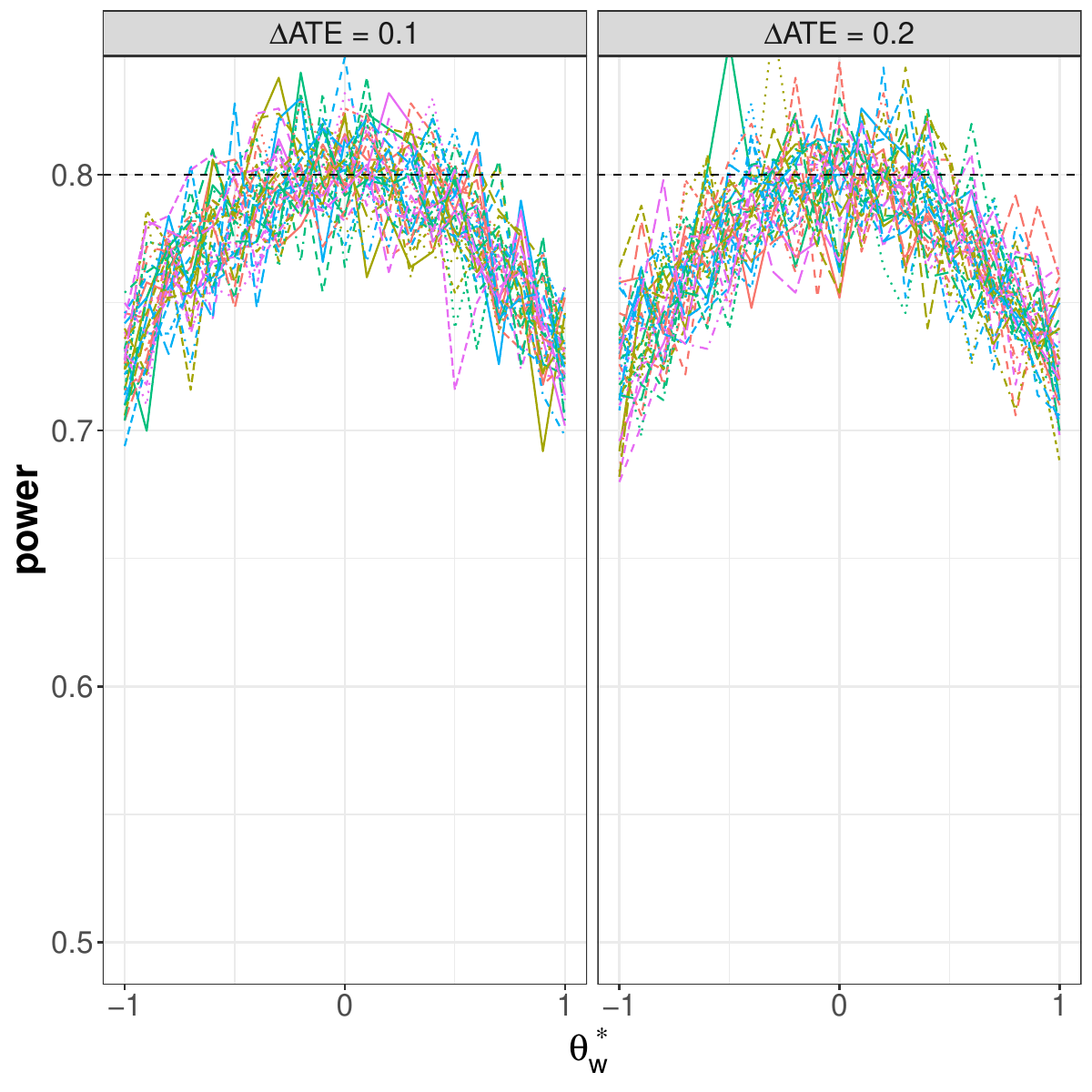}
        \caption{}
        \label{fig:gt_weekend true}
    \end{subfigure}
    \hfill
    \begin{subfigure}[b]{0.54\textwidth}  
        \centering 
        \includegraphics[width=\textwidth]{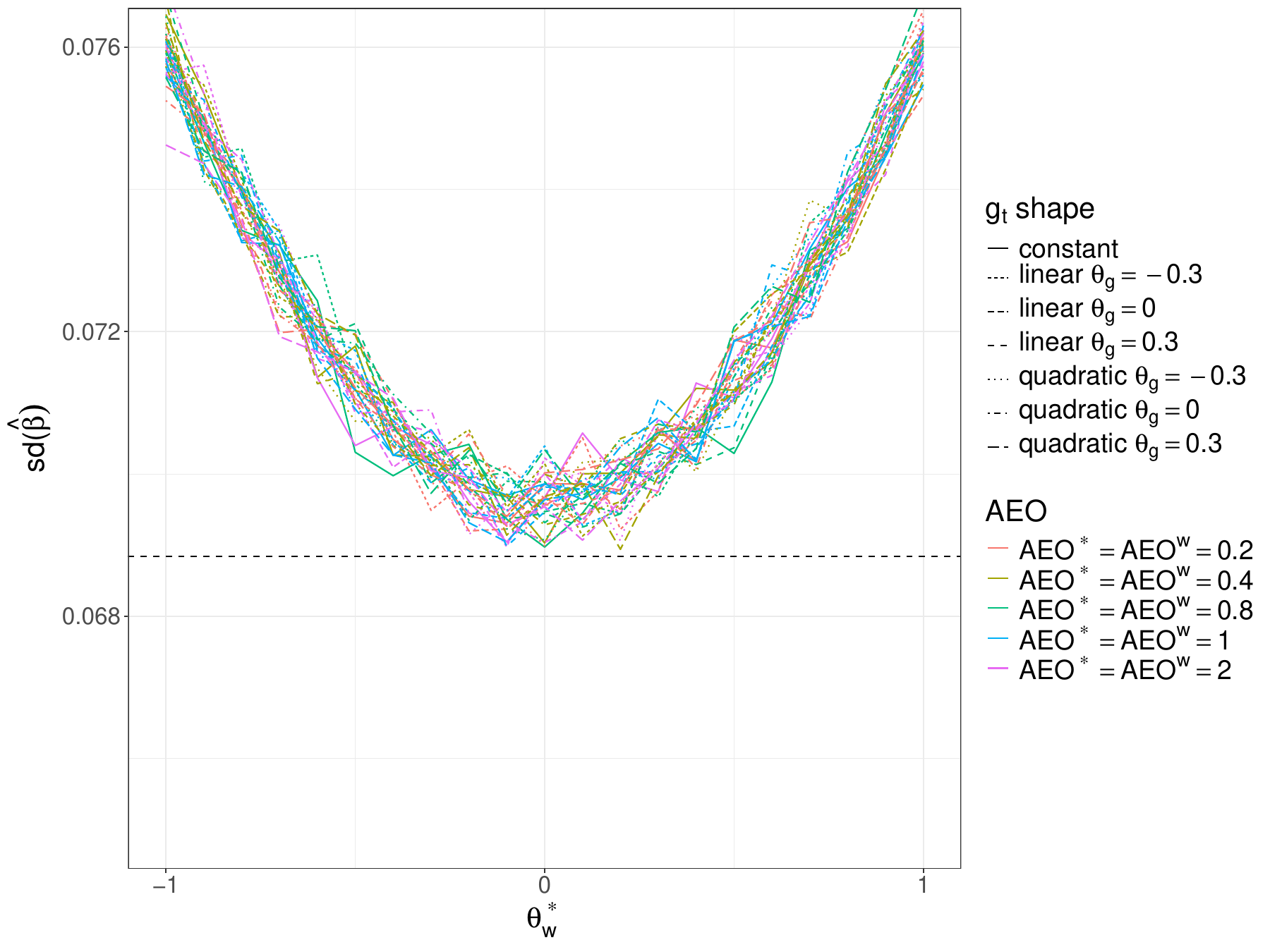}
        \caption{}  
        \label{fig:gt_weekend true se}
    \end{subfigure}
    \caption{Power when (WA-b) is violated in that there is a ``weekend'' effect on the true mean proximal outcome. \textbf{Panels (a):} the power. \textbf{Panel (b):} the standard deviation of $\hat\beta$ for the same setting as Panel (a), explaining the low power in Panel (a).}
\end{figure}

We further consider a variation of GM-0 with a ``weekend'' effect on the true mean proximal outcome, i.e., $Y_{it} = \one(A_{it} = 1) f_t^T \beta_1^* + \one(A_{it} = 2) f_t^T \beta_2^* + g_t^T \alpha^* + \text{weekend}_t \theta_w + \epsilon_{it}$, where $\text{weekend}_t = 1$ if the decision point is on a weekend and 0 otherwise. $\theta_w$ captures the strength of the weekend effect, and $\theta_w = 0$ implies that the mean outcome is correctly specified ($\eo^\w(t) = \eo^*(t)$; note that we use a linear $\eo^\w(t)$). The power decreases as the magnitude of the weekend effect ($|\theta_w|$) increases (\Cref{fig:gt_weekend true}). The low power is also due to the increased standard deviation of $\hat\beta$ (\Cref{fig:gt_weekend true se}), as argued in the previous paragraph.

\subsection{Power when (WA-c) is violated}
\label{subsec:simulation-violate-c}

Using GM-EV, we consider two scenarios where (WA-c) can be violated: one where $\var(Y_t \mid A_t, I_t = 1)$ depends linearly on $t$, and the other where $\var(Y_t \mid A_t, I_t = 1)$ depends on $A_t$. Recall that $\theta_r$ and $\theta_s$ capture the dependence in each setting, with detailed parameterization in \Cref{box:detail-gm-ev}.

\begin{figure}[htbp]
    \centering
    \begin{subfigure}{.31\textwidth}
      \centering
    \includegraphics[width=1\textwidth]{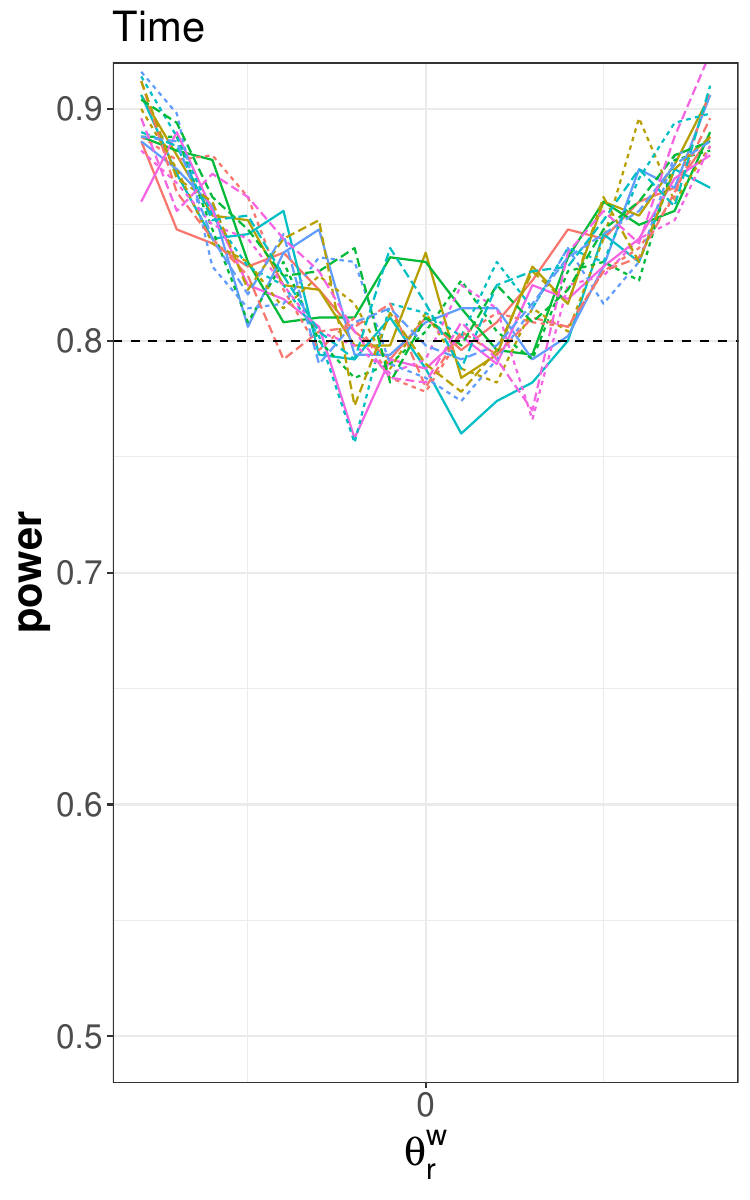}
    \caption{}
    \label{fig:const_var_miss_time_dependent}
    \end{subfigure}%
    \begin{subfigure}{.69\textwidth}
      \centering
      \includegraphics[width=1\textwidth]{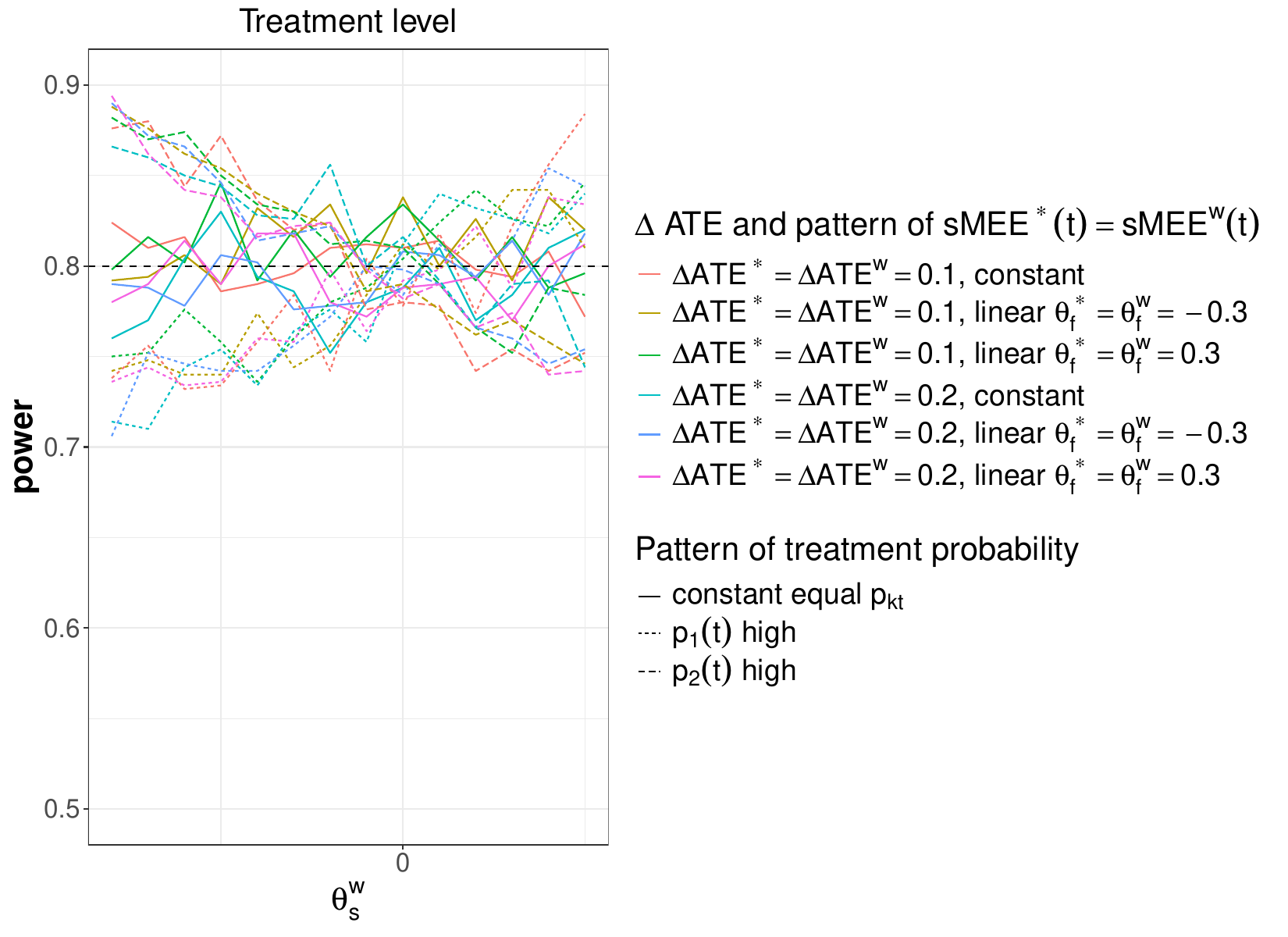}
      \caption{}
    \label{fig:const_var_miss_trt_lvl}
    \end{subfigure}
    \caption{Power when (WA-c) is violated. \textbf{Panels (a):} when $\var(Y_t \mid A_t, I_t = 1)$ depends linearly on $t$. \textbf{Panels (b):} when $\var(Y_t \mid A_t, I_t = 1)$ depends on $A_t$.}
\end{figure}

In the first scenario, where $\var(Y_t \mid A_t, I_t = 1)$ depends linearly on $t$, MRT is over-powered (\Cref{fig:const_var_miss_time_dependent}). In the second scenario, where $\var(Y_t \mid A_t, I_t = 1)$ depends on $A_t$, the power depends on whether the treatment assignment is balanced (i.e., each treatment level has the same assignment probability). If the treatment is balanced, the power is adequate. If the treatment assignment is inbalanced (i.e., the assignment probabilities for the treatment levels are not equal), this can lead to over-powering or under-powering (\Cref{fig:const_var_miss_trt_lvl}) for up to 10\% power in the simulations we considered.


\subsection{Power when (WA-d) is violated}
\label{subsec:simulation-violate-d}

Using GM-0, we consider two scenarios where (WA-d) can be violated: the magnitude of $\aaa$ is misspecified (so that $\aaa^* \neq \aaa^\w$ but the time-varying patterns of $\tau^*(t)$ and $\tau^\w(t)$ are the same) or the pattern of $\tau(t)$ is incorrect (so that $\aaa^* = \aaa^\w$ but the time-varying patterns of $\tau^*(t)$ and $\tau^\w(t)$ are different). Recall that $\aaa$ is the averaged-over-time expected availability defined in \eqref{M-eq:AA}.

\begin{figure}[htbp]
    \centering
    \includegraphics[width=1\textwidth]{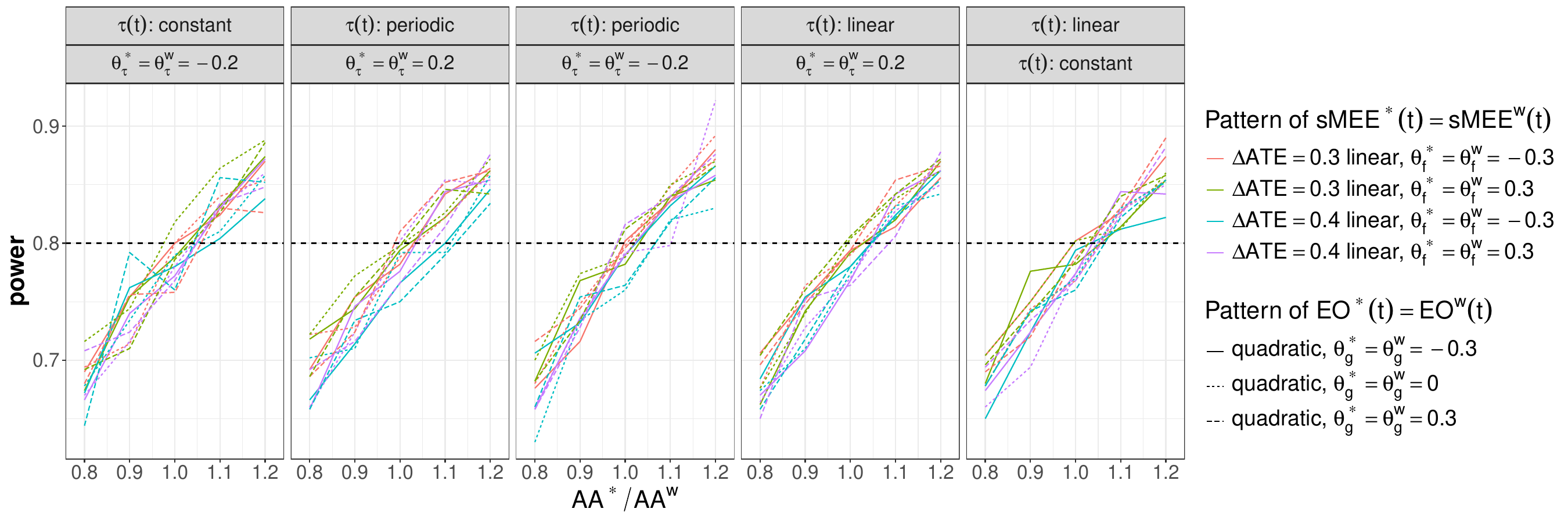}
    \caption{Power when (WA-d) is violated in that $\aaa^* \neq \aaa^\w$ but the time-varying patterns of $\tau^*(t)$ and $\tau^\w(t)$ are the same.}
    \label{fig:viol_avail_mag}
\end{figure}

In the first scenario, we set $\tau^\w(t)$ and $\tau^*(t)$ to be constants in $t$, so that $\tau^\w(t) = \aaa^\w$ and $\tau^*(t) = \aaa^*$ for all $t \in [T]$. We set $\aaa^\w = 0.3$ and varied $\aaa^*$ from 0.1 to 1. The MRT is overpowered if $\aaa^\w < \aaa^*$, and the MRT is underpowered if $\aaa^\w > \aaa^*$ (\Cref{fig:viol_avail_mag}).

\begin{figure}[htbp]
    \centering
        \vskip\baselineskip
    \begin{subfigure}[b]{1\textwidth}   
        \includegraphics[width=0.9\textwidth]{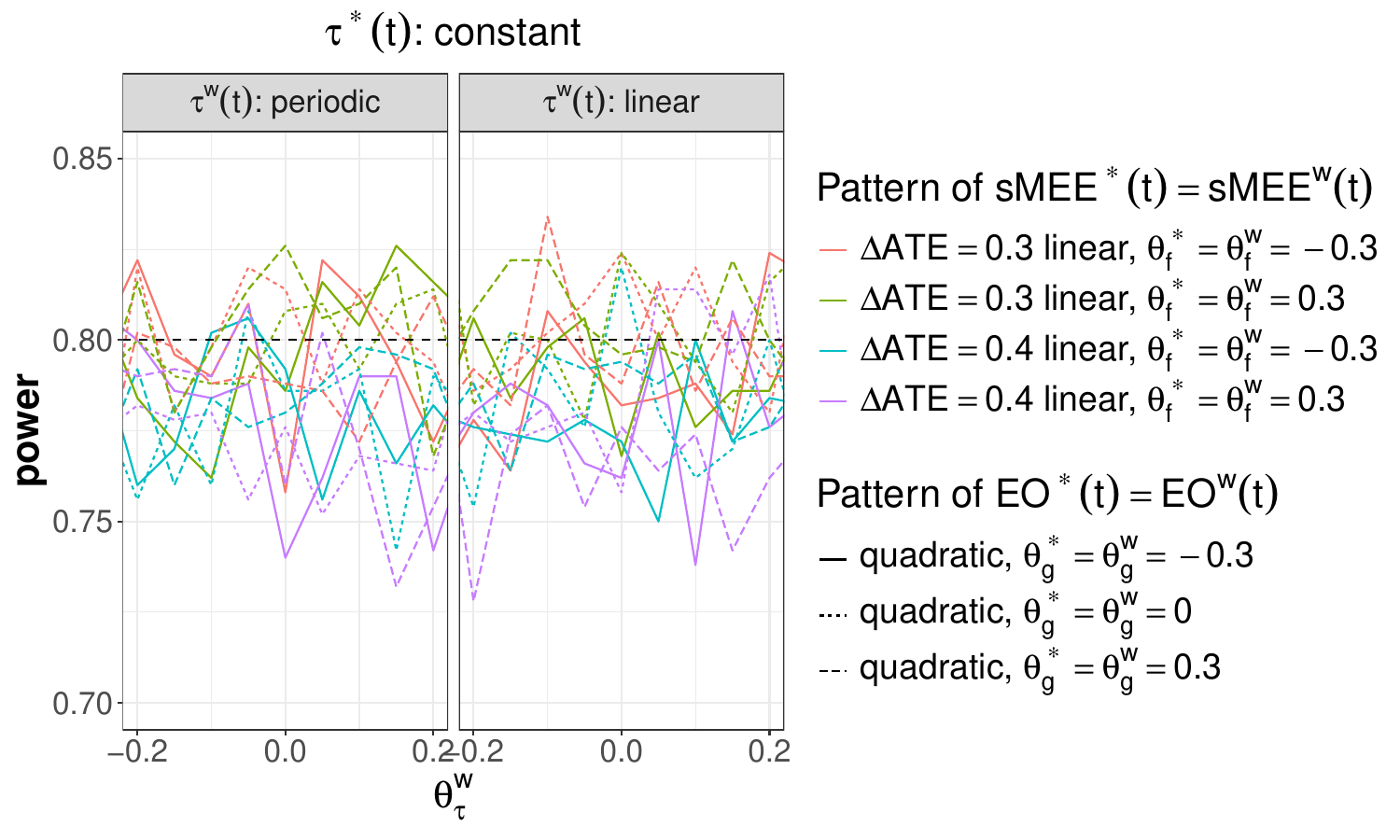}
        \caption{}
    \end{subfigure}
    
    \vskip\baselineskip
    \begin{subfigure}[b]{1\textwidth}   
        \centering
        \includegraphics[width=0.9\textwidth]{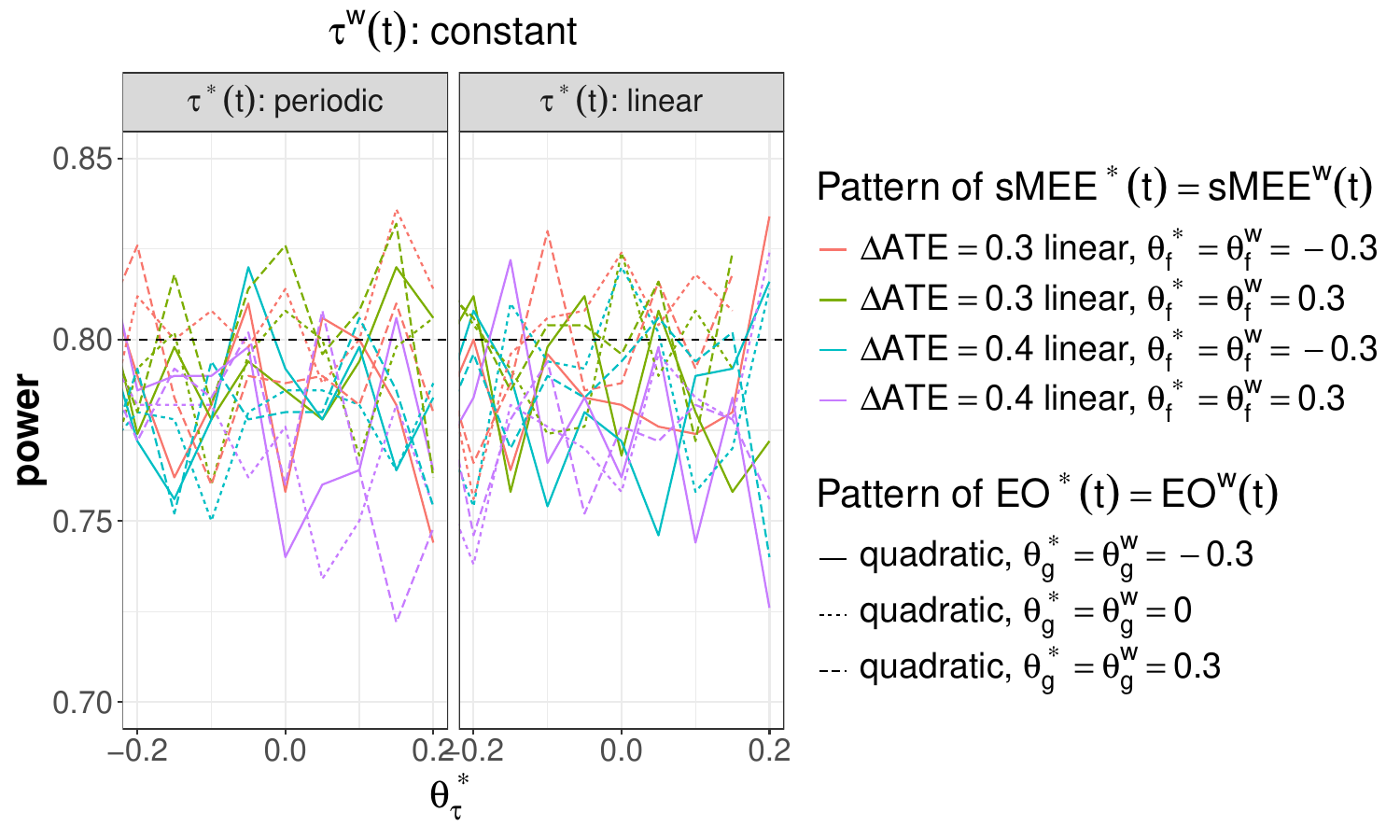}
        \caption{}
    \end{subfigure}
    \hfill
    \caption{Power when (WA-d) is violated in that $\aaa^* = \aaa^\w$ but the time-varying patterns of $\tau^*(t)$ and $\tau^\w(t)$ are different. \textbf{Panels (a):} when $\tau^*(t)$ is constant and $\tau^\w(t)$ is either periodic or linear in $t$. \textbf{Panels (b):} when when $\tau^\w(t)$ is constant and $\tau^*(t)$ is either periodic or linear in $t$.}
    \label{fig:pattern_viol}
\end{figure}

In the second scenario, we set $\aaa^* = \aaa^\w = 0.3$ and varied the time-varying pattern of $\tau^*(t)$ and $\tau^\w(t)$, so that one is constant in $t$ and the other is either linear or periodic in $t$. Regardless of which of $\tau^*(t)$ or $\tau^\w(t)$ is constant, the MRT is always adequately powered (\Cref{fig:pattern_viol}).

\subsection{Power when (WA-e) is violated}
\label{subsec:simulation-violate-e}

\begin{figure}[htbp]
    \centering
    \includegraphics[width=.8\textwidth]{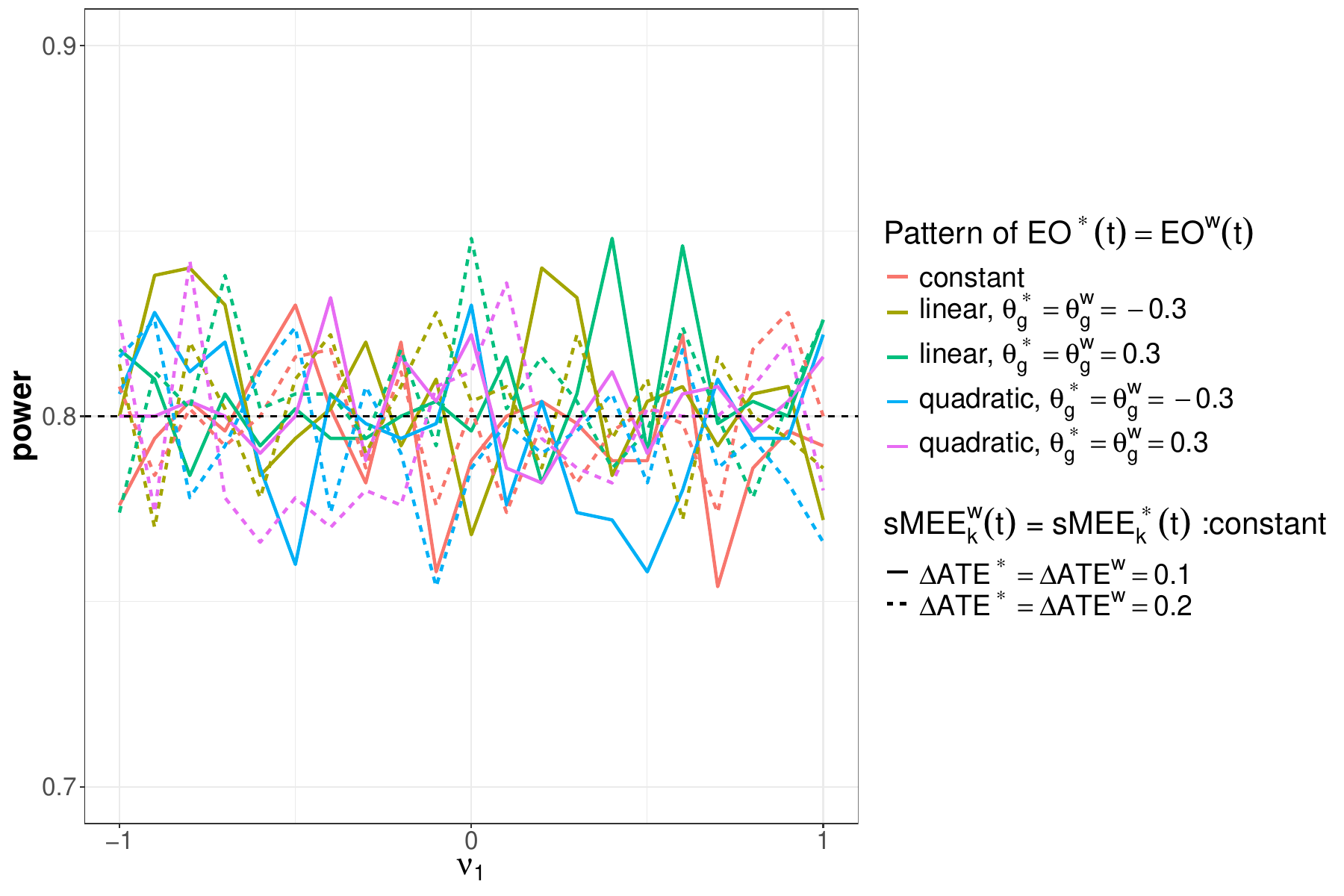}
    \caption{Power when (WA-e) is violated. A larger $|\nu_1|$ represents a larger magnitude of serial correlation in the proximal outcomes.}
    \label{fig:sc_viol}
\end{figure}

Using GM-SC, (WA-e) is violated as long as $\nu_1 \neq 0$. Recall that $\nu_1$ parameterizes the amount of serial correlation in the proximal outcomes. The power is always adequate regardless of the value of $\nu_1$ (\Cref{fig:sc_viol}).

\subsection{Power when (WA-f) is violated}
\label{subsec:simulation-violate-f}

\begin{figure}[htbp]
    \centering
    \includegraphics[width=1\textwidth]{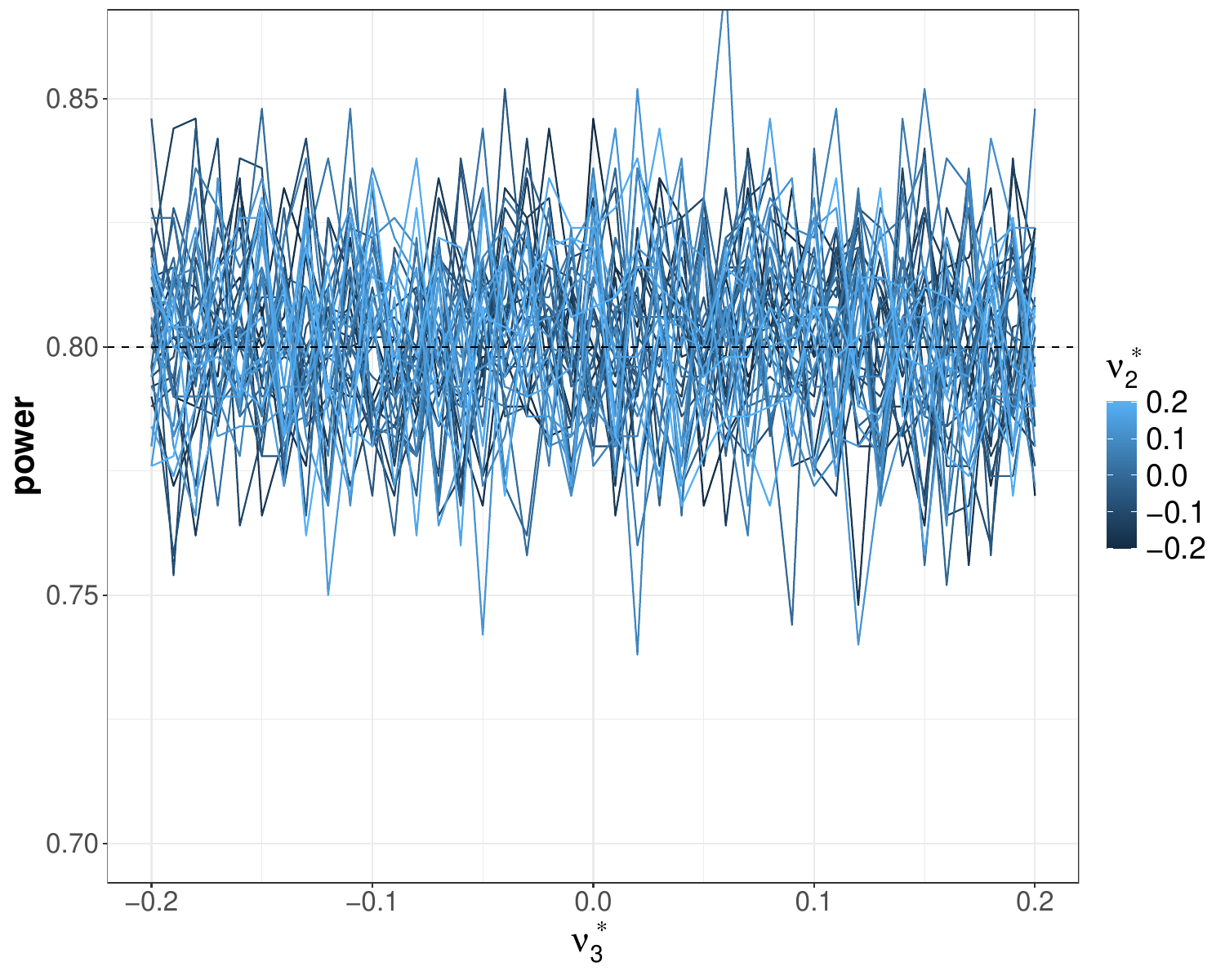}
    \caption{Power when (WA-f) is violated. A larger $|\nu_2|$ represents a larger effect of $A_{t-1}$ on $I_t$. A larger $|\nu_2|$ represents a larger dependence of $I_t$ on $Y_{t-1}$.}
    \label{fig:endo_avail}
\end{figure}

Using GM-EA, (WA-f) is violated as long as $\nu_2 \neq 0$ or $\nu_3 \neq 0$. Recall that $\nu_2$ parameterizes the effect of $A_{t-1}$ on $I_t$, and $\nu_3$ parameterizes the dependence of $I_t$ on $\epsilon_{t-1}$ (and in turn $Y_{t-1}$). The power is always adequate regardless of the value of $\nu_2$ and $\nu_3$ (\Cref{fig:endo_avail}).

\bibliography{references}
\bibliographystyle{plainnat}
\makeatletter\@input{xx.tex}\makeatother